\documentclass[prd,twocolumn,nofootinbib,aps,preprintnumbers,superscriptaddress,longbibliography]{revtex4-1}
\usepackage[dvipdfmx]{graphicx}
\usepackage{amssymb}
\usepackage{amsmath}  
\usepackage{eqnarray}            
\usepackage{mathtools}
\usepackage{cases}              
\usepackage{bm}
\usepackage{color}
\usepackage{url}
\usepackage{mathrsfs}

\usepackage{braket} % for braket
\usepackage{cancel} % for dirac slash

\allowdisplaybreaks

\newcommand{\fg}{\mathfrak{g}}
\newcommand{\h}{\hat}

\newcommand{\diagentry}[1]{\mathmakebox[1.8em]{#1}}
\newcommand{\xddots}{%
  \raise 4pt \hbox {.}
  \mkern 6mu
  \raise 1pt \hbox {.}
  \mkern 6mu
  \raise -2pt \hbox {.}
}

\newcommand\bovermat[2]{%
  \makebox[0pt][l]{$\smash{\overbrace{\phantom{%
    \begin{matrix}#2\end{matrix}}}^{\text{#1}}}$}#2}
    
\if0 
\left(
\begin{array}{ccc}
\\
\\
\end{array}
\right)
\fi 
\def\nn{\nonumber}

\def\l{\left}
\def\r{\right}
\def\be{\begin{equation}}
\def\ee{\end{equation}}
\def\bea{\begin{eqnarray}}
\def\eea{\end{eqnarray}}

\def\refheft{\cite{
Feruglio:1992wf,Burgess:1999ha,Giudice:2007fh,Grinstein:2007iv,Alonso:2012px,Buchalla:2012qq,Azatov:2012bz,Contino:2013kra,Jenkins:2013fya,Buchalla:2013rka,Alonso:2014rga,Guo:2015isa,Buchalla:2015qju,Alonso:2017tdy,Buchalla:2017jlu,Buchalla:2018yce}}
\def\refsmeft{\cite{Buchmuller:1985jz, Grzadkowski:2010es, Brivio:2017vri, Manohar:2018aog}}
\def\spa#1.#2{\left\langle#1\,#2\right\rangle} % angle braket
\def\spb#1.#2{\left[#1\,#2\right]} % box braket
\def\fsl#1{\setbox0=\hbox{$#1$}           % set a box for #1 
   \dimen0=\wd0                                 % and get its size
   \setbox1=\hbox{/} \dimen1=\wd1               % get size of /
   \ifdim\dimen0>\dimen1                        % #1 is bigger
      \rlap{\hbox to \dimen0{\hfil/\hfil}}      % so center / in box
      #1                                        % and print #1
   \else                                        % / is bigger
      \rlap{\hbox to \dimen1{\hfil$#1$\hfil}}   % so center #1
      /                                         % and print /
   \fi}                        
\begin{document}

\preprint{KYUSHU-HET-221}
\title{
Scalar and fermion on-shell amplitudes \\
in generalized Higgs effective field theory
}
\author{Ryo Nagai}
\email[E-mail: ]{nagai@het.phys.sci.osaka-u.ac.jp}
\affiliation{
Dipartimento di Fisica e Astronomia, Universita` degli Studi di Padova, Via Marzolo 8, 35131 Padova, Italy
}
\affiliation{
Istituto Nazionale di Fisica Nucleare (INFN), Sezione di Padova, Via Marzolo 8, 35131 Padova, Italy
}
\affiliation{
Department of Physics, 
Osaka University, Osaka 560-0043, Japan
}
\author{Masaharu Tanabashi}
\email[E-mail: ]{tanabash@eken.phys.nagoya-u.ac.jp}
\affiliation{
Department of Physics,
Nagoya University, Nagoya 464-8602, Japan
}
\affiliation{
Kobayashi-Maskawa Institute for the Origin of Particles and the
Universe, 
\\ Nagoya University, Nagoya 464-8602, Japan
}
\author{Koji Tsumura}
\email[E-mail: ]{tsumura.koji@phys.kyushu-u.ac.jp}
\affiliation{
Department of Physics, 
Kyushu University, 
744 Motooka, Nishi-ku, Fukuoka, 819-0395, Japan
}
\author{Yoshiki Uchida}
\email[E-mail: ]{uchida.yoshiki@phys.kyushu-u.ac.jp}
\affiliation{
Department of Physics,
Nagoya University, Nagoya 464-8602, Japan
}
\affiliation{
Department of Physics, 
Kyushu University, 
744 Motooka, Nishi-ku, Fukuoka, 819-0395, Japan
}
\date{\today}

\begin{abstract} 
Beyond standard model (BSM) particles should be included in effective field theory
in order to compute the scattering amplitudes involving these 
extra particles.
We formulate an extension of Higgs effective field theory which contains 
arbitrary number of scalar and fermion fields with arbitrary 
electric  and chromoelectric charges. 
The 
BSM Higgs sector is described by using the non-linear sigma model 
in a manner consistent with the spontaneous electroweak symmetry breaking. 
The chiral order counting rule is arranged consistently 
with the loop expansion.
The leading order Lagrangian is organized in accord with the chiral order
counting rule.
We use a geometrical language to describe the particle interactions.
The parametrization redundancy in the effective Lagrangian is resolved
by describing the on-shell scattering amplitudes only with the 
covariant quantities in the scalar/fermion field space.
We introduce a 
useful coordinate (normal coordinate), 
which simplifies the computations of the on-shell amplitudes significantly.
We show the high energy behaviors of the scattering amplitudes 
determine the ``curvature tensors'' in the scalar/fermion field
space.
The massive spinor-wavefunction formalism is shown to be useful 
in the computations of on-shell helicity amplitudes.

\end{abstract}

\maketitle
\newpage
\section{Introduction}

Four seemingly independent fundamental energy scales we know in the
elementary particle physics, 
the Planck scale $\simeq 1.2\times 10^{19}\, \mbox{GeV}$
(energy scale of gravitational interaction), 
the cosmological constant $\simeq (2.2\, \mbox{meV})^4$
(accelerated expansion of the universe), 
the weak scale $v\simeq 246\, \mbox{GeV}$
(masses of elementary particles), 
and the QCD scale $\simeq 300\, \mbox{MeV}$
(masses of hadrons).

Among these four known fundamental energy scales, 
the most well understood one is the QCD scale.
High energy hadronic particle collisions much above the QCD scale 
can be successfully investigated perturbatively, 
while the low energy hadron physics can be described in terms
of low energy effective field theories.
The QCD scale is generated dynamically through the
dimensional transmutation mechanism in the 
$SU(3)$ QCD gauge dynamics.
Since the scale generation is forbidden at the classical level, 
the QCD scale is stable against the quantum loop corrections.
Moreover, the global symmetry structure of QCD allows us to 
develop systematic expansions in these effective field theories.
Especially, the low energy pion physics can be described
in terms of the chiral perturbation theory~\cite{Weinberg:1978kz,Gasser:1983yg,Gasser:1984gg,Georgi:1994qn,Donoghue:1992dd}, 
in which
the pions are treated as pseudo-Nambu-Goldstone bosons
associated with the spontaneous breaking of the global chiral 
symmetry.
The low energy theorems in the pion scattering amplitudes are 
reproduced in the chiral perturbation theory at its leading order. 
It is also possible to include higher order corrections in a
systematic manner, by computing the quantum loop corrections
and by introducing higher order terms in the effective 
chiral Lagrangian arranged in accord with the chiral order 
counting rules.
Although the chiral perturbation does not converge above the 
resonance mass energy scale, the situation can be improved
by explicitly introducing the resonances such as the spin-1 
$\rho$ 
meson in the effective chiral Lagrangian~\cite{Bando:1984ej,Bando:1987br,Ecker:1988te}.
Actually, it is possible to formulate the chiral perturbation
theory even in the effective chiral Lagrangian including
the $\rho$ meson~\cite{Tanabashi:1993np,Harada:1992np,Harada:2003jx}.

On the other hand, 
the standard model (SM) of particle physics provides a 
consistent gauge theory framework to describe the physics associated 
with the weak scale.
A Higgs field and Higgs potential are introduced in the SM to generate 
the weak scale.
It has been shown that the 125GeV scalar particle discovered
at the LHC experiments can be identified successfully as a Higgs
particle associated with the SM Higgs field~\cite{LHCHIGGSXSWG}.
Unlike the QCD scale generation mechanism, however,
in the SM, the Higgs potential responsible for the weak scale 
is not classically forbidden
and the weak scale is subject to  huge quantum loop corrections.
Fine-tuning of parameters is required in the SM to explain the 
smallness of the weak scale compared with the Planck scale (naturalness
problem).
It is a common belief that the certain beyond-standard-model (BSM) 
new physics exist not far above the weak scale, guaranteeing the 
naturalness of the weak scale.
Unfortunately, however, it turned out that the current collider energy 
is not enough high to reveal the nature of the BSM physics.
Currently, we have no direct collider physics evidences supporting the 
existence of the BSM physics.
As compared with the QCD scale, current understanding 
of the weak scale physics is highly restricted in this sense, 
mainly due to
the lack of our knowledge on the physics far above the weak scale.

It should be useful if we establish
a weak scale-analogue 
to the chiral perturbation theory.
Assuming BSM is weakly interacting, 
the standard-model 
effective-field-theory (SMEFT)~\cite{Buchmuller:1985jz} along with recent reviews \cite{Brivio:2017vri, Manohar:2018aog}, in which 
the electroweak symmetry $SU(2)\times U(1)$ is realized linearly, 
can be used for such a purpose.
The SMEFT cannot be applied, however, for strongly interacting BSM,
in which heavy BSM particles do not decouple from the low energy
physics.
For strongly interacting BSM, 
we can use the Higgs effective-field-theory (HEFT)~\cite{Feruglio:1992wf,Burgess:1999ha,Giudice:2007fh,Grinstein:2007iv,Alonso:2012px,Buchalla:2012qq,Azatov:2012bz,Contino:2013kra,Jenkins:2013fya,Buchalla:2013rka,Alonso:2014rga,Guo:2015isa,Buchalla:2015qju,Alonso:2017tdy,Buchalla:2017jlu,Buchalla:2018yce},
in which 
the electroweak symmetry $SU(2)\times U(1)$ is realized nonlinearly.

Weak scale-analogues to the resonance chiral perturbation theory 
have also been studied.
Phenomenologies of the weak scale analogues to the spin-1 $\rho$ 
resonance have been investigated by using the resonance electroweak
chiral Lagrangian 
techniques~\cite{Casalbuoni:1985kq,Pappadopulo:2014qza,Abe:2015jra,Abe:2016fjs,Sanz-Cillero:2020szj}.
We have proposed the generalized Higgs-effective-field-theory (GHEFT) 
framework~\cite{Nagai:2019tgi}, 
in which arbitrary number of spin-0 resonances/particles 
with arbitrary electric 
charges 
are introduced in the HEFT Lagrangian.
In order to explain the naturalness of the weak scale, 
it is tempting to consider BSM scenarios having larger global symmetry
and thus with extra pseudo-Nambu-Goldstone particles.
Actually, in composite Nambu-Goldstone Higgs models~\cite{Panico:2015jxa},
global symmetries larger than the SM gauge group are introduced.
There exist extra pseudo-Nambu-Goldstone particles in non-minimal
composite Higgs models~\cite{Gripaios:2009pe, Mrazek:2011iu, Bertuzzo:2012ya, Frigerio:2012uc, DeCurtis:2018iqd, DeCurtis:2018zvh,DeCurtis:2019rxl}. 
We emphasize that 
these extra pseudo-Nambu-Goldstone particles can 
be successfully investigated 
in the GHEFT framework.
We here stress the difference between the weak scale $v\simeq 246$GeV
and the compositeness scale $\Lambda \simeq 4\pi f$ in the composite
Higgs scenarios.
It is known that $f$ needs to be several times larger than the weak
scale in order to keep the consistency with electroweak and flavor
precision constraints.
In the GHEFT framework, $\Lambda \simeq 4\pi f$ can be regarded as the scale
of perturbative unitarity violations, which can be pushed up to high 
energy scale independently of $v$ enough to meet these phenomenological
constraints. 
%, as we will shown in \S.~\ref{sec:amplitude}.
Although the electroweak symmetry is realized nonlinearly in our GHEFT
Lagrangian, our theory should be regarded to be valid below the compositeness
scale $4\pi f$ keeping the perturbative unitarity.
This fact motivates us to introduce TeV scale resonances in the GHEFT 
framework.

There is a difficulty in the studies of the 
effective field theories,
{\it{i.e.}},
non-uniqueness of its
parametrization methods.
Kamefuchi-O'Raifeartaigh-Salam (KOS) theorem~\cite{Kamefuchi:1961sb} 
tells us that there exist
equivalent class of seemingly different effective field theories
which describe the same physics.
As the KOS theorem suggests us, 
there exist many equivalent
formulation of the effective theories connected with each other 
through the changes of effective field variables (coordinates).
This makes us difficult to compare results computed 
in an effective field theory with results in seemingly different, 
but equivalent field theory which may be generated more directly 
from the UV physics.
Warsaw basis \cite{Grzadkowski:2010es} is often assumed to resolve the non-uniqueness 
in the SMEFT.
The Warsaw basis should be understood to be a 
symptomatic treatment
effective only at the lowest order, however.
It does not provide a systematic prescription to fix the issue 
beyond the leading oder.

The same problem exists in the electroweak resonance chiral perturbation
theories.
The existing studies of the electroweak resonance chiral perturbation
theories rely on particular field parameterizations.

In our previous paper on GHEFT~\cite{Nagai:2019tgi},
we have shown that the GHEFT  (electroweak resonance chiral 
perturbation theory) can be described by using the 
covariant tensors of the scalar manifolds, which allows us to parametrize
the particle scattering amplitudes and the quantum corrections
in a covariant manner under the changes of effective field
variables (coordinates) \cite{Alonso:2015fsp, Alonso:2016oah}.
It has been shown that the uses of the normal coordinate
simplify the computation of the scattering amplitudes significantly.
We have then shown that, once the perturbative unitarity at the
tree level is ensured, then one-loop finiteness is automatically 
guaranteed in the GHEFT framework.

There remains an issue we need to investigate in the electroweak
resonance chiral perturbation theory analysis.
As far as we know,
there is no studies on the electroweak resonance chiral perturbation
theory including fermionic spin-1/2 particles strongly coupled with 
the Higgs sector.
It should be emphasized, however, that the SM Higgs particle
couples with top-quark (fermion) most strongly.
The existence of BSM spin-1/2 particles is widely expected
in BSM models explaining the naturalness of the weak scale.
Actually, in the composite Higgs models, top-quark partner 
fermion is usually introduced to explain the mass of the top
quark. 

In the present paper, 
we generalize the findings we made in our previous paper to include
fermionic heavy particles in the GHEFT Lagrangian.
The particle scattering amplitudes are expressed using covariant
quantities of the bosonic and fermionic field coordinates.
The scattering amplitude formulas given in this paper can therefore
be easily compared with the formulas computed in other equivalent
formulation of the fermionic resonance electroweak chiral perturbation
theories.

This paper is organized as follows: In \S.~\ref{sec:GHEFT}, we introduce
extended GHEFT Lagrangian including the extra spin-1/2 fermionic
particles. We then provide a chiral order counting rule which 
allows us to perform a systematic expansion in the computation
of the scattering amplitudes in a manner similar to the well-known
chiral perturbation theory.
In \S.~\ref{sec:NC}, the normal coordinate technique is generalized to
include fermionic field coordinates.
We investigate tree-level spin-0 and spin-1/2 particles' 
scattering amplitudes in \S.~\ref{sec:amplitude}, applying the normal coordinate
technique.  
It is shown that these scattering amplitudes can be expressed
in terms of the covariant quantities of the GHEFT field manifold.
We conclude in \S.~\ref{sec:summary}.
A quick review on HEFT is given in Appendix~\ref{app:HEFT}.
Notation on the helicity eigenstate wavefunctions is summarized in
Appendix~\ref{app:helicitystate}.
Appendix~\ref{sec:higerod-nc} is for the explicit computations 
of higher order coefficients in the normal coordinate expansion,
and a proof of Bianchi-identity.

%%%%%%%%%%%%%%%%%%%%%%%
\section{Generalized Higgs Effective Field Theory}
\label{sec:GHEFT}
We need to incorporate new BSM particles in effective field theories 
(EFTs)
so as to compute production cross sections and decay widths involving 
these new BSM particles.
These new particles are not included in minimal EFTs
such as the standard-model effective-field-theory (SMEFT) \refsmeft\, and the 
Higgs effective-field-theory (HEFT) \refheft, however.
We have proposed, in our previous paper~\cite{Nagai:2019tgi},
the generalized Higgs-effective-field-theory (GHEFT) framework
in which arbitrary number of spin-0 resonances/particles 
with arbitrary electric charges are introduced.
In this section we further generalize our GHEFT framework to
incorporate BSM spin-1/2 fermions, as well as the 125\,GeV Higgs boson, 
BSM scalar particles, quarks and leptons.

The electroweak gauge symmetry, $G=SU(2)_W\times U(1)_Y$, is spontaneously broken to the electromagnetic $H=U(1)_{\rm em}$ at the electroweak symmetry breaking 
(EWSB) scale.
If the EWSB is triggered by new strong dynamics in BSM, 
the spontaneously broken symmetry $G$ should be realized non-linearly at 
the low-energy scale. 
Electroweakly charged particles, in such a case, 
transform non-linearly under the electroweak gauge symmetry.
We use 
the celebrated Callan-Coleman-Wess-Zumino (CCWZ) formalism~\cite{Coleman:1969sm,Callan:1969sn,Bando:1987br} 
to formulate the low-energy EFT Lagrangian in a manner 
consistent with the EWSB.
We note here that the CCWZ formalism can also be applied even if
the electroweak symmetry is broken by perturbative dynamics.

We then provide a chiral order counting rule in GHEFT which 
allows us to perform a systematic expansion in the computation
of the scattering amplitudes in a manner similar to the well-known
chiral perturbation theory~\cite{Weinberg:1978kz,Gasser:1983yg,Gasser:1984gg,Georgi:1994qn,Donoghue:1992dd}.

%%%%%%%%%%%%%%%%%%%%%%%
\subsection{Leading order GHEFT Lagrangian}
We start the discussion in the gaugeless limit ($g_W=g_Y=0$) for simplicity. 
The couplings with the SM gauge fields will be introduced at the end of this subsection.
The minimal EFT for strongly interacting EWSB is described by 
HEFT Lagrangian~\refheft{} in the gaugeless limit, 
\begin{align}
  \mathcal{L}_{\rm{HEFT}} &= \mathcal{L}_{\rm{HEFT, boson}}
  + \mathcal{L}_{\rm{HEFT, fermion}}
\label{eq:LHEFT2}
\end{align}
with the bosonic sector Lagrangian $\mathcal{L}_{\rm{HEFT,boson}}$ being
\bea
\mathcal{L}_{\rm{HEFT, boson}}
&=&
G(h)\,
\mbox{tr}[\partial^\mu U^\dag \partial_\mu U]
\nn\\
&+&
\frac{1}{2}G_Z(h)\,
\mbox{tr}[ U^\dag \partial^\mu U\tau^3]
\mbox{tr}[ U^\dag \partial_\mu U\tau^3]
\nn\\
&+&
\frac{1}{2}(\partial_\mu h)(\partial^\mu h)
-V(h)
\, . 
\label{eq:LHEFTboson}
\eea
The fermionic sector Lagrangian $\mathcal{L}_{\rm{HEFT, fermion}}$ is
given in appendix~\ref{app:HEFT}.
The bosonic sector HEFT Lagrangian~(\ref{eq:LHEFTboson})
should be regarded as a starting point of the GHEFT 
framework~\cite{Nagai:2019tgi}.
The 125\, GeV Higgs boson field is denoted by $h$, while 
$U$ is an exponential function of the Nambu-Goldstone (NG) boson 
fields, 
\be
U=
\xi_W\xi_Y
\,,
\ee
where
\bea
\xi_W(x)
&=&
\exp\l(i\sum_{a=1,2}\pi^a(x)\frac{\tau^a}{2}\r)
\,,\label{eq:xiW}\\
\xi_Y(x)
&=&
\exp\l(i\pi^3(x)\frac{\tau^3}{2}\r)
\,,\label{eq:xiY}
\eea
with $\tau^a$ and $\pi^a~(a=1,2,3)$ being the Pauli spin matrices and the NG boson fields. $G(h)$, $G_Z(h)$ and $V(h)$ are arbitrary functions of $h$,
which determine 
the interactions among 
the 125\, GeV Higgs field and the NG boson fields.
Custodial symmetry implies $G_Z(h)=0$.
We here do not impose $G_Z(h)=0$, however, to keep the generality.

For later convenience, 
we rewrite the HEFT Lagrangian (\ref{eq:LHEFTboson}) in terms of the 
CCWZ formalism, {\it{i.e.}}, by using the $G/H$ Lie-algebra 
valued Maurer-Cartan (MC) one-forms of the NG boson fields, 
\be
\alpha^a_{\perp\mu}
=
\mbox{tr}\biggl[
\frac{1}{i}\xi^\dag_W(\partial_\mu \xi_W)\tau^a
\biggr]
\,,~~~(a=1,2)
\,,
\ee
and
\be
\alpha^3_{\perp\mu}
=
\mbox{tr}\biggl[
\frac{1}{i}\xi^\dag_W(\partial_\mu \xi_W)\tau^3
\biggr]
+
\mbox{tr}\biggl[
\frac{1}{i}(\partial_\mu \xi_Y)\xi^\dag_Y\tau^3
\biggr]
\,.
\ee
The HEFT Lagrangian (\ref{eq:LHEFTboson}) is expressed as
\be
\mathcal{L}_{\rm{HEFT, boson}}
=
\frac{1}{2}{G}_{ab}(h)\alpha^a_{\perp\mu}\alpha^{b\mu}_{\perp}
+
\frac{1}{2}(\partial^\mu h)(\partial_\mu h)
-V(h)
\,,
\label{eq:LHEFTboson2}
\ee
where $G_{11}(h)=G_{22}(h)=G(h)$, $G_{33}(h)=G(h)+G_Z(h)$ and 
$G_{ab}(h)=0$ for $a\neq b$.
In the expression (\ref{eq:LHEFTboson2}) and hereafter,
summation $\displaystyle \sum_{a=1,2,3}$ is implied whenever an 
index $a$ is repeated in a product.

The CCWZ formalism allows us to systematically introduce 
BSM extra scalar particles in the low energy EFT.
Here we introduce extra $(n_R-1)$ BSM real 
scalars and $n_C$ BSM complex scalars 
in addition to the 125\,GeV Higgs boson. 
Therefore there are $n_s=n_R+2n_C$ real scalars in total. 
It is convenient to introduce a real scalar multiplet 
$\phi^I~(I=1,2,\cdots,n_s)$ as
\bea
\phi^I
&=&
(\,
\overbrace{\phi^1,\phi^2,\cdots,\phi^{n_N}}^{
n_R
},
\overbrace{\phi^{n_N+1}\,\cdots\,\phi^{n_s}}^{
2n_C
}
\,)
\,,
\eea
where we identify $\phi^1$ as the 125\,GeV Higgs boson, $\phi^1=h$. The $H=U(1)_{\rm{em}}$ transformation for the scalar multiplet is defined as
\be
\phi^I
\xrightarrow{H}
\biggl[\exp\biggl(iQ_\phi \theta_h\biggr)\biggr]^I_{\, \, \, J}
\phi^J
\,,
\ee
where $\theta_h$ is a real constant parameter, and the $(n_s\times n_s)$ matrix $Q_\phi$ is defined as
\begin{align}
Q_\phi=
\left(
\begin{array}{cccccc}
\bovermat{$n_R$}
{\diagentry{0} & 
\phantom{\diagentry{\ddots}} & 
\phantom{\diagentry{0}}}
&
\bovermat{$2n_C$}
{\phantom{\diagentry{-q_{1}\sigma_2}} & 
\phantom{\diagentry{\ddots}} & 
\phantom{\diagentry{-q_{1}\sigma_2}}}
\\
 & \diagentry{\ddots} &  &  &  & \\
 &  & \diagentry{0} &  &  & \\
 &  &  & \diagentry{-q_{1}\sigma_2} &  & \\
 &  &  &  & \diagentry{\ddots} & \\
 &  &  &  &  & \diagentry{-q_{n_C}\sigma_2\qquad}
\end{array}
\right)\,.
\end{align}
Here $\sigma_2=\tau^2$ and $q_i~(i=1,2,\cdots n_C)$ denotes the $U(1)_{\rm{em}}$ charges of the scalar fields. The $G=SU(2)_W\times U(1)_Y$ transformation of $\phi^I$ is given by
\be
\phi^I
\xrightarrow{G}
[\rho_\phi]^I_{~J}
\phi^J
\,,
~~~
\rho_\phi
=
\exp\biggl(iQ_\phi \theta_h(\pi,\fg_W,\fg_Y)\biggr)
\,,
\ee
where $\theta_h$ is a real function of group elements $\fg_W\in SU(2)_W$, $\fg_Y\in U(1)_Y$, and the NG boson fields ($\pi^a$). 
There may exist $SU(3)_C$ colored scalar particles such 
as the leptoquark scalars and the colored superpartner bosonic particles.
The flavor index $I$, $J$ are understood to include the color index
for these colored bosons.
It is straightforward to write down the $SU(3)_C$ transformation matrix 
for $\phi^I$.

Since the $G$ transformation matrix $\rho_\phi$ depends on the NG boson fields, the derivative of the scalar multiplet $\partial_\mu\phi^I$ transforms nonhomogeneously under $G$,
\be
\partial_\mu \phi^I
\xrightarrow{G}
[\rho_\phi]^I_{~J} (\partial_\mu \phi^J)
+
(\partial_\mu \rho_\phi )^I_{~J}\phi^J
\,.
\ee
Therefore, if $\phi^I$ contains the charged scalar (namely $\rho_\phi\neq{\bf{1}}$), the 
kinetic operator $(\partial_\mu\phi^I)(\partial^\mu\phi^I)$ is not invariant under the $G$-transformation. 
$G$-invariant kinetic terms for the charged scalar fields are formulated by introducing the covariant derivative on the $G/H$ coset space. 
The covariant derivative is defined as
\be
\mathcal{D}_\mu \phi^I
:=
\partial_\mu \phi^I
+i\,\mathcal{V}^3_\mu \,[Q_\phi]^I_{~J}\phi^J
\,,
\ee
where
\be
\mathcal{V}^3_\mu
:=
-\mbox{tr}
\biggl[
\frac{1}{i}(\partial_\mu \xi_Y)\xi^\dag_Y\tau^3
\biggr]
+
c\,\alpha^3_{\perp\mu}
\,,
\label{eq:covdel}
\ee 
with $c$ being an arbitrary constant. The $\mathcal{V}^3_\mu$ corresponds to the $H$ Lie-algebra valued MC one-form, which plays the role of the connection field on the $G/H$ coset space. It is straightforward to show that the covariant derivative $\mathcal{D}_\mu\phi^I$ homogeneously transforms under $G$,
\be
\mathcal{D}_\mu \phi^I
\xrightarrow{G}
[\rho_\phi]^I_{~J} (\mathcal{D}_\mu \phi^J)
\,.
\ee
The ``covariant'' kinetic term $(\mathcal{D}_\mu \phi^I)(\mathcal{D}^\mu \phi^I)$ respects the $G$-invariance. 
 
Using the $G$-covariant objects $\alpha^a_{\perp\mu}$, $\phi^I$, 
and $\mathcal{D}_\mu\phi^I$, 
we can systematically write down $G$ invariant Lagrangians.
As we will see later, 
the lowest order Lagrangian 
is written as \cite{Nagai:2019tgi}
\bea
\mathcal{L}_{\rm{GHEFT, boson}}
&=&
\frac{1}{2}G_{ab}(\phi)\alpha^a_{\perp\mu}\alpha^{b\mu}_{\perp}
\nn\\
&+&
G_{aI}(\phi)\alpha^a_{\perp\mu}(\mathcal{D}^\mu \phi^I)
\nn\\
&+&
\frac{1}{2}G_{IJ}(\phi)(\mathcal{D}^\mu \phi^I)(\mathcal{D}_\mu \phi^J)
\nn\\
&-&
V(\phi)
\, .
\label{eq:LGHEFT-boson}
\eea
$G_{ab}$, $G_{aI}$, $G_{IJ}$, and $V$ are functions of the scalar fields $\phi^I$, which homogeneously transform under the $G$ transformation. 
These functions determine the interactions 
among the scalar fields. 
Again, we do not impose the custodial symmetry in (\ref{eq:LGHEFT-boson}) 
to keep the generality. 
Once we specify the ultraviolet completion of the EFT, $G_{ab}$, $G_{aI}$, $G_{IJ}$, and $V$ are determined up to the uncertainty associated with the field redefinition.

We next discuss the fermion sector. 
We need to introduce, at least, SM quarks and leptons in our EFT framework. 
Moreover, 
the existence of BSM spin-1/2 particles is widely expected
in BSM models explaining the naturalness of the weak scale.
For this purpose, we incorporate $\hat{n}_M$ Majorana fermions and 
$\hat{n}_D$ Dirac fermions in the EFT Lagrangian (\ref{eq:LGHEFT-boson}). 
We describe these fermions by using $\hat{n}_f=\hat{n}_M+2\hat{n}_D$ 
two-component spinor 
fields $\psi^{\hat{i}}_\alpha~({\hat{i}}={\hat{1}},\cdots,{\hat{n}}_f)$,
\bea
\psi^{\hat{i}}_\alpha
&=&
(\,
\overbrace{\psi^{\hat{1}}_\alpha,\psi^{\hat{2}}_\alpha,\cdots,\psi^{\hat{n}_M}_\alpha}^{
\hat{n}_M
},
\overbrace{\psi^{\hat{n}_M+1}_\alpha ~~ \,\cdots\,\psi^{\hat{n}_f}_\alpha}^{2\hat{n}_D}
\,)
\,,
\label{eq:psimultiplet}
\eea
where $\alpha$ is the spinor index which takes 1 or 2. 
The Hermitian conjugate of $\psi^{\h{i}}_\alpha$ is denoted as 
$\psi^{\dag {\h{i}}^*}_{\dot{\alpha}}:=(\psi^{\h{i}}_\alpha)^\dag$.
The $U(1)_{\rm{em}}$ transformation for the fermion multiplet (\ref{eq:psimultiplet}) is defined as
\be
\psi^{\hat{i}}_\alpha
\xrightarrow{H}
\biggl[\exp\biggl(iQ_\psi \theta_h\biggr)\biggr]^{\hat{i}}_{~{\hat{j}}}
\psi^{\hat{j}}_\alpha
\,,
\ee
where $\theta_h$ is a real constant parameter, and the $(\hat{n}_f\times \hat{n}_f)$ matrix $Q_\psi$ is defined as
\begin{align}
Q_\psi=
\left(
\begin{array}{cccccc}
\bovermat{$\h{n}_M$}
{\diagentry{0} & 
\phantom{\diagentry{\ddots}} & 
\phantom{\diagentry{0}}}
&
\bovermat{$2\h{n}_D$}
{\phantom{\diagentry{q_{\h{1}}\sigma_3}} & 
\phantom{\diagentry{\ddots}} & 
\phantom{\diagentry{q_{\h{1}}\sigma_3}}}
\\
 & \diagentry{\ddots} &  &  &  & \\
 &  & \diagentry{0} &  &  & \\
 &  &  & \diagentry{q_{\h{1}}\sigma_3} &  & \\
 &  &  &  & \diagentry{\ddots} & \\
 &  &  &  &  & \diagentry{q_{\h{n}_D}\sigma_3\quad}
\end{array}
\right)\,.
\end{align}
Here $\sigma_3=\tau^3$ and $q_{\hat{i}}~({\hat{i}}={\hat{1}},{\hat{2}},\cdots {\hat{n}}_D)$ denotes the $U(1)_{\rm{em}}$ charges of the fermion fields. 
There certainly exist $SU(3)_C$ colored spin-1/2 particles.
The flavor index $\hat{i}$, $\hat{j}$ are understood to include the color index
for these colored fermions.
It is straightforward to write down the $SU(3)_C$ transformation matrix 
for $\psi^{\hat{i}}_\alpha$.

The $G=SU(2)_W\times U(1)_Y$ transformation of $\psi^{\hat{i}}_\alpha$ is given by
\be
\psi^{\hat{i}}_\alpha
\xrightarrow{G}
[\rho_\psi]^{\hat{i}}_{~{\hat{j}}}
\psi^{\hat{j}}_\alpha
\,,
~~~
\rho_\psi
=
\exp\biggl(iQ_\psi \theta_h(\pi,\fg_W,\fg_Y)\biggr)
\,,
\label{eq:fermiontr}
\ee
where $\theta_h$ is a real function of group elements $\fg_W\in SU(2)_W$, $\fg_Y\in U(1)_Y$, and the NG boson fields ($\pi^a$). We note that the derivative of the fermion field, $\partial_\mu \psi^{\h{i}}$, nonhomogeneously transforms under the $G$ transformation as the derivative of the scalar field $\partial_\mu\phi^I$ does. 
The covariant derivative can be defined as
\be
\mathcal{D}_\mu\psi^{\h{i}}_\alpha
:=
\partial_\mu \psi^{\h{i}}_\alpha
+i\,\mathcal{V}^3_\mu \,[Q_\psi]^{\h{i}}_{~{\h{j}}}\psi^{\h{j}}_\alpha
\,,
\ee
where the connection field $\mathcal{V}^3_\mu$ is defined in Eq.\,(\ref{eq:covdel}). It is easy to show that
the covariant derivative $\mathcal{D}_\mu \psi^{\h{i}}$ transforms homogeneously under the $G$ transformation,
\be
\mathcal{D}_\mu \psi^{\h{i}}_\alpha
\xrightarrow{G}
[\rho_\psi]^{\hat{i}}_{~{\hat{j}}}\,
(\mathcal{D}_\mu \psi^{\h{j}}_\alpha)\,.
\label{eq:Dfermiontr}
\ee

It is now straightforward to construct $G$-invariant Lagrangian for the scalar and fermion fields. We can systematically construct $G$-invariant operators by using the $G$-covariant objects, $\alpha^a_{\perp\mu}$, $\phi^I$, $\psi^{\h{i}}$, $\mathcal{D}_\mu\phi^I$, and $\mathcal{D}_\mu\psi^{\h{i}}$. 
Applying the chiral order counting rule which we will introduce in the next subsection \ref{sec:powercounting}, we write down the leading order Lagrangian of GHEFT as
\bea
\mathcal{L}_{\rm{GHEFT}}
&=&
\frac{1}{2}G_{ab}(\phi)\,\alpha^a_{\perp\mu}\,\alpha^{a\mu}_{\perp}
\nn\\
&+&
G_{aI}(\phi)\,\alpha^a_{\perp\mu}\,(\mathcal{D}^\mu \phi^I)
\nn\\
&+&
\frac{1}{2}G_{IJ}(\phi)(\mathcal{D}^\mu \phi^I)(\mathcal{D}_\mu \phi^J)
-V(\phi)
\nn\\
&+&\frac{i}{2}G_{{\h{i}}{\h{j}}^*}(\phi)
\biggl(
\psi^{\dag {\h{j}}^*}\,\bar{\sigma}^\mu \,(\mathcal{D}_\mu \psi^{\h{i}})
-
(\mathcal{D}_\mu\psi^{\dag {\h{j}}^*})\,\bar{\sigma}^\mu \, \psi^{\h{i}}
\biggr)
\nn\\
&+&V_{{\h{i}}{\h{j}}^*a}(\phi)
\,
\psi^{\dag {\h{j}}^*}\,
\bar{\sigma}^\mu \, 
\psi^{{\h{i}}} \,
\alpha^a_{\perp\mu} 
\nn\\
&+&
V_{{\h{i}}{\h{j}}^*I}(\phi)
\,
\psi^{\dag {\h{j}}^*}\,
\bar{\sigma}^\mu  \,
\psi^{{\h{i}}} \,
(\mathcal{D}_\mu \phi^I) 
\nn\\
&-&\frac{1}{2}M_{{\h{i}}{\h{j}}}(\phi)\,\psi^{\h{i}}\psi^{\h{j}}
-\frac{1}{2}M_{{\h{i}}^*{\h{j}}^*}(\phi)\,\psi^{\dag {\h{i}}^*}\psi^{\dag {\h{j}}^*}
\nn\\
&+&
\frac{1}{8}
S_{{\h{i}}{\h{j}}{\h{k}}{\h{l}}}(\phi)
(\psi^{{\h{i}}}\psi^{{\h{j}}})(\psi^{{\h{k}}}\psi^{{\h{l}}})
\nn\\
&+&
\frac{1}{8}
S_{{\h{i}}^*{\h{j}}^*{\h{k}}^*{\h{l}}^*}(\phi)
(\psi^{\dag {\h{i}}^*}\psi^{\dag {\h{j}}^*})
(\psi^{\dag {\h{k}}^*}\psi^{\dag {\h{l}}^*})
\nn\\
&+&
\frac{1}{4}
S_{{\h{i}}{\h{j}}{\h{k}}^*{\h{l}}^*}(\phi)
(\psi^{{\h{i}}}\psi^{{\h{j}}})
(\psi^{\dag {\h{k}}^*}\psi^{\dag {\h{l}}^*})
\,,
\label{eq:LGHEFT-sym}
\eea
where we use the spinor-index free notation for the fermion bilinear operators \cite{Dreiner:2008tw}, {\it{{\it{i.e.}}}}, 
\begin{align}
  (\psi^{\h{i}} \psi^{\h{j}}) 
  &:= \varepsilon^{\alpha \beta} \, \psi_{\beta}^{{\h{i}}} \, \psi_{\alpha}^{{\h{j}}} \, , 
  \\
  (\psi^{\dagger {\h{i}}^*} \psi^{\dagger {\h{j}}^*} )
  &:= \psi_{\dot{\alpha}}^{\dagger {\h{i}}^*} \, \varepsilon^{\dot{\alpha} \dot{\beta}} \, 
      \psi_{\dot{\beta}}^{\dagger {\h{j}}^*} \, , 
  \\
 \psi^{\dagger {\h{j}}^*} \, \bar{\sigma}^\mu \, \psi^{{\h{i}}}
  &:= \psi_{\dot{\alpha}}^{\dagger {\h{j}}^*} \, 
      (\bar{\sigma}^\mu)^{\dot{\alpha}\alpha}\, \psi_{\alpha}^{{\h{i}}} 
      \,,
  \\
  \psi^{\hat{i}} \sigma^{\mu\nu} \psi^{\hat{j}}
  &:= \varepsilon^{\gamma\alpha} \psi^{\hat{i}}_\alpha 
      (\sigma^{\mu\nu})_{\gamma}{}^{\beta} \psi^{\hat{j}}_\beta \, , 
  \\
  \psi^{\dagger \hat{i}^*} \bar{\sigma}^{\mu\nu} \psi^{\dagger \hat{j}^*}
  &:= \psi^{\dagger \hat{i}^*}_{\dot{\alpha}}
      (\bar{\sigma}^{\mu\nu})^{\dot{\alpha}}{}_{\dot{\beta}} 
     \varepsilon^{\dot{\beta}\dot{\gamma}}
     \psi^{\dagger \hat{j}^*}_{\dot{\gamma}} \, , 
\end{align}
with $\varepsilon^{12} = -\varepsilon^{21} = -\varepsilon_{12} = \varepsilon_{21} = 1$. 
The spinor matrices 
$\bar{\sigma}^\mu$, $\sigma^\mu$, $\sigma^{\mu\nu}$, $\bar{\sigma}^{\mu\nu}$ are 
defined as 
 \begin{align}
 (\bar{\sigma}^\mu)^{\dot{\alpha} \alpha}
  &:=({\bf{1}}^{\dot{\alpha}\alpha},
-(\sigma^a)^{\dot{\alpha}\alpha} ) \, , 
\\
  (\sigma^\mu)_{\alpha\dot{\alpha}}
  &:= \varepsilon_{\alpha\beta}
     \varepsilon_{\dot{\alpha}\dot{\beta}}
     (\bar{\sigma}^\mu)^{{\dot{\beta}\beta}} \, , 
\\
  (\sigma^{\mu\nu})_{\alpha}{}^\beta
  &:= \dfrac{i}{4} \left[
    (\sigma^\mu)_{\alpha\dot{\gamma}}
    (\bar{\sigma}^\nu)^{\dot{\gamma}\beta}
   -(\sigma^\nu)_{\alpha\dot{\gamma}}
    (\bar{\sigma}^\mu)^{\dot{\gamma}\beta}
   \right] \, , 
\\
  (\bar{\sigma}^{\mu\nu})^{\dot{\alpha}}{}_{\dot{\beta}}
  &:= \dfrac{i}{4} \left[
    (\bar{\sigma}^\mu)^{\dot{\alpha}\gamma}
    (\sigma^\nu)_{\gamma\dot{\beta}}
   -(\bar{\sigma}^\nu)^{\dot{\alpha}\gamma}
    (\sigma^\mu)_{\gamma\dot{\beta}}
   \right] \, , 
 \end{align}
where ${\bf{1}}$ and $\sigma^a~(a=1,2,3)$ 
denote a $2\times 2$ unit matrix and the Pauli spin matrices, respectively.
Since $\phi^I$  transforms homogeneously under $G$-transformation, 
the functions $G_{ab}$, $G_{aI}$, $G_{IJ}$, 
$G_{{\h{i}}{\h{j}}^*}$, $V_{{\h{i}}{\h{j}}^*a}$, $V_{{\h{i}}{\h{j}}^*I}$, 
$M_{{\h{i}}{\h{j}}}$, $M_{{\h{i}}^*{\h{j}}^*}$, $S_{{\h{i}}{\h{j}}{\h{k}}{\h{l}}}$, 
$S_{{\h{i}}^*{\h{j}}^*{\h{k}}^*{\h{l}}^*}$, and $S_{{\h{i}}{\h{j}}{\h{k}}^*{\h{l}}^*}$ 
also transform homogeneously under $G$.
They are also assumed to satisfy the index-exchange symmetry,
\bea
&&
G_{ab}(\phi)
=G_{ba}(\phi) \, , 
\\
&&
G_{IJ}(\phi)
=G_{JI}(\phi) \, , 
\\
&&
M_{{\h{i}}{\h{j}}}(\phi)
=
M_{{\h{j}}{\h{i}}}(\phi)
\,,\\
&&
M_{{\h{i}}^*{\h{j}}^*}(\phi)
=
M_{{\h{j}}^*{\h{i}}^*}(\phi)
\,,\\
&&
S_{{\h{i}}{\h{j}}{\h{k}}{\h{l}}}(\phi)
=
S_{{\h{j}}{\h{i}}{\h{k}}{\h{l}}}(\phi)
=
S_{{\h{i}}{\h{j}}{\h{l}}{\h{k}}}(\phi)
=
S_{{\h{k}}{\h{l}}{\h{i}}{\h{j}}}(\phi)
\,,\\
&&
S_{{\h{i}}{\h{j}}{\h{k}}^*{\h{l}}^*}(\phi)
=
S_{{\h{j}}{\h{i}}{\h{k}}^*{\h{l}}^*}(\phi)
=
S_{{\h{i}}{\h{j}}{\h{l}}^*{\h{k}}^*}(\phi)
\,.
\eea
The Hermiticity of the Lagrangian requires
\bea
&&
[G_{ab}(\phi)]^*
= G_{ab}(\phi) \, , 
\\
&&
[G_{aI}(\phi)]^*
= G_{aI}(\phi) \, , 
\\
&&
[G_{IJ}(\phi)]^*
= G_{IJ}(\phi) \, , 
\\
&&
[V(\phi)]^*
=V(\phi) \, , 
\\
&&[G_{{\h{i}}{\h{j}}^*}(\phi)]^*
=
G_{{\h{j}}{\h{i}}^*}(\phi)
\,,
\label{eq:G-sym}
\\
&&[V_{{\h{i}}{\h{j}}^*a}(\phi)]^*
=
V_{{\h{j}}{\h{i}}^*a}(\phi)
\,,\\
&&[V_{{\h{i}}{\h{j}}^*I}(\phi)]^*
=
V_{{\h{j}}{\h{i}}^*I}(\phi)
\,,
\label{eq:V-sym}\\
&&[M_{{\h{i}}{\h{j}}}(\phi)]^*
=
M_{{\h{j}}^*{\h{i}}^*}(\phi)
\,,\\
&&
[S_{{\h{i}}{\h{j}}{\h{k}}{\h{l}}}(\phi)]^*
=
S_{{\h{i}}^*{\h{j}}^*{\h{k}}^*{\h{l}}^*}(\phi)
\,,\\
&&
[S_{{\h{i}}{\h{j}}{\h{k}}^*{\h{l}}^*}(\phi)]^*
=
S_{{\h{k}}{\h{l}}{\h{i}}^*{\h{j}}^*}(\phi)
\,.
\eea
These functions determine the interactions 
among the scalar bosons and the spin-1/2 fermions.
The operator 
$\tilde{G}_{{\h{i}}{\h{j}}^*}(\phi)\mathcal{D}_\mu(\psi^{\dag {\h{j}}^*}\bar{\sigma}^\mu \psi^{\h{i}} )$ is absent in the Lagrangian (\ref{eq:LGHEFT-sym}), 
because it can be eliminated by adding the total derivative operator 
$\partial_\mu (\tilde{G}_{{\h{i}}{\h{j}}^*}(\phi)\psi^{\dag {\h{j}}^*}\bar{\sigma}^\mu \psi^{\h{i}})$ and 
redefining $V_{{\h{i}}{\h{j}}^*a}$ and $V_{{\h{i}}{\h{j}}^*I}$ appropriately. 

As we will see in the next subsection~\ref{sec:powercounting},
four-fermion operators should be introduced in the leading order 
Lagrangian (\ref{eq:LGHEFT-sym}), 
while we do not introduce operators like
\begin{align}
 (\psi^{\hat{i}}\sigma^{\mu\nu} \psi^{\hat{j}})
  \,[\alpha^a_{\perp\mu}, \alpha^b_{\perp\nu}]  \, ,
\end{align}
which seemingly possess lower mass dimensions.
The four-fermion operators 
\begin{align*}
& 
(\psi^{\dagger \hat{i}^*} \bar{\sigma}^\mu \psi^{\hat{i}})
 (\psi^{\dagger \hat{j}^*} \bar{\sigma}_\mu \psi^{\hat{j}}) \, ,
\quad
(\psi^{\hat{i}} \sigma^{\mu\nu} \psi^{\hat{j}}) 
(\psi^{\hat{k}} \sigma_{\mu\nu} \psi^{\hat{l}}) \, , 
\\
&
(\psi^{\dagger \hat{i}^*} \bar{\sigma}^{\mu\nu} \psi^{\dagger\hat{j}^*}) 
 (\psi^{\dagger \hat{k}^*} \bar{\sigma}_{\mu\nu} \psi^{\dagger\hat{l}^*}) \, ,
\quad
(\psi^{\dagger \hat{i}^*} \bar{\sigma}^{\mu\nu} \psi^{\dagger\hat{j}^*}) 
 (\psi^{\hat{i}} \sigma_{\mu\nu} \psi^{\hat{j}})  
\end{align*}
are Fierz rearranged to the standard forms
\begin{align*}
 (\psi^{\dagger \hat{i}^*} \psi^{\dagger \hat{j}^*}) 
(\psi^{\hat{i}} \psi^{\hat{j}}) \, , 
\,\,
(\psi^{\hat{i}} \psi^{\hat{j}}) 
(\psi^{\hat{k}} \psi^{\hat{l}}) \, , 
\, \,
(\psi^{\dagger \hat{i}^*} \psi^{\dagger \hat{j}^*}) 
(\psi^{\dagger \hat{k}^*} \psi^{\dagger \hat{l}^*})
\end{align*}
in the Lagrangian~(\ref{eq:LGHEFT-sym})

The HEFT Lagrangian~(\ref{eq:LHEFT2})
can be reproduced by restricting the particle contents and the structures 
of the coupling functions. 
We summarize 
the relationship between GHEFT~(\ref{eq:LGHEFT-sym}) and 
HEFT~(\ref{eq:LHEFT2}) 
in appendix \ref{app:HEFT}. 

The minimal electroweak gauge interactions are 
introduced to the EFT Lagrangian by 
replacing $\partial_\mu\xi_W$ and $\partial_\mu\xi_Y$ with the 
covariant derivatives;
\bea
D_\mu\xi_W
&=&
\partial_\mu \xi_W-ig_W W^a_\mu\frac{\tau^a}{2}\xi_W\,,
\label{eq:DxiW}\\
D_\mu\xi_Y
&=&
\partial_\mu \xi_Y+ig_Y \xi_YB_\mu\frac{\tau^3}{2}\,,
\label{eq:DxiY}
\eea
with $W^a_\mu~(a=1,2,3)$, $B_\mu$, 
$g_W$ and $g_Y$ being the 
$SU(2)_W$ and $U(1)_Y$ 
gauge fields and 
gauge coupling strengths.
It is also straightforward to introduce minimal QCD 
interactions with gluons by gauging the bosonic indices $I$, $J$
and fermionic indices $\hat{i}$, $\hat{j}$, $\hat{i}^*$, $\hat{j}^*$
in an appropriate manner.
We can also include non-minimal gauge interactions through 
operators like
$g_V (\psi^{\hat{i}}\sigma^{\mu\nu} \psi^{\hat{j}}) V_{\mu\nu}$ with $V_{\mu\nu}$, 
$g_V$ being the field strength and the coupling strength of the gauge boson.
As we will discuss in the next subsection, however, 
these operators do not appear at the leading order 
in the chiral order counting rule.

%%%%%%%%%%%%%%%%%%%%%
\subsection{Chiral order counting rule}
\label{sec:powercounting}

Low energy effective theories are not renormalizable.
They therefore contain infinitly many free parameters.
In order to compute scattering amplitudes keeping certain predictability
in effective theories,  
we need to introduce an order counting rule which 
enables us to distinguish phenomenologically relevant parameters 
from irrelevant ones.
If the underlying physics behind the effective theory is 
a perturbative theory, the operators in the effective theory
can simply be organized by their mass dimensions.
Higher dimensional operators decouple from the low energy
physics quickly and the associated parameters are suppressed 
by the inverse power of the cutoff scale.
The SMEFT \refsmeft~is constructed based on this idea.

This idea cannot be applied to non-perturbative physics, however.
The chiral perturbation theory describing
the low energy pion scattering amplitudes in the hadron physics
is a well-known example~\cite{Weinberg:1978kz,Gasser:1983yg,Gasser:1984gg,Georgi:1994qn,Donoghue:1992dd}.
In the chiral perturbation theory, 
the operators in the effective theory are not organized by their 
mass dimensions. They are organized by the number of derivatives,
instead.
The chiral order counting rule in the chiral perturbation theory is
known to be consistent with the expansion in terms of the
energy in the scattering amplitudes,  with the expansion
in terms of loops, and also with the expansion in terms of light 
quark masses.

How can we organize the chiral order counting rule in our effective 
Lagrangian?
The rule should be consistent with the expansion in terms 
of the energy and also with the expansion in terms of 
loops.
In order to construct such a chiral order counting rule, 
we next study divergence structure in the 
radiative corrections, and we justify that 
(\ref{eq:LGHEFT-sym}) is regarded as the leading order Lagrangian 
in the loop expansion.

We first consider an amputated connected $L$-loop Feynman diagram 
${\mathscr D}$ made only from the interactions in the 
Lagrangian~(\ref{eq:LGHEFT-sym}).
The diagram possesses $I_\phi$ scalar internal propagators and
$I_\psi$ fermion internal propagators.
The vertices in the diagram are labelled by an integer $n=1, 2, \cdots, 
N_v$, with $N_v$ being the total number of vertices in the diagram.
The superficial degree of divergence for ${{\mathscr D}}$ is 
denoted by $d({{\mathscr D}})$.  
It can be calculated from
\be
\int (d^4 p)^{L} \,
p^{(\# \partial)} 
\l(\frac{1}{p^2}\r)^{I_\phi}\l(\frac{1}{p}\r)^{I_\psi}\, , 
\quad
(\# \partial) := \sum_{n=1}^{N_v} (\# \partial)_n \, .
\ee
Here the $n$-th vertex appears from the operator with
$(\# \partial)_n$ derivatives and
$2\times (\# \psi\psi)_n$ fermions.
We also introduce spurions for later convenience.
The $n$-th vertex operator is assigned to have $(\# s)_n$ spurion fields.
These numbers for the operators in the Lagrangian (\ref{eq:LGHEFT-sym}) 
are listed in table~\ref{tab:numbers}.

\begin{table}
  \centering
  \begin{tabular}{c|ccccccccc}
     & $G_{ab}$ & $G_{aI}$ & $G_{IJ}$ & $V$ & $G_{\hat{i}\hat{j}^*}$ & $V_{\hat{i}\hat{j}^* a}$ & $V_{\hat{i}\hat{j}^* I}$ & $M_{\hat{i}\hat{j}}$ & $S_{\hat{i}\hat{j}\hat{k}\hat{l}}$ \\
  \hline
  $(\# \partial)_n$ & 2 & 2 & 2 & 0 & 1 & 1 & 1 & 0 & 0 \\   
  $(\# \psi\psi)_n$ & 0 & 0 & 0 & 0 & 1 & 1 & 1 & 1 & 2 \\
  $(\# s)_n$        & 0 & 0 & 0 & 2 & 0 & 0 & 0 & 1 & 0 
  \end{tabular}
  \caption{
  \label{tab:numbers}
  The number of derivatives, the number of fermion bilinears, 
  and the number of
  spurions for operators in the lowest order Lagrangian.
  }
\end{table}

We obtain
\begin{equation}
  d({\mathscr D}) = 4L + \sum_{n=1}^{N_v} (\# \partial)_n  - 2 I_\phi - I_\psi \, .
\end{equation}
We next expand the diagram ${\mathscr D}$ in terms of the external momentum 
$p$,
\begin{align}
  {{\mathscr D}} 
  &= \sum_{(\# p)=0,1,2,\cdots} {{\mathscr D}}_{(\# p)} \, p^{(\# p)} 
  \nonumber\\
  &= {{\mathscr D}}_0 + {{\mathscr D}}_1\,p + {{\mathscr D}}_2 \,p^2 + \cdots \, .
\end{align}
The superficial degree of divergence for ${{\mathscr D}}_{(\# p)}$ is thus
\begin{equation}
  d({{\mathscr D}}_{(\# p)})
  = 4L + \sum_{n=1}^{N_v} (\# \partial)_n  - (\# p) - 2 I_\phi - I_\psi  \, .
\label{eq:superficial}
\end{equation}
The number of scalar propagators $I_\phi$ can be removed from
Eq.\,(\ref{eq:superficial}) by using the graph-theoretical Euler 
formula 
\begin{equation}
  L + N_v - I_\phi - I_\psi = 1   \, .
\end{equation}
We find
\begin{equation}
  d({{\mathscr D}}_{(\# p)})
  = 2L + \sum_{n=1}^{N_v} \left[
      (\# \partial)_n  -2 
   \right] - (\# p) 
   + 2 +  I_\psi  \, .
\label{eq:superficial2}
\end{equation}

We next turn to the renormalization of the effective theory.
We assume that the effective theory is non-anomalous.
The divergences associated with $d({{\mathscr D}}_{(\# p)}) \ge 0$ 
can thus be subtracted
by introducing local operator counter terms ${\cal O}$.
The number of derivatives, 
the number of fermions, 
and the number of spurions in ${\cal O}$ are computed as
\begin{align}
 & (\# \partial)_{\cal O} = (\# p) 
 \,,\nonumber\\
 & (\# \psi\psi)_{\cal O} = \sum_{n=1}^{N_v}   (\# \psi\psi)_n  -  I_\psi   
 \,, \nonumber\\
 &  (\# s)_{\cal O} = \sum_{n=1}^{N_v}  (\# s)_n \, .
\end{align}

Using the relations above, the inequality $d({{\mathscr D}}_{(\# p)}) \ge 0$ 
can be rewritten as
\begin{align}
  2L & + \sum_{n=1}^{N_v} \left[
    (\# \partial)_n  + (\# \psi\psi)_n + (\# s)_n -2   
  \right]  
  \nn\\
  &
\ge 
(\# \partial)_{\cal O} + (\# \psi\psi)_{\cal O} + (\# s)_{\cal O} -2 
  \, .
\label{eq:superficial3}
\end{align}

We define ``chiral dimension'' of the operator ${\cal O}$ as
\begin{equation}
  C({\cal O})
  :=   (\# \partial)_{\cal O} + (\# \psi\psi)_{\cal O} + (\# s)_{\cal O} \, .
\label{eq:chira-dimension-counting}
\end{equation}
The spurion field dependences 
$(\# s)_n$ in the table~\ref{tab:numbers} 
are determined so as to keep $C(n)=2$ in the lowest order Lagrangian. 
Here we define $C(n)$  as the chiral dimension for 
operator from which $n$-th vertex arises in the Feynman diagram
${{\mathscr D}}$.

The counter terms ${\cal O}$ we need to introduce to subtract 
the divergences
in the diagram ${{\mathscr D}}_{(\# p)}$ therefore satisfy an inequality
\begin{equation}
  2 L + \sum_{n=1}^{N_v} \left[
   C(n) - 2 
  \right]  \ge C({\cal O}) -2 \, .
\label{eq:condition}
\end{equation}
Here the equality corresponds to the logarithmic divergence.
Since $C(n)=2$ for operators in the lowest order Lagrangian, 
we obtain
\begin{equation}
  2 L + 2 \ge C({\cal O}) \, .
\label{eq:chiral-order-condition2}
\end{equation}
The divergences in the $L$ loop diagram made from the 
lowest order Lagrangian can thus be subtracted by using finite
number of counter terms having ``chiral dimensions'' less than or equal to 
$2L+2$.

We need to pay a special attention to the four-fermion 
operators~\cite{Nyffeler:1999ap,Hirn:2004ze,Hirn:2005fr,Hirn:2005sj} 
in the Lagrangian~(\ref{eq:LGHEFT-sym}).
These four-fermion operators would arise at leading-order for instance
from the exchange of a heavy resonance with a strong coupling.
The coefficients of the four-fermion operators {\em do not decouple}
and appear at the leading order in the low energy effective theory 
due to the strong interaction with the heavy resonance\footnote{
The four-fermion operators 
have been ignored in the HEFT approach \refheft , however.
This is because that, in the HEFT, quarks and leptons  are assumed to 
couple with the heavy resonance only perturbatively.
Therefore, the number of spurions $(\# s)_n$ for the
four-fermion operators is assigned to be $(\# s)_n>0$ 
in the HEFT approach.
They therefore can be treated as 
the next-to-leading order Lagrangian in the HEFT \cite{Buchalla:2013eza}.
The assumptions made in the HEFT approach need not to hold, however, 
if we do not assume underlying UV physics behind our effective theory.
}.
Moreover, 
the SM quarks and leptons may be composite states arising from 
new strong dynamics.  
The exchange of common constituent in the composite state
naturally produce large coefficient $(4\pi)^2/\Lambda^2$ for the 
composite four-fermion 
operators~\cite{Eichten:1983hw,Hagiwara:1985wt,Baur:1987ga}.
Even if the SM quarks and leptons are assumed to be elementary fermions,
there may also exist relatively light partner fermions in the
strongly interacting EWSB sector.
The four-fermion operators involving these strongly interacting
partner fermions appear at the leading-order in our effective Lagrangian.
These are the reasons why we {\em did not} assign $(\# s)_n > 0$
for the four-fermion operators in the table~\ref{tab:numbers}.
We also note that the assignments of the spurion field dependences of $V$ and $M_{\h{i}\h{j}}$ terms in the table~\ref{tab:numbers} are determined to balance the chiral dimensions of mass- and kinetic-terms in scalar and fermion propagators.
 
There exist one-loop divergence which can be subtracted by 
the counter term 
$s\, \psi^{\hat{i}}\sigma^{\mu\nu} \psi^{\hat{j}}\,[\alpha^a_{\perp\mu}, 
\alpha^b_{\perp\nu}]$.
Note that, due to the chirality-flip structure of the operator, 
the divergence appears only with the chirality flipping spurion field $s$.  
%See Table~\ref{tab:numbers}.
The chiral dimension of the counter term is therefore counted as 4.
The appearance of the one-loop divergence associated with this operator 
is consistent with the expectation 
from the chiral order counting rule (\ref{eq:chiral-order-condition2}).
It should also be stressed that, if we included the operator
$s\, \psi^{\hat{i}}\sigma^{\mu\nu} \psi^{\hat{j}}\,[\alpha^a_{\perp\mu}, 
\alpha^b_{\perp\nu}]$ in the lowest order Lagrangian, 
we were not able to perform systematic expansion in the 
computation of the amplitudes
based on the chiral order counting rule.

It is now straightforward to construct a systematic expansion of the
amplitudes based on the chiral order counting rule.
Note here that the inequality (\ref{eq:condition}) holds even 
in general $L$-loop diagram with $C(n)\ge 2$.
It therefore assures us to obtain
finite amplitude by applying the standard
subtraction procedure with these counter terms.
The loop expansion should therefore be performed simultaneously 
with the expansion in terms of the chiral dimension (\ref{eq:chira-dimension-counting}).

If we restrict ourself to the operators with $(\#\psi\psi)_{\cal O}=0$, 
this result is well-known in the context of the 
chiral perturbation theory (low-energy effective theory for QCD 
pion)~\cite{Weinberg:1978kz,Gasser:1983yg,Gasser:1984gg,Georgi:1994qn,Donoghue:1992dd}. 
Our finding therefore can be regarded as a fermionic 
generalization of the chiral perturbation theory.

Finally let us give a comment on the chiral order counting of the gauge sector. We remark that, in order to make the gauge boson kinetic Lagrangian
\bea
\mathcal{L}_{\rm{gauge,kin}}
=
-\frac{1}{4}W^a_{\mu\nu}W^{a\mu\nu}
-\frac{1}{4}B_{\mu\nu}B^{\mu\nu}\,,
\eea
to be at the leading-order (chiral dimension 2), 
we need to assign the chiral dimensions of the field strengths as
\bea
&& C(W^a_{\mu\nu})=C(B_{\mu\nu})=1\,.\label{eq:chiraldim-Vmunu}
\eea
Since $C(\partial)=1$, 
(\ref{eq:chiraldim-Vmunu}) implies
\bea
&&C(W^a_{\mu})=C(B_{\mu})=0\,.\label{eq:chiraldim-Vmu}
\eea
Furthermore, since the gauge bosons are introduced as Eqs.~(\ref{eq:DxiW}) and (\ref{eq:DxiY}), the chiral dimension of the gauge coupling parameters should be
\bea
&&C(g_W)=C(g_Y)=1\,.
\eea

Computing the one-loop diagrams with an external gauge line, 
we find there exist divergences in the operator, 
\begin{equation}
  g_V s (\psi^{\hat{i}} \sigma^{\mu\nu} \psi^{\hat{j}}) V_{\mu\nu}  \, .
\label{eq:dipole-operator}
\end{equation}
The chiral dimension of (\ref{eq:dipole-operator}) is 
\begin{equation}
  C(g_V s (\psi^{\hat{i}} \sigma^{\mu\nu} \psi^{\hat{j}}) V_{\mu\nu}  )
  = 4 \, .
\end{equation}
The appearance of the one-loop divergence associated with this operator 
is consistent with the expectation 
from the chiral order counting rule (\ref{eq:chiral-order-condition2}).
It should also be stressed that, if we included the operator
$g_V s (\psi^{\hat{i}} \sigma^{\mu\nu} \psi^{\hat{j}}) V_{\mu\nu}$  
in the lowest order Lagrangian, 
we were not able to perform systematic expansion in the 
computation of the amplitudes
based on the chiral order counting rule.

%%%%%%%%%%%%%%%%%%%%%%%
\subsection{Geometrical form}

The scalar fields in the leading order GHEFT Lagrangian (\ref{eq:LGHEFT-sym})
consist of NG boson fields $\pi^a$ and the non-NG boson fields $\phi^I$.
It is convenient to introduce a scalar field multiplet notation $\phi^i$, 
not distinguishing the NG bosons from the non-NG bosons, 
\be
\phi^i
=
(\pi^a,\phi^I)
=
(\pi^1,\pi^2,\pi^3,\phi^1,\cdots,\phi^{n_s})
\,.
\ee
Using the scalar multiplet $\phi^i$, the EFT Lagrangian (\ref{eq:LGHEFT-sym}) can be expressed in a geometrical form:
\bea
\mathcal{L}_{\rm{GHEFT}}
&=&
\frac{1}{2}g_{ij}(\phi)(\partial_\mu \phi^i)(\partial^\mu \phi^j)-V(\phi)
\nn\\
&+&
\frac{i}{2}g_{{\h{i}}{\h{j}}^*}(\phi)
\biggl(
\psi^{\dag {\h{j}}^*}\bar{\sigma}^\mu (\partial_\mu \psi^{\h{i}})
-
(\partial_\mu\psi^{\dag {\h{j}}^*})\bar{\sigma}^\mu  \psi^{\h{i}}
\biggr)
\nn\\
&+&
v_{{\h{i}}{\h{j}}^*i}(\phi)
\,
\psi^{\dag {\h{j}}^*}\bar{\sigma}^\mu \psi^{\h{i}} (\partial_\mu \phi^i)
\nn\\
&-&
\frac{1}{2}
M_{{\h{i}}{\h{j}}}(\phi)\,
\psi^{\h{i}}\psi^{\h{j}}
-\frac{1}{2}
M_{{\h{i}}^*{\h{j}}^*}(\phi)\,
\psi^{\dag {\h{i}}^*}\psi^{\dag {\h{j}}^*}
\nn\\
&+&
\frac{1}{8}
S_{{\h{i}}{\h{j}}{\h{k}}{\h{l}}}(\phi)
(\psi^{{\h{i}}}\psi^{{\h{j}}})(\psi^{{\h{k}}}\psi^{{\h{l}}})
\nn\\
&+&
\frac{1}{8}
S_{{\h{i}}^*{\h{j}}^*{\h{k}}^*{\h{l}}^*}(\phi)
(\psi^{\dag {\h{i}}^*}\psi^{\dag {\h{j}}^*})
(\psi^{\dag {\h{k}}^*}\psi^{\dag {\h{l}}^*})
\nn\\
&+&
\frac{1}{4}
S_{{\h{i}}{\h{j}}{\h{k}}^*{\h{l}}^*}(\phi)
(\psi^{{\h{i}}}\psi^{{\h{j}}})
(\psi^{\dag {\h{k}}^*}\psi^{\dag {\h{l}}^*})
\,,
\label{eq:gheft1}
\eea
where $g_{{\h{i}}{\h{j}}^*}$ and $v_{{\h{i}}{\h{j}}^*i}$ satisfy
\bea
&&[g_{{\h{i}}{\h{j}}^*}(\phi)]^*
=
g_{{\h{j}}{\h{i}}^*}(\phi)
\,,\\
&&[v_{{\h{i}}{\h{j}}^*i}(\phi)]^*
=
v_{{\h{j}}{\h{i}}^*i}(\phi)
\,.
\eea
The coefficients $g_{{\h{i}}{\h{j}}^*}$ and $v_{{\h{i}}{\h{j}}^*i}$ are calculated from 
$G_{{\h{i}}{\h{j}}^*}$, $V_{{\h{i}}{\h{j}}^*a}$, and $V_{{\h{i}}{\h{j}}^*I}$ as
\bea
g_{{\h{i}}{\h{j}}^*}
&=&
G_{{\h{i}}{\h{j}}^*}
\,,\\
v_{{\h{i}}{\h{j}}^*1}
&=&
V_{{\h{i}}{\h{j}}^*1}
-
\frac{1}{2}
V_{{\h{i}}{\h{j}}^*3}\pi^2
\nn\\
&-&
\frac{1}{6}
V_{{\h{i}}{\h{j}}^*1}
\pi^2\pi^2
+
\frac{1}{6}
V_{{\h{i}}{\h{j}}^*2}\pi^1\pi^2
+
\mathcal{O}((\pi)^3)
\,,\\
v_{{\h{i}}{\h{j}}^*2}
&=&
V_{{\h{i}}{\h{j}}^*2}
+
\frac{1}{2}
V_{{\h{i}}{\h{j}}^*3}\pi^1
\nn\\
&-&
\frac{1}{6}V_{{\h{i}}{\h{j}}^*2}\pi^1\pi^1
+
\frac{1}{6}
V_{{\h{i}}{\h{j}}^*1}\pi^1\pi^2
+
\mathcal{O}((\pi)^3)
\,,\\
v_{{\h{i}}{\h{j}}^*3}
&=&
V_{{\h{i}}{\h{j}}^*3}
+
\frac{1}{2}
(
G_{\hat{i}'\hat{j}^*}[Q_\psi]^{\hat{i}'}_{~\hat{i}}
+
G_{\hat{i}\hat{j}'^*}[Q_\psi]^{\hat{j}'^*}_{~\hat{j}^*}
)
\nn\\
&-&
iV_{{\h{i}}{\h{j}}^*I}[Q_\phi]^I_{~J}\phi^J
+
\mathcal{O}((\pi)^3)
\,,\\
v_{{\h{i}}{\h{j}}^*I}
&=&
V_{{\h{i}}{\h{j}}^*I}
\,.
\eea
and the scalar metric tensor $g_{ij}$ are calculated from $G_{ab}$, $G_{aI}$, and $G_{IJ}$ \cite{Nagai:2019tgi}.

It may also be illuminating to point out a similarity 
between our GHEFT Lagrangian
(\ref{eq:gheft1}) and the supersymmetric non-linear sigma model 
Lagrangian,
\begin{align}
  {\cal L}
  &= g_{ii^*} 
     (\partial_\mu \phi^i) \eta^{\mu\nu} (\partial_\nu \phi^{\dagger i^*})
    -g^{ii^*} P_{, i} P^\dagger_{, i^*}
  \nonumber\\
  &\quad
    +\dfrac{i}{2} g_{ii^*}\, \left(
        \psi^{\dagger i^*} \bar{\sigma}^\mu (\partial_\mu \psi^i)
       -(\partial_\mu \psi^{\dagger i^*}) \bar{\sigma}^\mu \psi^i
     \right)
  \nonumber\\
  &\quad
    +\dfrac{i}{2} (\psi^{\dagger i^*} \bar{\sigma}^\mu \psi^i) \, 
     \left( g_{ii^* , j} \partial_\mu \phi^j 
           -g_{ii^*, j^*} \partial_\mu \phi^{\dagger j^*}
     \right)
  \nonumber\\
  &\quad
   -\left(
      P_{, ij}-P_{,i'} g^{i' i^*} g_{ii^*, i} 
    \right)  (\psi^i \psi^j)
    \nn\\
     &\quad
   -\left(
      P^\dagger_{, i^*j^*}-P^\dagger_{,i^{\prime *}} g^{i i^{\prime *}} g_{ii^*, j^*} 
    \right)  (\psi^{\dagger i^*} \psi^{\dagger j^*})
  \nonumber\\
  &\quad
  +\dfrac{1}{4} (g_{ii^* , jj^*} - g^{i' i^{\prime *}} g_{i' i^*, j^*} g_{ii^{\prime *}, j})
   (\psi^{\dagger i^*} \psi^{\dagger j^*})
   (\psi^i \psi^j) \, ,
\label{eq:susynlsm}
\end{align}
with $P(\phi)$ being the superpotential and the K\"{a}hler metric 
$g_{ij^*}(\phi, \phi^\dagger)$ is computed from the K\"{a}hler potential $K(\phi, \phi^\dagger)$,
\begin{align}
  g_{ij^*}(\phi, \phi^\dagger)
  &=  \dfrac{\partial^2 K}{\partial \phi^i \partial \phi^{\dagger j^*}} \, .
\end{align}
Here we use a comma-derivative notation 
\begin{align}
  g_{ii^* , j} := \dfrac{\partial}{\partial \phi^j} g_{ii^*} \, , 
  \quad 
  g_{ii^* , j^*} := \dfrac{\partial}{\partial \phi^{\dagger j^*}} g_{ii^*} \, , 
  \quad 
  \cdots \, , 
\end{align}
to keep the expression as simple as possible.
Not only the scalar and fermion kinetic terms, but also the 
counter parts to $v_{\hat{i}\hat{j}^* i}(\phi)$ and four-fermion terms in the 
GHEFT Lagrangian
(\ref{eq:gheft1}) are expressed in terms of K\"{a}hler 
manifold geometry.
Even though the scalar manifold of our GHEFT Lagrangian (\ref{eq:gheft1}) 
does not possess K\"{a}hler properties, as we will show later,
the particle scattering amplitudes in the GHEFT Lagrangian
can also be described in terms of the covariant tensors 
of the manifold.
%%%%%%%%%%%%%%%%%%%%%%%
\section{Normal coordinate}
\label{sec:NC}

\subsection{Field coordinate transformations}
Kamefuchi-O'Raifeartaigh-Salam (KOS) theorem \cite{Kamefuchi:1961sb} tells us that 
seemingly different effective Lagrangians connected through
the field coordinate transformations can describe the
identical scattering amplitudes.
Effective Lagrangian therefore cannot be determined uniquely.
We here summarize the field coordinate transformation properties
in the effective Lagrangian.

We consider a field transformation which keeps the chiral dimension of the fields. Under such a redefinition of the field coordinates 
$\psi^{\h{i}}$, $\psi^{\dagger {\h{i}}^*}$, $\phi^i$:
\begin{align}
%  &\phi^i \to
%  f^i{}_j(\phi) \phi^j \, ,\nn\\
  &\phi^i \to
  f^i{}(\phi)  \, ,\nn\\
  &  \psi^{\h{i}} \to
  f^{\, {\h{i}}}{}_{{\h{j}}}(\phi) \, \psi^{\h{j}} \, , \nn\\
  &  \psi^{\dagger {\h{i}}^*} \to
  f^{* \, {\h{i}}^*}{}_{{\h{j}}^*}(\phi) \, \psi^{\dagger {\h{j}}^*} \, ,
\label{eq:redef}
\end{align}
the functions $g_{ij}(\phi)$, $g_{{\h{i}}{\h{j}}^*}(\phi)$, $v_{{\h{i}}{\h{j}}^* i}(\phi)$,
$M_{{\h{i}}{\h{j}}}(\phi)$, $M_{{\h{i}}^*{\h{j}}^*}(\phi)$, 
$S_{{\h{i}}{\h{j}}{\h{k}}{\h{l}}}(\phi)$,  $S_{{\h{i}}^* {\h{j}}^* {\h{k}}^* {\h{l}}^*}(\phi)$,  
and
$S_{{\h{i}}{\h{j}} {\h{k}}^* {\h{l}}^*}(\phi)$ in Eq.\,(\ref{eq:gheft1})
transform as
\begin{align}
  g_{ij}(\phi)
  &\to g_{i'j'}(f(\phi)) \, f^{i'}{}_{,\, i}(\phi) \, f^{j'}{}_{,\, j}(\phi)  \, , 
\label{eq:transf1}
  \\
  g_{{\h{i}}{\h{j}}^*}(\phi)
  &\to g_{{\h{i}}'{\h{j}}'^*}(f(\phi)) \, f^{{\h{i}}'}{}_{{\h{i}}}(\phi) \, f^{*{\h{j}}'^*}{}_{{\h{j}}^*}(\phi) \, ,
  \\
  v_{{\h{i}}{\h{j}}^*i}(\phi)
  &\to v_{{\h{i}}'{\h{j}}'^* i'}(f(\phi)) \, f^{i'}{}_{,\, i}(\phi) \, 
       f^{{\h{i}}'}{}_{{\h{i}}}(\phi) \, f^{*{\h{j}}'^*}{}_{{\h{j}}^*}(\phi) 
  \nonumber\\
  & \quad
   +\dfrac{i}{2} g_{{\h{i}}'{\h{j}}'^*}(f(\phi)) \, \biggl( 
      f^{{\h{i}}'}{}_{{\h{i}},\, i}(\phi) \, f^{*{\h{j}}'^*}{}_{{\h{j}}^*}(\phi) \nn\\
      &\qquad  \qquad \qquad \qquad
    - f^{{\h{i}}'}{}_{\h{i}}(\phi) \, f^{*{\h{j}}'^*}{}_{{\h{j}}^*, \, i}(\phi) 
    \biggr)  \, ,
\end{align}
\begin{align}
  M_{{\h{i}}{\h{j}}}(\phi)
  &\to M_{{\h{i}}'{\h{j}}'}(\phi)
   f^{{\h{i}}'}{}_{{\h{i}}}(\phi) \, 
   f^{{\h{j}}'}{}_{{\h{j}}}(\phi) \, ,
  \\
  M_{{\h{i}}^* {\h{j}}^*}(\phi)
  &\to M_{{\h{i}}'^* {\h{j}}'^*}(\phi)
   f^{* {\h{i}}'^*}{}_{{\h{i}}}(\phi) \, 
   f^{* {\h{j}}'^*}{}_{{\h{j}}}(\phi) \, ,
\end{align}
and
\begin{align}
  \lefteqn{S_{{\h{i}}{\h{j}}{\h{k}}{\h{l}}}(\phi)}\nonumber\\
  &\to S_{{\h{i}}'{\h{j}}'{\h{k}}'{\h{l}}'}(\phi) \, 
   f^{{\h{i}}'}{}_{{\h{i}}}(\phi) \, 
   f^{{\h{j}}'}{}_{{\h{j}}}(\phi) \, 
   f^{{\h{k}}'}{}_{{\h{k}}}(\phi) \, 
   f^{{\h{l}}'}{}_{{\h{l}}}(\phi) \, ,
  \\
 \lefteqn{ S_{{\h{i}}^* {\h{j}}^* {\h{k}}'^* {\h{l}}'^*}(\phi)}\nonumber\\
  &\to S_{{\h{i}}'^* {\h{j}}'^* {\h{k}}'^* {\h{l}}'^*}(\phi) \, 
   f^{* {\h{i}}'^*}{}_{\!\!\!{\h{i}}^*}(\phi) \, 
   f^{* {\h{j}}'^*}{}_{\!\!\!{\h{j}}^*}(\phi) \, 
   f^{* {\h{k}}'^*}{}_{\!\!\!{\h{k}}^*}(\phi) \, 
   f^{* {\h{l}}'^*}{}_{\!\!\!{\h{l}}^*}(\phi) \, ,
  \\
  \lefteqn{S_{{\h{i}} {\h{j}} {\h{k}}^* {\h{l}}^*}(\phi)}\nonumber\\
  &\to S_{{\h{i}}' {\h{j}}' {\h{k}}'^* {\h{l}}'^*}(\phi) \, 
   f^{{\h{i}}'}{}_{{\h{i}}}(\phi) \, 
   f^{{\h{j}}'}{}_{{\h{j}}}(\phi) \, 
   f^{* {\h{k}}'^*}{}_{\!\!\!{\h{k}}'^*}(\phi) \, 
   f^{* {\h{l}}'^*}{}_{\!\!\!{\h{l}}'^*}(\phi) \, .
\label{eq:transf2}
\end{align}
The model parametrization in the GHEFT Lagrangian (\ref{eq:gheft1})
is, therefore, not unique.
Seemingly different Lagrangians can describe the same scattering 
amplitudes, if these Lagrangians are connected with each other
through the field
redefinitions (\ref{eq:redef}).
The GHEFT Lagrangian (\ref{eq:gheft1}) therefore contains the 
redundancy in its model parametrization, which leads to a lot of
inconveniences in its phenomenological analysis.
Clearly, we need a method which can uniquely identify the 
class of GHEFT Lagrangians which describe the same scattering amplitudes.

Note that the field redefinitions (\ref{eq:redef})
can be regarded as a general coordinate transformation
in the field space manifold.
Therefore, the scattering amplitudes are expected to be 
described in terms of covariantly transforming tensors under these 
general coordinate transformations.
We can consider more general field redefinitions than Eq.\,(\ref{eq:redef}) such as $\phi \to \phi + \psi \psi $ and $\phi \to \phi + \partial_\mu \phi\, \partial^\mu \phi$. 
These field transformations, however, violate the chiral-order counting rule. 
Amplitudes computed with finite order in the chiral order counting are 
affected by these field redefinitions. 
We therefore restrict ourselves within the field coordinate transformations
given in Eq.\,(\ref{eq:redef}).

In our previous paper \cite{Nagai:2019tgi}, we have explicitly shown that the scalar 
scattering
amplitudes are described in terms of the Riemann curvature
tensor $R_{i_1 i_2 i_3 i_4}(\phi)$ in the scalar field space and the scalar potential $V(\phi)$ 
and their covariant derivatives.
We have also shown that the use of the Riemann Normal Coordinate (RNC)
can reduce the computational tasks significantly.
In the present paper, we generalize these findings to the 
fermionic GHEFT Lagrangian (\ref{eq:gheft1}).

\subsection{Scalar sector}

As we have shown in Ref.\,\cite{Nagai:2019tgi}, the use of RNC
significantly reduces the computational task of the
scalar boson scattering amplitudes.  
This is because of the fact that, in RNC, 
the Taylor expansion of the field metric tensor $g_{ij}(\phi)$ 
is expressed  in terms of covariant quantities, {\it{i.e.}}, the Riemann
curvature tensor and its covariant derivatives.
On the other hand, although the RNC is defined by using the geodesics on the manifold, there is no direct connection between the computation
of the scattering amplitudes and the geodesic equations on the 
scalar manifold.
Moreover, the GHEFT Lagrangian we gave in Eq.\,(\ref{eq:gheft1}) contains
a complex valued fermion metric $g_{{\h{i}}{\h{j}}^*}(\phi)$, in addition to the
real valued scalar manifold metric $g_{ij}(\phi)$ and
the meaning of the geodesic equations in 
fermionic metric $g_{{\h{i}}{\h{j}}^*}(\phi)$ is not clear \cite{Higashijima:2002fq}.
It should therefore be illuminating,
before going to the fermionic sector of the GHEFT Lagrangian, 
to reconsider the derivation of the normal coordinate in the scalar manifold
in a manner not relying on the geodesic equations.
Here the normal coordinate is defined as a coordinate in which 
Taylor expansion coefficients of metric tensor around the vacuum 
are all expressed in terms of covariant tensors.

We introduce contravariant and covariant vectors, 
\begin{equation}
  v^i(\phi) \, , \qquad
  a_i(\phi) \, ,   
\end{equation}
which transform as
\begin{align}
  v^i(\phi) \to v^{i'}(f(\phi)) \, f^i_{,\, i'}(\phi) \, , \quad
  a_i(\phi) \to a_{i'}(f(\phi)) \, f^{i'}_{,\, i}(\phi) \, .
\end{align}
The covariant derivative of $v^i$ and $a_i$ are expressed
by using the ``semi-colon'' covariant-derivative notation,
\begin{align}
  v^i_{;\, j} 
  = v^i_{,\, j} + v^{i'} \Gamma^i_{i'j} \, ,  \qquad
  a_{i;\, j}
  = a_{i,\, j} - a_{i'} \Gamma^{i'}_{ij} \, ,
\end{align}
with the bosonic Affine connection $\Gamma^i_{jk}$ being
defined as
\begin{equation}
  \Gamma^i_{jk} = \dfrac{1}{2} g^{ii'} \left(
    g_{i'j, \, k} + g_{i'k, \, j} - g_{jk, \, i'}  
  \right)  \, ,
\end{equation}
with $g^{ii'}$  being the inverse of the metric tensor,
\begin{equation}
  g^{ii'} \,  g_{i'j} = \delta^i_j \, .
\end{equation}
We consider
\begin{align}
  a_{i;\, jk}-a_{i;\, kj} 
  &= a_{i''} \left(
       \Gamma^{i''}_{ik,\, j} - \Gamma^{i''}_{ij,\, k}
      +\Gamma^{i''}_{i'j} \Gamma^{i'}_{ik}
      -\Gamma^{i''}_{i'k} \Gamma^{i'}_{ij}
     \right)
  \nonumber\\   
  &= a_{i''} R^{i''}{}_{ijk} \, , 
\end{align}
where we define Riemann curvature tensor as
\begin{align}
  R^{i''}{}_{ijk} 
  &= 
       \Gamma^{i''}_{ik,\, j} - \Gamma^{i''}_{ij,\, k}
      +\Gamma^{i''}_{i'j} \Gamma^{i'}_{ik}
      -\Gamma^{i''}_{i'k} \Gamma^{i'}_{ij} \, .
\end{align}
We also introduce
\begin{align}
  R_{ii'jk} &:= g_{ii''} R^{i''}{}_{i'jk}
\end{align}
for later convenience.

We assume the field values at the vacuum are $\phi^i=0$.  
If the eigenvalues of the scalar manifold metric $g_{ij}$ are 
all positive definite{\footnote{
This condition guarantees the absence of ghost particles in GHEFT framework. 
See \cite{Abe:2020ikj, Abe:2018rwb} for the discussion 
on the pseudo unitarity in theories with ghost particles.
}}, we are able to find a normal coordinate in 
which $g_{ij}(\phi)$ is written as
\begin{align}
  g_{ij}(\phi)
  &= \delta_{ij} 
    + \dfrac{1}{2} G_{ijk_1 k_2} \, \phi^{k_1} \phi^{k_2} 
  \nonumber\\
  & \quad
    + \dfrac{1}{3!} G_{ij k_1 k_2 k_3 } \, \phi^{k_1} \phi^{k_2} \phi^{k_3}
    \nonumber\\
  & \quad
    + \dfrac{1}{4!} G_{ij k_1 k_2 k_3 k_4 } \, \phi^{k_1} \phi^{k_2} \phi^{k_3} \phi^{k_4} + \cdots 
  \nonumber\\
  &= \delta_{ij} 
    + \dfrac{1}{2} G_{ij(k_1 k_2)} \, \phi^{k_1} \phi^{k_2} 
  \nonumber\\
  & \quad
    + \dfrac{1}{3!} G_{ij (k_1 k_2 k_3)} \, \phi^{k_1} \phi^{k_2} \phi^{k_3}
      \nonumber\\
  & \quad
    + \dfrac{1}{4!} G_{ij (k_1 k_2 k_3 k_4)} \, \phi^{k_1} \phi^{k_2} \phi^{k_3} \phi^{k_4} + \cdots
\label{eq:metric-taylor}
\end{align}
with coefficients $G_{ijk_1 k_2}$, $G_{ijk_1 k_2 k_3}$, $\cdots$
being expressed in terms of covariant tensors at the vacuum
\begin{align}
{R}_{ijk_1k_2}\biggl|_{0} 
 & :={R}_{ijk_1k_2}\biggl|_{\phi^i=0}
\, ,\nonumber\\
{R}_{ijk_1k_2; \, k_3}\biggl|_{0}
  &:={R}_{ijk_1 k_2 ; \, k_3}\biggl|_{\phi^i=0}
\, , \nonumber\\
 & \vdots \, 
   \label{eq:4}
\end{align}
Here the indices between the parentheses are understood to be symmetrized, 
{\it{i.e.}}, 
\begin{align}
  G_{ij(k_1 k_2)} &:= \dfrac{1}{2} \left( 
     G_{ijk_1 k_2} + G_{ij k_2 k_1}
  \right)  \, , \\
  G_{ij(k_1 k_2 k_3)} &:= \dfrac{1}{3!} \left( 
     G_{ijk_1 k_2 k_3} + G_{ij k_2 k_3 k_1} + G_{ij k_3 k_1 k_2}  \r.\nn\\
     &
     \l.
    +\,G_{ijk_1 k_3 k_2} + G_{ij k_2 k_1 k_3} + G_{ij k_3 k_2 k_1}  
  \right)  \, .
\end{align}
Since the metric tensor $g_{ij}$ is symmetric under the $i\leftrightarrow j$
exchange, the coefficients $G_{ijk_1 k_2}$, $G_{ijk_1 k_2 k_3}$, $\cdots$
need to satisfy 
\begin{equation}
  G_{ijk_1 k_2} = G_{jik_1k_2} \, , \quad
  G_{ijk_1 k_2 k_3 } = G_{jik_1k_2 k_3} \, , \quad
  \cdots \, .
\end{equation}
The Riemann curvature tensor can be computed as
\begin{align}
  R_{i' ijk}
  &= \dfrac{1}{2} \left(
       G_{i' k(ij)} - G_{ik(i'j)} - G_{i'j(ik)} + G_{ij(i'k)}
     \right)
  \nonumber\\
  &\quad
    +\dfrac{1}{2} \left(
        G_{i'k(ijk_1)} - G_{ik(i'jk_1)} \r.\nn\\
        &\quad \qquad \l.- G_{i'j(ik k_1)} + G_{ij(i'k k_1)}
     \right) \phi^{k_1}
  \nonumber\\
  &\quad
   +\dfrac{1}{4} \left(
        G_{i'k(ijk_1 k_2)} - G_{ik(i'jk_1 k_2)} \r.\nn\\
        &\quad \qquad \l.- G_{i'j(ik k_1 k_2)} + G_{ij(i'k k_1 k_2)}
    \right) \phi^{k_1} \phi^{k_2}
  \nonumber\\
  &\quad
   -\dfrac{1}{4} \left(
        G_{i' j'(jk_1)} + G_{jj'(i' k_1)} - G_{i'j(j'k_1)}
    \right) \delta^{j' j''} \times
  \nonumber\\
  &\quad\quad \times
    \left(
        G_{j''i (k k_2)} + G_{j'' k (i k_2)} - G_{ik(j'' k_2)}
    \right) \phi^{k_1} \phi^{k_2}
  \nonumber\\
  &\quad
   +\dfrac{1}{4} \left(
        G_{i' j'(k k_1)} + G_{k j'(i' k_1)} - G_{i' k (j'k_1)}
    \right) \delta^{j' j''} \times
  \nonumber\\
  &\quad\quad \times
    \left(
        G_{j''i (j k_2)} + G_{j'' j (i k_2)} - G_{i j (j'' k_2)}
    \right) \phi^{k_1} \phi^{k_2}
  \nonumber\\
  & \quad + \cdots .
\label{eq:curvature-expansion}
\end{align}
We therefore obtain an expression for the Riemann curvature tensor at 
the vacuum, 
\begin{align}
R_{i12j} \biggr|_0 
  &= \dfrac{1}{2} \left(
       G_{ij(12)} - G_{1j(i2)} - G_{i2(1j)} + G_{12(ij)}
     \right) \, ,
\label{eq:zero-R}
\end{align}
and thus
\begin{align}
  R_{i(12)j} \biggr|_0 
  &= \dfrac{1}{4} \left(
       2G_{ij(12)} - G_{1j(i2)} - G_{1i(j2)} \r.\nn\\
        &\qquad  \l.- G_{i2(1j)} - G_{j2(1i)} + 2G_{12(ij)}
     \right) \, .
\label{eq:zero-R1}
\end{align}
We here introduced a shorthand 
abbreviation with which $1$, $2$, $\cdots$
are understood to be $k_1$, $k_2$, $\cdots$, respectively.
Note that, as we explained earlier, 
the coefficient $G_{ij(kl\cdot \, \cdot)}$ should be expressed
in terms of the Riemann curvature tensor
and its covariant derivatives at the vacuum.
The form of $G_{ij(12)}$ is uniquely determined as
\begin{equation}
  G_{ij(12)} = a R_{i(12)j} \biggr|_0 \, ,
\label{eq:assumption1}
\end{equation}
with $a$ being a constant, 
thanks to the Riemann tensor symmetry
\bea
&  R_{1234} + R_{1243} = 0 \, , \nn\\
&  R_{1234} + R_{2134} = 0 \, , \nn\\
&  R_{1234} - R_{3412} = 0 \, ,
\label{eq:r-symmetry}
\eea
Actually, all other index structures even under the $1 \leftrightarrow 2$ 
exchange can be reduced to the form of Eq.\,(\ref{eq:assumption1}):
\begin{align}
  R_{ij12} + R_{ij21} &= 0 \, , 
  \\
  R_{i1j2} + R_{i2j1} 
  &= - R_{i12j} - R_{i21j} 
  \nonumber\\
  &= - 2R_{i(12)j} \ , 
  \\
  R_{i12j} + R_{i21j} 
  &= 2R_{i(12)j} \ .
\end{align}
Plugging Eq.\,(\ref{eq:assumption1}) into the RHS of Eq.\,(\ref{eq:zero-R1}), 
we obtain
\begin{align}
  R_{i(12)j} \biggr|_0
  &= \dfrac{3}{2} a R_{i(12)j} \biggr|_0 
\end{align}
and therefore we find
\begin{align}
  a = \dfrac{2}{3} \, .
\end{align}
The coefficient $G_{ij(12)}$ is now determined as
\begin{align}
  G_{ij(12)} &=  \dfrac{2}{3} R_{i(12)j} \biggr|_0 \, .
\label{eq:zero-result}
\end{align}

Note that, in our derivation of Eq.\,(\ref{eq:zero-result}), 
we used the $1\leftrightarrow 2$ symmetrized 
condition (\ref{eq:zero-R1}) only.
We did not use the original condition (\ref{eq:zero-R}).
Since Eq.\,(\ref{eq:zero-R}) contains more information than 
its symmetrized form Eq.\,(\ref{eq:zero-R1}), we should check 
whether Eq.\,(\ref{eq:zero-result}) does satisfy the
original condition (\ref{eq:zero-R}) or not.

Plugging Eq.\,(\ref{eq:zero-result}) in the RHS of
Eq.\,(\ref{eq:zero-R}), we see
\begin{align}
\lefteqn{
\dfrac{1}{2} (
       G_{ij(12)}   - G_{1j(i2)} - G_{i2(1j)} + G_{12(ij)}
     )}\nn\\
&= \dfrac{1}{3} 
  \left(
       2 R_{i12j} + R_{i21j} - R_{ij12} 
     \right)  
\biggr|_0
\nonumber\\
&= R_{i12j} \biggr|_0  \, , 
\label{eq:zero-R-check}
\end{align}
and Eq.\,(\ref{eq:zero-R})  is actually satisfied with our result 
Eq.\,(\ref{eq:zero-result}).
In the last line computation of Eq.\,(\ref{eq:zero-R-check}), we used 
the Bianchi identity
\begin{equation}
  R_{i123} + R_{i231} + R_{i312} = 0 \, . 
\label{eq:r-bianchi}
\end{equation}
Since $R_{i12j}  \ne R_{i21j}$ in general manifolds, 
the Bianchi identity plays an essential role for the consistency of 
the normal coordinates.

The higher order terms in the Taylor expansion of $g_{ij}$ in the
normal coordinate are computed in appendix~\ref{sec-taylor-gij}. 
We find the function $g_{ij}(\phi)$ can be expanded
in terms of the covariant tensors as
\begin{align}
  g_{ij}(\phi)
  &= \delta_{ij} 
    + \dfrac{1}{2} G_{ij(k_1 k_2)} \, \phi^{k_1} \phi^{k_2} 
  \nonumber\\
  & \quad
    + \dfrac{1}{3!} G_{ij (k_1 k_2 k_3)} \, \phi^{k_1} \phi^{k_2} \phi^{k_3}
     \nonumber\\
  & \quad
    + \dfrac{1}{4!} G_{ij (k_1 k_2 k_3 k_4)} \, \phi^{k_1} \phi^{k_2} \phi^{k_3} \phi^{k_4} + \cdots
\label{eq:taylor-gij}
\end{align}
with
\begin{align}
  G_{ij(12)} &= \dfrac{2}{3} R_{i(12)j} \biggr|_0 \, , 
  \\
  G_{ij(123)} &= \dfrac{1}{3} \left[
      R_{i(12)j; 3} + R_{i(23)j; 1} + R_{i(31)j; 2}  
  \right]  \biggr|_0 \, ,
\\
  G_{ij(1234)}
  &= \dfrac{1}{5} \left[
       R_{i(12)j; (34)}  + R_{i(34)j; (12)}  \r.\nn\\
       &\l.
      \phantom{\dfrac{1}{5} \biggl\{}
     +\, R_{i(13)j; (24)}  + R_{i(24)j; (13)}  \r.\nn\\
       & \l.
      \phantom{\dfrac{1}{5} \biggl\{}
     +\, R_{i(14)j; (23)}  + R_{i(23)j; (14)} 
     \right] \biggr|_0
  \nonumber\\
  & \quad 
   + \dfrac{8}{45}
          g^{i' j'} \left[
       R_{i'(12)i} R_{j'(34)j} + R_{i'(13)i} R_{j'(24)j} \r.\nn\\
       &\l.
       \phantom{\dfrac{8}{45} \biggl\{}
       +\, R_{i'(14)i} R_{j'(23)j}
      +R_{i'(34)i} R_{j'(12)j}   \r.\nn\\
       & \l.
       \phantom{\dfrac{8}{45} \biggl\{}
       +\, R_{i'(24)i} R_{j'(13)j}+R_{i'(23)i} R_{j'(14)j}
     \right] \biggr|_0 .
\end{align}

The Taylor expansion of the potential term $V(\phi)$ can also be given 
in a similar manner.
We obtain 
\begin{align}
  V_{, 12} \biggr|_0
  &= V_{; 12} \biggr|_0 \, , 
  \\
  V_{, 123} \biggr|_0
  &= V_{; 123} \biggr|_0 \, , 
  \\
  V_{, 1234} \biggr|_0
  &= V_{; 1234} \biggr|_0 
    -\dfrac{2}{3} \left[
        V_{;1i} R^i{}_{(23)4}
       +V_{;2i} R^i{}_{(13)4}
       \r.\nn\\
       &\quad \l.+\,
       V_{;3i} R^i{}_{(12)4}
       +V_{;4i} R^i{}_{(12)3}
     \right] \biggr|_0 \, ,
  \\
  V_{, 12345} \biggr|_0 
  &= V_{; 12345} \biggr|_0 
   - \dfrac{2}{3} \biggl\{ 
       V_{;34i} R^i{}_{(12)5}
      +V_{;35i} R^i{}_{(12)4}
         \nn\\
       &\quad  
       \phantom{-\dfrac{2}{3} \biggl\{}
      +V_{;45i} R^i{}_{(12)3}
      +V_{;24i} R^i{}_{(13)5}
  \nonumber\\
  & \quad
       \phantom{-\dfrac{2}{3} \biggl\{}  
      +V_{;25i} R^i{}_{(13)4}
      +V_{;14i} R^i{}_{(23)5}
       \nonumber\\
  & \quad
       \phantom{-\dfrac{2}{3} \biggl\{}
     +V_{;15i} R^i{}_{(23)4}
      +V_{;23i} R^i{}_{(14)5}
       \nonumber\\
  & \quad
       \phantom{-\dfrac{2}{3} \biggl\{}
      +V_{;13i} R^i{}_{(24)5}
      +V_{;12i} R^i{}_{(34)5}
   \biggr\} \biggr|_0
  \nonumber\\
  & \quad +\dfrac{1}{6} \biggl\{
         V_{;1i} (R^i{}_{(45)(2; \, 3)} - 5 R^i{}_{(23)(4; \, 5)})
          \nonumber\\
  & \quad \phantom{+\dfrac{1}{6} \biggl\{}
        +V_{;2i} (R^i{}_{(45)(1; \, 3)} - 5 R^i{}_{(13)(4; \, 5)})
  \nonumber\\
  & \quad \phantom{+\dfrac{1}{6} \biggl\{}
      +V_{;3i} (R^i{}_{(45)(1; \, 2)} - 5 R^i{}_{(12)(4; \, 5)})
        \nonumber\\
  & \quad \phantom{+\dfrac{1}{6} \biggl\{}
      +V_{;4i} (R^i{}_{(35)(1; \, 2)} - 5 R^i{}_{(12)(3; \, 5)})
  \nonumber\\
  & \quad \qquad  
      +V_{;5i} (R^i{}_{(34)(1; \, 2)} - 5 R^i{}_{(12)(3; \, 4)})
    \biggr\} \biggr|_0 
\end{align}
and therefore
\begin{align}
  V_{, 12} \biggr|_0 
  &= V_{; (12)} \biggr|_0 \, , 
  \\
  V_{, 123} \biggr|_0 
  &= V_{; (123)} \biggr|_0 \, , 
  \\
  V_{, 1234} \biggr|_0 
  &= V_{; (1234)} \biggr|_0 \, , 
  \\
  V_{, 12345} \biggr|_0 
  &= V_{; (12345)} \biggr|_0 \, , 
  \\
  & \vdots
\nonumber
\end{align}
in the normal coordinate.
The potential term in the Lagrangian (\ref{eq:gheft1}) can also be
expanded in terms of the covariant tensors
\begin{align}
  V(\phi)
  &= V \biggr|_0 
      + \dfrac{1}{2} V_{; (k_1 k_2)} \biggr|_0 \phi^{k_1} \phi^{k_2} 
        \nn\\
      &\qquad
      + \dfrac{1}{3!} V_{; (k_1 k_2 k_3)} \biggr|_0 \phi^{k_1} \phi^{k_2} \phi^{k_3} 
      \nn\\
      &\qquad
     + \dfrac{1}{4!} V_{; (k_1 k_2 k_3 k_4)} \biggr|_0 \phi^{k_1} \phi^{k_2} \phi^{k_3} \phi^{k_4}
  \nonumber\\
  & \qquad
      + \dfrac{1}{5!} V_{; (k_1 k_2 k_3 k_4 k_5)} \biggr|_0 
\phi^{k_1} \phi^{k_2} \phi^{k_3} \phi^{k_4} \phi^{k_5}
  \nn\\
      &\qquad
      + \cdots
\end{align}
in the normal coordinate.

\subsection{Fermion bilinear sector}
We next move to the fermion bilinear sector in the 
GHEFT Lagrangian (\ref{eq:gheft1}). The fermion bilinear sector depends on the three kinds of coupling functions, $g_{\h{i}\h{j}^*}$, $v_{\h{i}\h{j}^*i}$, and $M_{\h{i}\h{j}}$.
We define a normal coordinate on the fermion field space so that the coupling functions are expanded in terms of the covariantly transforming tensors.

Before computing the expansion coefficients, we introduce the covariant quantities on the fermion field transformation. 
We first define the ``Affine connection'' as
\begin{align}
  \Gamma^{\h{i}}_{j{\h{k}}}
  &:= \dfrac{1}{2} g^{{\h{i}} {\h{l}}^*} \left[
        g_{{\h{k}}{\h{l}}^*, j} + g_{j{\h{l}}^*, {\h{k}}} - g_{j{\h{k}}, {\h{l}}^*} 
     \right] \, , 
\label{eq:affine-f1}
  \\
  \Gamma^{{\h{i}}^*}_{j{\h{k}}^*}
  &:= \dfrac{1}{2} g^{{\h{l}} {\h{i}}^*} \left[
        g_{{\h{l}}{\h{k}}^*, j} + g_{j{\h{k}}, {\h{l}}^*} - g_{j{\h{k}}^*, {\h{l}}} 
     \right] \, , 
\label{eq:affine-f2}
\end{align}
where $g^{{\h{i}} {\h{j}}^*}$ is defined as the inverse of $g_{{\h{i}} {\h{j}}^*}$, {\it{i.e.}}, 
\begin{align}
  g^{{\h{i}} {\h{k}}^*} g_{{\h{j}} {\h{k}}^*} = \delta^{\h{i}}_{\h{j}} \, , \qquad
  g^{{\h{k}}{\h{i}}^*} g_{{\h{k}}{\h{j}}^*} = \delta^{{\h{i}}^*}_{{\h{j}}^*} \, .
\end{align}
Got inspiration from the supersymmetric non-linear sigma model Lagrangian (\ref{eq:susynlsm}), we introduce $g_{i{\h{j}}^*, {\h{i}}}$ and $g_{i{\h{i}}, {\h{j}}^*}$ satisfying
\begin{align}
  v_{{\h{i}}{\h{j}}^* i}(\phi)
  = \dfrac{i}{2} \biggl(
       g_{i{\h{j}}^*, {\h{i}}}(\phi)
      -g_{i{\h{i}}, {\h{j}}^*}(\phi)
     \biggr) \,,
    \label{eq:defv}
\end{align}
which allows us to study supersymmetric theories in the GHEFT framework.
There is an ambiguity in the decomposition (\ref{eq:defv}) which will be discussed later.
We introduce a function $\chi^{\h{i}}(\phi)$ and its covariant derivative
\begin{align}
  \chi^{\h{i}}_{;i} &:= \chi^{\h{i}}_{, i} + \chi^{{\h{i}}'} \Gamma^{\h{i}}_{i{\h{i}}'} \, .
  \label{eq:covdel-psi}
\end{align}
It is easy to show that the derivative (\ref{eq:covdel-psi}) covariantly transforms under the field transformation (\ref{eq:redef}). 
Moreover, the covariant derivative on the fermionic metric satisfies
\begin{align}
  g_{{\h{i}}{\h{j}}^* ; i }
  := g_{{\h{i}}{\h{j}}^*, i} - g_{{\h{i}}' {\h{j}}^*} \Gamma^{{\h{i}}'}_{i{\h{i}}} - g_{{\h{i}}{\h{j}}'^*} \Gamma^{{\h{j}}'^*}_{i{\h{j}}^*}
  = 0 \, .
\end{align}
The formulas (\ref{eq:affine-f1}) and (\ref{eq:affine-f2}) are therefore 
considered to be ``Affine connections''.

The covariant derivative of $\chi^{\h{i}}_{;i}(\phi)$ is also defined as usual
\begin{align}
  \chi^{\h{i}}_{;ij} &:= (\chi^{\h{i}}_{;i})_{,j}
                + \chi^{{\h{i}}'}_{;i} \Gamma^{\h{i}}_{i{\h{i}}'}
                - \chi^{\h{i}}_{;i'} \Gamma^{i'}_{ij} \, .
\end{align}
We therefore obtain
\begin{align}
  \chi^{\h{i}}_{; ij} 
  -\chi^{\h{i}}_{; ji} 
  &= - \chi^{{\h{i}}'} \left[
          \Gamma^{{\h{i}}'}_{j{\h{i}}', i} - \Gamma^{{\h{i}}}_{i{\h{i}}', j}
         +\Gamma^{\h{i}}_{i{\h{i}}''} \Gamma^{{\h{i}}''}_{j{\h{i}}'} - \Gamma^{\h{i}}_{j{\h{i}}''} \Gamma^{{\h{i}}''}_{i{\h{i}}'}
       \right]
  \nonumber\\
 & = - \chi^{{\h{i}}'} R^{\h{i}}{}_{{\h{i}}' ij} \, .
 \label{eq:half-fermionic-curvature2}
\end{align}
Here we define the ``Riemann curvature'' tensor $R^{\h{i}}{}_{{\h{j}}kl}$ as
\begin{align}
  R^{\h{i}}{}_{{\h{j}}kl}
  &:= \Gamma^{\h{i}}_{l{\h{j}}, k} - \Gamma^{\h{i}}_{k{\h{j}},l}
    +\Gamma^{\h{i}}_{k{\h{l}}} \Gamma^{{\h{l}}}_{l{\h{j}}} 
    -\Gamma^{\h{i}}_{l{\h{l}}} \Gamma^{{\h{l}}}_{k{\h{j}}} \, .
\label{eq:half-fermionic-curvature}
\end{align}
Note that the definition of ``Riemann curvature tensor'' leads 
\begin{align}
   R^{\h{i}}{}_{{\h{j}}kl} +  R^{\h{i}}{}_{{\h{j}}lk} = 0\, .
%  R_{{\h{i}}{\h{j}}^* ij} + R_{{\h{i}}{\h{j}}^* ji} = 0 \, .
\end{align}
It is easy to show that $R^{\h{i}}{}_{{\h{j}}kl}$ %$R_{{\h{i}}{\h{j}}^* ij}$ 
transforms covariantly 
under the coordinate transformation given in 
Eq.\,(\ref{eq:redef}).
For the latter convenience, we also define
\begin{align}
  R_{{\h{j}}^* {\h{i}} kl} &:= g_{{\h{l}}{\h{j}}^*} R^{\h{l}}{}_{{\h{i}}kl} \, ,\\
  R_{{\h{i}} {\h{j}}^* kl} &:= - R_{{\h{j}}^* {\h{i}} kl} \, .
\end{align}
%Here $g^{{\h{i}} {\h{j}}^*}$ is defined as the inverse of $g_{{\h{i}} {\h{j}}^*}$, {\it{i.e.}}, 
%\begin{align}
%  g^{{\h{i}} {\h{k}}^*} g_{{\h{j}} {\h{k}}^*} = \delta^{\h{i}}_{\h{j}} \, , \qquad
%  g^{{\h{k}}{\h{i}}^*} g_{{\h{k}}{\h{j}}^*} = \delta^{{\h{i}}^*}_{{\h{j}}^*} \, ,
%\end{align}

We are now ready to compute the expansion coefficients of the coupling functions in the normal coordinate.
The normal coordinate on the fermion field space is defined so that the coupling functions $g_{\hat{i}\hat{j}^*}$ and $v_{\hat{i}\hat{j}^*i}$ are expanded in terms of the covariantly transforming tensors.
We first focus on $g_{\h{i}\h{j}^*}$. 
Thanks to the Hermiticity of $g_{\h{i}\h{j}^*}$, and since
$g_{\h{i}\h{j}^*}$ does note depend on the fermion fields, 
it is always possible to take a fermion coordinate satisfying
\begin{align}
   g_{{\h{i}}{\h{j}}^*}(\phi)
   &= \delta_{{\h{i}}{\h{j}}^*}  \, .
\end{align}
The expansion of the fermionic metric is therefore trivial. 

We next consider the expansion of $v_{\h{i}\h{j}^*i}$. 
Neglecting the anomaly factor only appearing in the loop level, 
we are allowed to take a coordinate satisfying
\begin{align}
  v_{{\h{i}}{\h{j}}^* i}(\phi)
  &= A_{{\h{i}}{\h{j}}^* ij}(\phi) \, \phi^j \, , 
\label{eq:normal}
\end{align}
with
\begin{align}
  A_{{\h{i}}{\h{j}}^* ij}(\phi)
  &= -A_{{\h{i}}{\h{j}}^* ji}(\phi) \, .
\label{eq:anti-sym-temp}
\end{align}
We resolve the ambiguity in Eq.\,(\ref{eq:defv}) as
\begin{align}
  g_{i{\h{j}}^*,\, {\h{i}}}(\phi)
  = - i A_{{\h{i}}{\h{j}}^* ij}(\phi) \, \phi^j \, , \quad
  g_{i{\h{i}}, \,{\h{j}}^*}(\phi)
  =  i A_{{\h{i}}{\h{j}}^* ij}(\phi) \, \phi^j \, .
\end{align}
It is now straightforward to obtain
\begin{align}
  \Gamma^{\h{i}}_{i{\h{j}}} 
  = - ig^{{\h{i}}{\h{k}}^*} A_{{\h{j}}{\h{k}}^* ij} \phi^j \, , \quad
  \Gamma^{{\h{i}}^*}_{i{\h{j}}^*} 
  =  ig^{{\h{k}}{\h{i}}^*} A_{{\h{k}}{\h{j}}^* ij} \phi^j \, .
\label{eq:fermion-conn2}
\end{align}
We next determine the expansion coefficient of $A_{\hat{i}\hat{j}ij}$. 
Combining Eq.\,(\ref{eq:half-fermionic-curvature}) and Eq.\,(\ref{eq:fermion-conn2}), 
we obtain the master formula for the determination of the coefficients;
\begin{align}
  R_{{\h{i}}{\h{j}}^* ij}
  &= i (A_{{\h{i}}{\h{j}}^* jk} \phi^k )_{, \,i}
    -i (A_{{\h{i}}{\h{j}}^* ik} \phi^k )_{, \,j}
  \nonumber\\
  & \qquad + A_{{\h{i}}' {\h{j}}^* ik_1} g^{{\h{i}}' {\h{j}}'^*} A_{{\h{i}} {\h{j}}'^* jk_2} \phi^{k_1} \phi^{k_2}
   \nonumber\\
  & \qquad 
           - A_{{\h{i}}' {\h{j}}^* jk_1} g^{{\h{i}}' {\h{j}}'^*} A_{{\h{i}} {\h{j}}'^* ik_2} \phi^{k_1} \phi^{k_2} \, .
\label{eq:fermion-curvature}
\end{align}
Plugging the vacuum condition $\phi^i = 0$ in Eq.\,(\ref{eq:fermion-curvature}), 
we obtain
\begin{align}
  R_{{\h{i}}{\h{j}}^* ij} \biggr|_0
  &= i A_{{\h{i}}{\h{j}}^* ji} \biggr|_0 - i A_{{\h{i}}{\h{j}}^* ij} \biggr|_0 \, .
\label{eq:temp1}
\end{align}
Since the function $A_{{\h{i}}{\h{j}}^* ij}(\phi)$ is anti-symmetric under the
exchange of $i\leftrightarrow j$, Eq.\,(\ref{eq:temp1}) can be expressed as
\begin{align}
  R_{{\h{i}}{\h{j}}^* ij} \biggr|_0
  &= -2i A_{{\h{i}}{\h{j}}^* ij} \biggr|_0
\end{align}
and we thus obtain
\begin{align}
  A_{{\h{i}}{\h{j}}^* ij} \biggr|_0
  &= \dfrac{i}{2} R_{{\h{i}}{\h{j}}^* ij} \biggr|_0 \, .
\label{eq:fermion-a1}
\end{align}
Combining Eq.\,(\ref{eq:fermion-a1}) with
Eq.\,(\ref{eq:fermion-conn2}), 
we obtain formulas for the fermionic Affine connections and their
derivatives at the vacuum, 
\begin{align}
  &\Gamma^{\h{i}}_{i{\h{j}}} \biggr|_0
  = 0 \, ,&
  &\Gamma^{{\h{i}}^*}_{i{\h{j}}^*} \biggr|_0
  = 0 \, , 
\label{eq:fermionic-affine02} 
  \\
  &\Gamma^{\h{i}}_{i{\h{j}}, \, j} \biggr|_0
  = \dfrac{1}{2} g^{{\h{i}}{\h{k}}^*} R_{{\h{j}}{\h{k}}^* ij} \biggr|_0 \, ,&
  &\Gamma^{{\h{i}}^*}_{i{\h{j}}^*, \, j} \biggr|_0 
  = -\dfrac{1}{2} g^{{\h{k}}{\h{i}}^*} R_{{\h{k}}{\h{j}}^* ij} \biggr|_0 \, .
\label{eq:fermionic-affine12} 
\end{align}

The higher order terms in the Taylor expansion of $v_{{\h{i}}{\h{j}}^*i}$ in the
normal coordinate are computed in appendix~\ref{sec-taylor-vnni}. 
We find
\begin{align}
  g_{{\h{i}}{\h{j}}^*}(\phi) &= \delta_{{\h{i}}{\h{j}}^*}
\end{align}
and the function $v_{{\h{i}}{\h{j}}^* i}(\phi)$ can be expanded in terms of 
the covariant tensors as
\begin{align}
  v_{{\h{i}}{\h{j}}^* i}(\phi) 
  &= A_{{\h{i}}{\h{j}}^* i k_1} \phi^{k_1}
    +\dfrac{1}{2!} A_{{\h{i}}{\h{j}}^* i k_1 k_2} \phi^{k_1} \phi^{k_2}
     \nonumber\\
  & \qquad 
    +\dfrac{1}{3!} A_{{\h{i}}{\h{j}}^* i k_1 k_2 k_3} \phi^{k_1} \phi^{k_2} \phi^{k_3}
    +\cdots \, ,
\label{eq:taylor-expansion-vnni}
\end{align}
with 
\begin{align}
  A_{{\h{i}}{\h{j}}^* i1} 
  &= \dfrac{i}{2} R_{{\h{i}}{\h{j}}^* i1} \biggr|_0 \, , 
  \\
  A_{{\h{i}}{\h{j}}^* i12}
  &= \dfrac{i}{3} R_{{\h{i}}{\h{j}}^* i(1 ; \, 2)} \biggr|_0 \, , 
  \\
  A_{{\h{i}}{\h{j}}^* i123}
  &= \dfrac{i}{4} R_{{\h{i}}{\h{j}}^* i(1 ; \, 23)} \biggr|_0 
  +\dfrac{i}{36} \left[
      R_{{\h{i}}{\h{j}}^* i' 1} R^{i'}{}_{(23)i}
      \r.\nn\\
      &\l.\qquad
     +R_{{\h{i}}{\h{j}}^* i' 2} R^{i'}{}_{(31)i}
     +R_{{\h{i}}{\h{j}}^* i' 3} R^{i'}{}_{(12)i}
    \right] \biggr|_0 \, ,
  \\
  &\vdots 
  \nonumber
\end{align}
in the normal coordinate.

We next move to $M_{{\h{i}}{\h{j}}}(\phi)$.
Evaluating the Affine connection in the normal coordinate, 
we obtain Eqs.~(\ref{eq:fermionic-affine02}), (\ref{eq:fermionic-affine12}), and  
\begin{align}
  \Gamma^{\h{i}}_{1{\h{j}}, 23} \biggr|_0
  &= -\dfrac{2}{3} R^{\h{i}}{}_{{\h{j}}1(2; \, 3)} \biggr|_0 \, .
\end{align}
It is now easy to evaluate
\begin{align}
  M_{{\h{i}}{\h{j}}, \, 1} \biggr|_0
  &= M_{{\h{i}}{\h{j}}; \, 1} \biggr|_0 \, , 
  \\
  M_{{\h{i}}{\h{j}}, \, 12} \biggr|_0 
  &= M_{{\h{i}}{\h{j}};\,  12} \biggr|_0 
   -\dfrac{1}{2} \left[
     M_{{\h{i}}'{\h{j}}} R^{{\h{i}}'}{}_{{\h{i}}12}
   + M_{{\h{i}}{\h{j}}'} R^{{\h{j}}'}{}_{{\h{j}}12} 
   \right] \biggr|_0 \, , 
  \\
  M_{{\h{i}}{\h{j}}, \, 123} \biggr|_0
  &= M_{{\h{i}}{\h{j}}; \, 123} \biggr|_0
  \nn\\
  &\!\!\!\!\!\!\!\!
  -\dfrac{1}{2} \left[
        M_{{\h{i}}' {\h{j}}; 1} R^{{\h{i}}'}{}_{{\h{i}}23}
       +M_{{\h{i}}' {\h{j}}; 2} R^{{\h{i}}'}{}_{{\h{i}}13}
       +M_{{\h{i}}' {\h{j}}; 3} R^{{\h{i}}'}{}_{{\h{i}}12}
     \right] \biggr|_0
  \nonumber\\
  &\!\!\!\!\!\!\!\!
    -\dfrac{1}{2} \left[
        M_{{\h{i}} {\h{j}}'; 1} R^{{\h{j}}'}{}_{{\h{j}}23}
       +M_{{\h{i}} {\h{j}}'; 2} R^{{\h{j}}'}{}_{{\h{j}}13}
       +M_{{\h{i}} {\h{j}}'; 3} R^{{\h{j}}'}{}_{{\h{j}}12}
     \right] \biggr|_0
     \nonumber\\
  &\!\!\!\!\!\!\!\!
  -\dfrac{2}{3} M_{{\h{i}}{\h{j}}; i} R^i{}_{(12)3} \biggr|_0
  \nonumber\\
  &\!\!\!\!\!\!\!\!
   -\dfrac{2}{3} \left[
       M_{{\h{i}}'{\h{j}}} R^{{\h{i}}'}{}_{{\h{i}}1(2; 3)} + M_{{\h{i}}{\h{j}}'} R^{{\h{j}}'}{}_{{\h{j}}1(2; 3)} 
    \right] \biggr|_0 
\end{align}
and therefore
\begin{align}
  M_{{\h{i}}{\h{j}}, \, 1} \biggr|_0
  &= M_{{\h{i}}{\h{j}}; \, 1} \biggr|_0 \, , 
  \\
  M_{{\h{i}}{\h{j}}, \, 12} \biggr|_0 
  &= M_{{\h{i}}{\h{j}};\,  (12)} \biggr|_0  \, ,
  \\
  M_{{\h{i}}{\h{j}}, \, 123} \biggr|_0
  &= M_{{\h{i}}{\h{j}}; \, (123)} \biggr|_0 \, .
\end{align}
The fermion mass term in the Lagrangian (\ref{eq:gheft1}) can also be
expanded in terms of the covariant tensors
\begin{align}
  M_{{\h{i}}{\h{j}}}(\phi)
  &= M_{{\h{i}}{\h{j}}} \biggr|_0 
      + M_{{\h{i}}{\h{j}};\,  k_1} \biggr|_0 \phi^{k_1} 
      + \dfrac{1}{2!} M_{{\h{i}}{\h{j}}; (k_1 k_2)} \biggr|_0 \phi^{k_1} \phi^{k_2}
      \nn\\
      &
      + \dfrac{1}{3!} M_{{\h{i}}{\h{j}}; (k_1 k_2 k_3)} \biggr|_0 \phi^{k_1} \phi^{k_2} \phi^{k_3} 
      + \cdots
      \label{eq:M-exp}
\end{align}
in the normal coordinate.

It is worth emphasizing that the introduction of the metric-like
objects $g_{i\h{j}}$ and $g_{i\h{j}^*}$ in (\ref{eq:defv}) allows us to express
covariant formulas (\ref{eq:taylor-expansion-vnni}) and (\ref{eq:M-exp})
in compact forms.
These metric-like objects, which mix the scalars and fermions,
may be understood as ``convenient abbreviations'' in the present 
non-supersymmetric case.
They can be regarded as ``metrics'', however, in a real sense
if we embed the theory in supersymmetric models.

\subsection{Holomorphic four-fermion sector}
We consider holomorphic four-fermion operators
\begin{equation}
  {\cal O}^{(({\h{i}_1}{\h{i}_2})({\h{i}_3}{\h{i}_4}))}
  := (\psi^{{\h{i}}_1}_\alpha \, \varepsilon^{\alpha\beta} \, \psi^{{\h{i}}_2}_\beta ) \, 
     (\psi^{{\h{i}}_3}_\gamma \, \varepsilon^{\gamma\delta} \,  \psi^{{\h{i}}_4}_\delta ) \, .
\label{eq:four-fermion-type}
\end{equation}
We put the indices $\h{i}_1$, $\h{i}_2$, $\h{i}_3$, $\h{i}_4$ in parentheses
so as to emphasize the index-exchange symmetry, 
\begin{equation}
  {\cal O}^{(({\h{1}}{\h{2}})({\h{3}}{\h{4}}))}
  = {\cal O}^{(({\h{1}}{\h{2}})({\h{4}}{\h{3}}))}
  = {\cal O}^{(({\h{2}}{\h{1}})({\h{3}}{\h{4}}))}
  = {\cal O}^{(({\h{3}}{\h{4}})({\h{1}}{\h{2}}))} \, ,
\label{eq:riemann-like}
\end{equation}
with $\h{i}_1$, $\h{i}_2$, $\cdots$ being
abbreviated by $\h{1}$, $\h{2}$, $\cdots$.
Furthermore, 
multiplying the fermion fields
$\psi^{{\h{1}}}_\alpha \, \psi^{{\h{2}}}_\beta \, \psi^{{\h{3}}}_\gamma \, \psi^{{\h{4}}}_\delta$
to the Schouten identity
\begin{equation}
  \varepsilon^{\alpha\beta} \, \varepsilon^{\gamma\delta}
  +  \varepsilon^{\alpha\gamma} \, \varepsilon^{\delta\beta}
  +  \varepsilon^{\alpha\delta} \, \varepsilon^{\beta\gamma}
  \equiv 0 \, , 
  \label{eq:Schouten}
\end{equation}
we obtain a Bianchi-like identity
\begin{equation}
  {\cal O}^{(({\h{1}}{\h{2}})({\h{3}}{\h{4}}))}
  +   {\cal O}^{(({\h{1}}{\h{3}})({\h{4}}{\h{2}}))}
  +   {\cal O}^{(({\h{1}}{\h{4}})({\h{2}}{\h{3}}))}
  \equiv 0 \, .
\label{eq:bianchi-like}
\end{equation}

Using the Bianchi-like identity (\ref{eq:bianchi-like}), we are able to show
\begin{equation}
  {\cal O}^{(({\h{1}}{\h{1}})({\h{1}}{\h{1}}))}
+ {\cal O}^{(({\h{1}}{\h{1}})({\h{1}}{\h{1}}))}
+ {\cal O}^{(({\h{1}}{\h{1}})({\h{1}}{\h{1}}))} \equiv 0 \, , 
\end{equation}
therefore,
\begin{equation}
  {\cal O}^{(({\h{1}}{\h{1}})({\h{1}}{\h{1}}))} \equiv 0 \, .
\label{eq:cond1}
\end{equation}
In a similar manner, we find
\begin{align}
  {\cal O}^{(({\h{1}}{\h{2}})({\h{2}}{\h{2}}))} &\equiv 0  \, , 
\label{eq:cond2}
  \\
  {\cal O}^{(({\h{1}}{\h{1}})({\h{2}}{\h{2}}))} &\equiv -2 {\cal O}^{(({\h{1}}{\h{2}})({\h{1}}{\h{2}}))} \, .
\label{eq:cond3}
  \\
  {\cal O}^{(({\h{1}}{\h{1}})({\h{2}}{\h{3}}))} &\equiv - 2 {\cal O}^{(({\h{1}}{\h{2}})({\h{1}}{\h{3}}))} \, , 
\label{eq:cond4}
  \\
  {\cal O}^{(({\h{1}}{\h{2}})({\h{3}}{\h{4}}))} &\equiv - {\cal O}^{(({\h{1}}{\h{3}})({\h{2}}{\h{4}}))} - {\cal O}^{(({\h{1}}{\h{4}})({\h{2}}{\h{3}}))} \, .
\label{eq:cond5}
\end{align}

We next count the independent degrees of freedom (DOF) of the four fermion operators.
The number of independent DOF satisfying the
condition (\ref{eq:riemann-like}) is 
\begin{equation}
  \dfrac{1}{2} \left[
    \dfrac{1}{2} N (N+1)
  \right] \, \left[
    \dfrac{1}{2} N (N+1) + 1
  \right]  \, .
\end{equation}
Among them, the four-fermion operators having the identical
flavor index automatically vanish as shown in Eq.\,(\ref{eq:cond1}),
which reduces the DOF by $N$.
In a similar manner, the operator identities 
(\ref{eq:cond2}), (\ref{eq:cond3}), (\ref{eq:cond4}), and
(\ref{eq:cond5}) 
reduce
the DOF by
\begin{align}
&  N(N-1) \, , \qquad
  \dfrac{1}{2} N(N-1) \, , \nn\\
&  \dfrac{1}{2} N(N-1)(N-2) \, , \qquad
  \dfrac{1}{4!} N(N-1)(N-2)(N-3) \, , 
  \nonumber
\end{align}
accordingly.  We therefore find the DOF of the four-fermion operator
${\cal O}^{((12)(34))}$ is given by
\begin{align}
&  \dfrac{1}{2} \left[
    \dfrac{1}{2} N (N+1)
  \right] \, \left[
    \dfrac{1}{2} N (N+1) + 1
  \right]  \nn\\
  &
  -  N
  -N(N-1) 
  -\dfrac{1}{2} N(N-1) 
  -\dfrac{1}{2} N(N-1)(N-2) 
  \nonumber\\
  &
  -\dfrac{1}{4!} N(N-1)(N-2)(N-3) 
 =\dfrac{1}{12} N^2(N^2-1) \, ,
\label{eq:DOF-O}
\end{align}
which accords the DOF of the $N$-dimensional Riemann curvature tensor.

We are now ready to consider the holomorphic 
four-fermion interactions of the type shown in Eq.\,(\ref{eq:four-fermion-type}), 
\begin{align}
  {\cal L} 
  &\ni 
  \dfrac{1}{8} S_{{\h{i}}_1 {\h{i}}_2 {\h{i}}_3 {\h{i}}_4} 
  (\psi^{{\h{i}}_1}_\alpha \, \varepsilon^{\alpha\beta} \, \psi^{{\h{i}}_2}_\beta ) \, 
     (\psi^{{\h{i}}_3}_\gamma \, \varepsilon^{\gamma\delta} \,  \psi^{{\h{i}}_4}_\delta ) 
\, .
  \nonumber\\
  &= \dfrac{1}{8} S_{{\h{i}_1}{\h{i}_2} {\h{i}_3}{\h{i}_4}} 
     {\cal O}^{(({\h{i}_1}{\h{i}_2})({\h{i}_3}{\h{i}_4}))} \, .  
\end{align}
Thanks to the index-exchange symmetry (\ref{eq:riemann-like}), we are able 
to show
\begin{equation}
  S_{{\h{i}_1}{\h{i}_2} {\h{i}_3}{\h{i}_4}} 
     {\cal O}^{(({\h{i}_1}{\h{i}_2})({\h{i}_3}{\h{i}_4}))} 
  = S_{(({\h{i}_1}{\h{i}_2}) ({\h{i}_3}{\h{i}_4}))}
     {\cal O}^{(({\h{i}_1}{\h{i}_2})({\h{i}_3}{\h{i}_4}))} \, ,
\end{equation}
with
\begin{align*}
  S_{((\h{1}  \h{2}) (\h{3} \h{4}))}
  &:= \dfrac{1}{2} \left[
        S_{(\h{1}  \h{2}) (\h{3} \h{4})}
       +S_{(\h{3} \h{4}) (\h{1}  \h{2}) }
     \right] \, ,
  \\
S_{(\h{1} \h{2})(\h{3}\h{4})}
  &:= \dfrac{1}{4} \left[
        S_{\h{1}  \h{2} \h{3} \h{4}}
       +S_{\h{2}  \h{1} \h{3} \h{4}}
       +S_{\h{1}  \h{2} \h{4} \h{3}}
       +S_{\h{2}  \h{1} \h{4} \h{3}}
      \right] \, ,
\end{align*}
which, of course, satisfies the index-exchange symmetry
\begin{align}
  S_{(({\h{1}}{\h{2}})({\h{3}}{\h{4}}))} 
= S_{(({\h{2}}{\h{1}})({\h{3}}{\h{4}}))} 
= S_{(({\h{1}}{\h{2}})({\h{4}}{\h{3}}))} 
= S_{(({\h{3}}{\h{4}})({\h{1}}{\h{2}}))}  \, .
\label{eq:symmetry-s}
\end{align}
Therefore the DOF of $S_{(({\h{1}}{\h{2}})({\h{3}}{\h{4}}))}$ is
counted as
\begin{align}
    \dfrac{1}{2} \left[
    \dfrac{1}{2} N (N+1)
  \right] \, \left[
    \dfrac{1}{2} N (N+1) + 1
  \right]  \, ,
\end{align}
which is larger than the DOF of the operator 
${\cal O}^{(({\h{1}}{\h{2}})({\h{3}}{\h{4}}))}$ 
as counted in Eq.\,(\ref{eq:DOF-O}).
The $S_{(({\h{1}}{\h{2}})({\h{3}}{\h{4}}))}$ parametrization 
therefore contains redundancy.
It is desired to describe the four-fermion interactions in a
non-redundant parametrization.
For such a purpose, we rewrite the 
four-fermion interactions as
\begin{align}
 \lefteqn{ S_{(({\h{i}_1}{\h{i}_2})({\h{i}_3}{\h{i}_4}))} 
 {\cal O}^{(({\h{i}_1}{\h{i}_2})({\h{i}_3}{\h{i}_4}))}} \nn\\
  &= \dfrac{2}{3} \left[
      S_{(({\h{i}_1}{\h{i}_2})({\h{i}_3}{\h{i}_4}))} 
    - S_{(({\h{i}_1}{\h{i}_3})({\h{i}_2}{\h{i}_4}))}
     \right] {\cal O}^{(({\h{i}_1}{\h{i}_2})({\h{i}_3}{\h{i}_4}))} \, ,
\label{eq:rewrite}
\end{align}
where we used the
Bianchi-like identity (\ref{eq:bianchi-like}) and 
the index-exchange symmetry (\ref{eq:symmetry-s}).

We are now ready to introduce a non-redundant parametrization for
holomorphic four-fermion interactions,
\begin{align}
  R_{{\h{1}}{\h{4}}{\h{2}}{\h{3}}} := 
      S_{({\h{1}}{\h{2}})({\h{3}}{\h{4}})} - S_{({\h{1}}{\h{3}})({\h{2}}{\h{4}})}
  \, .
\label{eq:new-coeff}
\end{align}
which satisfies the index-exchange symmetries
\begin{align}
  &R_{{\h{1}}{\h{2}}{\h{3}}{\h{4}}}
  +
  R_{{\h{1}}{\h{2}}{\h{4}}{\h{3}}} =0\, , \\
  &
  R_{{\h{1}}{\h{2}}{\h{3}}{\h{4}}}
  +
  R_{{\h{2}}{\h{1}}{\h{3}}{\h{4}}} =0\, , \\
  &
  R_{{\h{1}}{\h{2}}{\h{3}}{\h{4}}}
  - 
  R_{{\h{3}}{\h{4}}{\h{1}}{\h{2}}} =0\, ,
\end{align}
and the Bianchi identity
\begin{align}
  R_{{\h{1}}{\h{2}}{\h{3}}{\h{4}}}
 +R_{{\h{1}}{\h{3}}{\h{4}{\h{2}}}}
 +R_{{\h{1}}{\h{4}}{\h{2}}{\h{3}}} 
  = 0 \, .
\end{align}
The DOF of $R_{\h{1}\h{2}\h{3}\h{4}}$ is
\begin{align}
&  \dfrac{1}{2} \left[
    \dfrac{1}{2} N (N-1)
  \right] \, \left[
    \dfrac{1}{2} N (N-1) + 1
  \right]  
 \nn\\
 &
  -\dfrac{1}{4!} N(N-1)(N-2)(N-3) 
 =\dfrac{1}{12} N^2(N^2-1) \, ,
\label{eq:DOF-A}
\end{align}
which coincides with the DOF of the operators 
${\cal O}^{(({\h{1}}{\h{2}})({\h{3}}{\h{4}}))}$.
The parametrization
\begin{align}
  {\cal L}_{\mbox{\scriptsize four-fermion}} = \dfrac{1}{12} 
  R_{{\h{i}_1}{\h{i}_4}{\h{i}_2}{\h{i}_3}} 
  \,{\cal O}^{(({\h{i}_1}{\h{i}_2})({\h{i}_3}{\h{i}_4}))} 
\end{align}
therefore describes the holomorphic four-fermion interactions 
in a non-redundant manner.

\subsection{Non-holomorphic four-fermion sector}
We next consider non-holomorphic four-fermion operators
\begin{align}
  {\cal O}^{({\h{i}_1}{\h{i}_2})({\h{i}}_3^* {\h{i}}_4^*)}
  := (\psi_\alpha^{{\h{i}}_1} \varepsilon^{\alpha\beta} \psi_\beta^{{\h{i}}_2}) \, 
    (\psi_{\dot{\alpha}}^{\dagger {\h{i}}_3^*} \varepsilon^{\dot{\alpha}\dot{\beta}} \psi_{\dot{\beta}}^{\dagger {\h{i}}_4^*})  \, .
\end{align}
Again, we put the indices in parentheses 
in order to emphaize the index-exhange symmetry
\begin{align}
  {\cal O}^{({\h{1}}{\h{2}})({\h{3}}^* {\h{4}}^*)}
  = {\cal O}^{({\h{1}}{\h{2}})({\h{4}}^* {\h{3}}^*)}
  = {\cal O}^{({\h{2}}{\h{1}})({\h{3}}^* {\h{4}}^*)} \, ,
\label{eq:condition1}
\end{align}
with $\h{i}_1$, $\h{i}_2$, $\h{i}_3^*$, $\h{i}_4^*$ being
abbreviated by
$\h{1}$, $\h{2}$, $\h{3}^*$, $\h{4}^*$.

The DOF of the four fermion operators satisfying the 
conditions (\ref{eq:condition1}) is
\begin{align}
  \left[ \dfrac{1}{2} N (N+1) \right]^2 \, . 
\label{eq:dof-non-holomorphic}
\end{align}
Note 
\begin{align}
  [{\cal O}^{({\h{1}}{\h{2}})({\h{3}}^* {\h{4}}^*)}]^\dagger
  = {\cal O}^{({\h{3}}{\h{4}})({\h{1}}^* {\h{2}}^*)} \, .
\label{eq:condition2}
\end{align}
The DOF as counted in (\ref{eq:dof-non-holomorphic}) is therefore
regarded as the degrees of freedom counted 
in {\em real} parameters.
This is in contrast to the DOF of holomorphic four-fermion
operators (\ref{eq:DOF-O}) counted in {\em complex} parameters.

Nonholomorphic four-fermion interactions in the lowest order 
GHEFT Lagrangian can be expressed as
\begin{align}
  {\cal L} 
  & \ni -\dfrac{1}{4} S_{{\h{i}}_1 {\h{i}}_2 {\h{i}}_3^* {\h{i}}_4^*} 
    (\psi_\alpha^{{\h{i}}_1} \varepsilon^{\alpha\beta} \psi_\beta^{{\h{i}}_2}) \, 
    (\psi_{\dot{\alpha}}^{\dagger {\h{i}}_3^*} \varepsilon^{\dot{\alpha}\dot{\beta}} \psi_{\dot{\beta}}^{\dagger {\h{i}}_4^*})  
  \nonumber\\
  &= -\dfrac{1}{4} S_{{\h{i}_1}{\h{i}_2} {\h{i}}_3^* {\h{i}}_4^*} \, 
     {\cal O}^{({\h{i}_1}{\h{i}_2})({\h{i}}_3^* {\h{i}}_4^*)} \, .
\end{align}
Thanks to the index-exchange symmetry (\ref{eq:condition1}), we 
are able to show
\begin{align}
  S_{\h{i}_1 \h{i}_2 \h{i}_3^* \h{i}_4^*} 
  {\cal O}^{(\h{i}_1 \h{i}_2) (\h{i}_3^* \h{i}_4^*)}
  = 
  S_{(\h{i}_1 \h{i}_2) (\h{i}_3^* \h{i}_4^*)} 
  {\cal O}^{(\h{i}_1 \h{i}_2) (\h{i}_3^* \h{i}_4^*)} \, ,
\end{align}
with
\begin{align*}
  S_{(\h{1} \h{2})(\h{3}^*\h{4}^*)}
  &:= \dfrac{1}{4} \left[
        S_{\h{1}  \h{2} \h{3}^* \h{4}^*}
       +S_{\h{2}  \h{1} \h{3}^* \h{4}^*}
       +S_{\h{1}  \h{2} \h{4}^* \h{3}^*}
       +S_{\h{2}  \h{1} \h{4}^* \h{3}^*}
      \right] \, .
\end{align*} 
It is easy to show
\begin{align}
  S_{(\h{1} \h{2})(\h{3}^*\h{4}^*)}
  = S_{(\h{2}\h{1} )(\h{3}^*\h{4}^*)}
  = S_{(\h{1} \h{2})(\h{4}^*\h{3}^*)} \, .
\end{align}
Since the Hermiticity of the Lagrangian requires
\begin{align}
  [S_{({\h{1}}{\h{2}})({\h{3}}^* {\h{4}}^*)}]^* = S_{({\h{3}}{\h{4}})({\h{1}}^* {\h{2}}^*)} 
  \, ,
\end{align}
we find the number of DOF of $S_{(\h{1} \h{2})(\h{3}^*\h{4}^*)}$ 
is $N^2 (N+1)^2/4$ {\em real} parameters, 
which agrees with the DOF of 
the nonholomorphic four-fermion operators (\ref{eq:dof-non-holomorphic}).
Therefore, the non-holomorphic four-fermion interactions can be
parametrized by using $S_{({\h{1}}{\h{2}})({\h{3}}^* {\h{4}}^*)}$ in 
a non-redundant manner.

We finally remark that the nonholomorphic four fermion operators appear in the supersymmetric non-linear sigma model as
\begin{equation}
  {\cal L} \ni
  \dfrac{1}{4} R_{\h{i}_1 \h{i}_3^* \h{i}_2 \h{i}_4^*} 
  (\psi^{\h{i}_1} \psi^{\h{i}_2})
  (\psi^{\dagger \h{i}^*_3} \psi^{\dagger \h{i}^*_4})
  \,,
\end{equation}
where
\begin{equation}
  R_{\h{i}_1 \h{i}_3^* \h{i}_2 \h{i}_4^*} 
  = g_{\h{i}_1 \h{i}_3^* , \, \h{i}_2 \h{i}_4^*}
   -g^{\h{i}' \h{i}^{\prime *}} g_{\h{i}' \h{i}_3^*, \, \h{i}_4^*}
    g_{\h{i}_1 \h{i}^{\prime *}, \, \h{i}_2} \, ,
\end{equation}
with $g_{\hat{i}\hat{j}^*}$ being the K\"{a}hler metric. We therefore define
\begin{equation}
  R_{\h{1}\h{3}^* \h{2} \h{4}^*} := S_{(\h{1}\h{2})(\h{3}^* \h{4}^*)} \, ,  
  \label{eq:Rnonh}
\end{equation}
for the non-holomorphic four-fermion couplings even in non-supersymmetric GHEFT.

%%%%%%%%%%%%%%%%%%%%%%%
\section{On-shell amplitudes}
\label{sec:amplitude}

The purpose of the GHEFT Lagrangian is to compute the production
cross sections and the decay widths involving the new BSM 
particles.
As we have shown in the previous section, 
the non-uniqueness of the parametrization in the effective Lagrangian
associated with the KOS theorem~\cite{Kamefuchi:1961sb}
can be resolved by using the normal 
coordinate.
The scattering amplitudes can now be computed straightforwardly
in the normal coordinate as functions of the covariant tensors.
Applying the normal coordinate in the GHEFT, 
the on-shell amplitudes are expressed by covariant 
quantities on the coupling functions evaluated at the vacuum.

In this section,
we explicitly compute tree-level on-shell helicity amplitudes 
applying the normal coordinate in the lowest order GHEFT Lagrangian.
In the computation of the on-shell amplitudes, we ignore the gauge boson 
contributions for simplicity. 
The computation on the on-shell amplitudes including spin-1 gauge bosons
will be published elsewhere.
The high-energy behavior of the longitudinally polarized 
gauge boson scattering amplitudes can be computed even in the
gaugeless limit,  
thanks to equivalence theorem 
between the longitudinally polarized gauge boson scattering amplitudes 
and the corresponding would-be NG boson 
amplitudes~\cite{Cornwall:1974km,Chanowitz:1985hj,Gounaris:1986cr,He:1993yd,He:1993qa}. 
In what follows, we also study the high-energy behaviors of 
the on-shell amplitudes and discuss their implications.

\subsection{Notation}
We express an $N$-particle invariant amplitude generally as
\begin{equation}
  {\cal A}_N(12\cdots  N)  \, .
\end{equation}
Generalized Mandelstam variables and particle masses are
\begin{equation}
  s_{ij} := (p_i + p_j)^2 \,  , \qquad
  m_i := \sqrt{p_i^2} \, .   
\end{equation}
with the momentum of $i$-th particle $p_i$ is understood to be outgoing.
For an example, the amplitude involving two-fermions and 
one-scalar is denoted as
\begin{equation}
\mathcal{A}_3(123) = 
\mathcal{A}_3
(
{\bf{{1}}}^{\lambda_{{1}}}, {\bf{{2}}}^{\lambda_{{2}}}, 3)
\,,
\label{eq:amplitude-in-bold}
\end{equation}
where 
${\bf 1}^{\lambda_1}$, ${\bf 2}^{\lambda_2}$
denote the momentum, helicity and flavor quantum numbers 
for the on-shell fermions, and
$3$ denotes the momentum and flavor quantum numbers for the 
on-shell bosons.
$\lambda_i$ labels the helicity of the fermion state.

If the fermion masses could be neglected in the amplitude, 
we were able to use the celebrated spinor-helicity 
formalism in the massless limit~\cite{Dixon:2013uaa}.
The masses of heavy particles including BSM particles
cannot be neglected, however, 
in the GHEFT framework.
We therefore 
employ the Dreiner-Haber-Martin (DHM) notation~\cite{Dreiner:2008tw}
for the two-component fermion wavefunctions in the 
amplitudes. 
The fermion wavefunction carrying three momentum $\vec{p}$ is 
expressed by two-component spinors,
\be
x_\alpha (\vec{p},\lambda)\,,~~~~~
y_\alpha (\vec{p},\lambda)\,,~~~~~
\ee
with $x_\alpha (\vec{p},\lambda)$ and
$y_\alpha (\vec{p},\lambda)$ being positive and negative frequency
wavefunctions, respectively.
$\lambda=\pm 1$ labels the little group representation index 
for the massive spin-1/2 fermion \cite{Arkani-Hamed:2017jhn}.
The explicit forms of the spinor wavefunctions 
are summarized in appendix \ref{app:helicitystate}. 

For later convenience, we introduce square/angle bras and kets denoting 
massive spinor wavefunctions,
\begin{align}
( \, [{\bf 1}^{\lambda_1} \, )^\alpha  
  & := y^\alpha(\vec{p}_1, \lambda_1) \, , 
  \label{eq:defmassivebraket1}
  \\
(\, {\bf 1}^{\lambda_1}] \, )_\alpha   
  & := y_\alpha(\vec{p}_1, \lambda_1) \, , 
  \label{eq:defmassivebraket2}
\\
(\, \langle {\bf 1}^{\lambda_1} \, )_{\dot{\alpha}} 
  &:= x^\dagger_{\dot{\alpha}} (\vec{p}_1 , \lambda_1) \, , 
  \label{eq:defmassivebraket3}
  \\
(\, {\bf 1}^{\lambda_1} \rangle \, )^{\dot{\alpha}} 
  &:= x^{\dagger{\dot{\alpha}}} (\vec{p}_1 , \lambda_1) \, ,
  \label{eq:defmassivebraket4}
\end{align}
where 
\bea
&&
x^\alpha(\vec{p},\lambda)
:=
\varepsilon^{\alpha\beta}\,
x_\beta(\vec{p},\lambda)
\,,\\
&&
y^\alpha(\vec{p},\lambda)
:=
\varepsilon^{\alpha\beta}\,
y_\beta(\vec{p},\lambda)
\,,\\
&&
x^\dag_{\dot{\alpha}}(\vec{p},\lambda)
:=
\varepsilon_{\dot{\alpha}\dot{\beta}}\,
x^{\dag\dot{\beta}} (\vec{p},\lambda)
\,,\\
&&
y^\dag_{\dot{\alpha}}(\vec{p},\lambda)
:=
\varepsilon_{\dot{\alpha}\dot{\beta}}\,
y^{\dag\dot{\beta}} (\vec{p},\lambda)
\,.
\eea
Bracket notations for $x_\alpha$ and $y^\dagger_{\dot{\alpha}}$
do not need to be introduced, 
since the amplitudes can be expressed 
without using $x_\alpha$ and $y^\dagger_{\dot{\alpha}}$. 
The inner products among these spinor wavefunctions are expressed as
\bea
\spb{{\bf{1}}^{\lambda_1}}.{{\bf{2}}^{\lambda_2}}
&=&
y^\alpha(\vec{p}_1,\lambda_1)\,
y_\alpha(\vec{p}_2,\lambda_2)
\,,
\label{eq:defspb}
\\
\spa{{\bf{1}}^{\lambda_1}}.{{\bf{2}}^{\lambda_2}}
&=&
x^\dag_{\dot{\alpha}}(\vec{p}_1,\lambda_1)\,
x^{\dag{\dot{\alpha}}}(\vec{p}_2,\lambda_2)
\, ,
\label{eq:defspa}
\eea
In the massless limit,
these brackets reduce to the
massless angle/square brackets,
\begin{align}
  \spa{\bf{1}^{\lambda_1}}.{\bf{2}^{\lambda_2}} & \to 
  \left\{
    \begin{array}{cc}
      \spa{1}.{2}  & \mbox{ for }  \lambda_1 = \lambda_2 = -1 \, , \\
      0 & \mbox{ otherwise}  
    \end{array}
    \right. 
    \label{eq:massless1}
\\
  \spb{\bf{1}^{\lambda_1}}.{\bf{2}^{\lambda_2}} & \to 
  \left\{
    \begin{array}{cc}
      \spb{1}.{2}  & \mbox{ for }  \lambda_1 = \lambda_2 = +1 \, , \\
      0 & \mbox{ otherwise}  
    \end{array}
    \right. 
     \label{eq:massless2}
\end{align}
Here the massless spinor wavefunctions are denoted by non-bold bras and kets
$\langle 1$, $\langle 2$, $[ 1$, $[ 2$, 
$1 \rangle$, $2 \rangle$, $1]$, $2]$. 
See Ref.~\cite{Dixon:2013uaa} for the detail of the massless spinor formalism.
Note that the index $\lambda=\pm 1$ in $\langle {\bf 1}^{\lambda}$
and $[ {\bf 1}^{\lambda}$ corresponds to the helicity of the outgoing
state.

Here we briefly summarize the properties of massless spinor wavefunctions.
The massless spinor wavefunctions satisfy 
the exchange (anti)symmetries,
\begin{align}
 & \spa{1}.{2} = -\spa{2}.{1} \, ,\nn\\
 & \spb{1}.{2} = -\spb{2}.{1} \, , \nn\\
 & \langle 1 \bar{\sigma}^\mu 2 ]
  = [2 \sigma^\mu 1 \rangle \, ,
\end{align}
and the Fierz identity
\begin{align}
  \langle 1 \bar{\sigma}^\mu 2 ] \, 
   [3 \sigma^\mu 4 \rangle
  = 2 \spa{1}.{4}  \spb{3}.{2} \, . 
\end{align}
Eq.\,(\ref{eq:Schouten}) leads to the Schouten identities,
\begin{align}
&
\spb{1}.{2}
\spb{3}.{4}
+
\spb{1}.{3}
\spb{4}.{2}
+
\spb{1}.{4}
\spb{2}.{3}
=
0\,,\\
&
\spa{1}.{2}
\spa{3}.{4}
+
\spa{1}.{3}
\spa{4}.{2}
+
\spa{1}.{4}
\spa{2}.{3}
=
0\,.
\end{align}
The matrices $\sigma^\mu$ and $\bar{\sigma}^\mu$ can be decomposed
into spinor products as
  \begin{equation}
    (\, p\, ])_\alpha \, (\langle \, p)_{\dot{\alpha}} 
    = p_\mu  \, ({\sigma}^\mu)_{\alpha\dot{\alpha}} \, ,
    \quad
    (\, p \, \rangle )^{\dot{\alpha}} \, ([\, p)^{\alpha} 
    = p_\mu  \, (\bar{\sigma}^\mu)^{\dot{\alpha}\alpha} \, .
\label{eq:massless-decomposition}
  \end{equation}
The complex conjugates are given by
\begin{align}
  (\spa{1}.{2})^*
  = \spb{2}.{1}  \, , \qquad
  (\spb{1}.{2})^*
  = \spa{2}.{1}  \, .
\label{eq:spinor-conjugate}
\end{align}
Combining Eq.\,(\ref{eq:massless-decomposition})
and Eq.\,(\ref{eq:spinor-conjugate}), we are able to show
\begin{align}
  |\spa{1}.{2}|^2
  =|\spb{1}.{2}|^2 = 2 p_1 \cdot p_2 
  = (p_1+p_2)^2 \, .
\end{align}

Similarly, the massive spinor wavefunctions satisfy
the exchange (anti)symmetries,
\begin{align}
&
  \spa{{\bf 1}^{\lambda_1}}.{{\bf 2}^{\lambda_2}} = -\spa{{\bf 2}^{\lambda_2}}.{{\bf 1}^{\lambda_1}} \, ,\nn\\
&  
  \spb{{\bf 1}^{\lambda_1}}.{{\bf 2}^{\lambda_2}} = -\spb{{\bf 2}^{\lambda_2}}.{{\bf 1}^{\lambda_1}} \, , \nn\\
  &
  \langle {\bf 1}^{\lambda_1} \bar{\sigma}^\mu {\bf 2}^{\lambda_2} ]
  = [{\bf 2}^{\lambda_2} \sigma^\mu {\bf 1}^{\lambda_1} \rangle \, ,
\end{align}
and the Fierz identity
\begin{align}
  \langle {\bf 1}^{\lambda_1} \bar{\sigma}^\mu {\bf 2}^{\lambda_2} ] \, 
   [{\bf 3}^{\lambda_3} \sigma_\mu {\bf 4}^{\lambda_4} \rangle
  = 2 \spa{{\bf 1}^{\lambda_1}}.{{\bf 4}^{\lambda_4}}  \spb{{\bf 3}^{\lambda_3}}.{{\bf 2}^{\lambda_2}} \, . 
\end{align}
The Schouten identites among the massive spinor
wavefunctions are
\begin{align}
&
\spb{{\bf{1}}^{\lambda_1}}.{{\bf{2}}^{\lambda_2}}
\spb{{\bf{3}}^{\lambda_3}}.{{\bf{4}}^{\lambda_4}}
\nn\\
&\qquad +
\spb{{\bf{1}}^{\lambda_1}}.{{\bf{3}}^{\lambda_3}}
\spb{{\bf{4}}^{\lambda_4}}.{{\bf{2}}^{\lambda_2}}
\nn\\
&\qquad \qquad +
\spb{{\bf{1}}^{\lambda_1}}.{{\bf{4}}^{\lambda_4}}
\spb{{\bf{2}}^{\lambda_2}}.{{\bf{3}}^{\lambda_3}}
=
0\,,
\label{eq:bracket-schouten1}
\\
&
\spa{{\bf{1}}^{\lambda_1}}.{{\bf{2}}^{\lambda_2}}
\spa{{\bf{3}}^{\lambda_3}}.{{\bf{4}}^{\lambda_4}}
\nn\\
&\qquad +
\spa{{\bf{1}}^{\lambda_1}}.{{\bf{3}}^{\lambda_3}}
\spa{{\bf{4}}^{\lambda_4}}.{{\bf{2}}^{\lambda_2}}
\nn\\
&\qquad \qquad +
\spa{{\bf{1}}^{\lambda_1}}.{{\bf{4}}^{\lambda_4}}
\spa{{\bf{2}}^{\lambda_2}}.{{\bf{3}}^{\lambda_3}}
=
0\,.
\label{eq:bracket-schouten2}
\end{align}
The completeness relations are expressed as
\begin{align}
  &\left(
  \sum_{\lambda=\pm 1} {\bf 1}^\lambda ] \, \lambda \, \langle {\bf 1}^{-\lambda}
  \right)_{\alpha\dot{\alpha}} 
  = \left( p_{1\mu} \sigma^\mu \right)_{\alpha\dot{\alpha}} \, , 
  \label{eq:completeness1}
\\
  &\left(
  \sum_{\lambda=\pm 1} {\bf 1}^\lambda ] \, \lambda \, [ {\bf 1}^{-\lambda}
  \right)_{\alpha}^{\, \, \, \, \beta}
  = m_1 \delta_\alpha^\beta \, , 
\label{eq:completeness2}
\\[2ex]
  &\left(
  \sum_{\lambda=\pm 1} {\bf 1}^{-\lambda} \rangle \, \lambda \, [{\bf 1}^{\lambda}
  \right)^{\dot{\alpha}\alpha} 
  = \left( p_{1\mu} \bar{\sigma}^\mu \right)^{\dot{\alpha} \alpha} \, , 
  \label{eq:completeness3}
  \\
  &\left(
  \sum_{\lambda=\pm 1} {\bf 1}^{-\lambda} \rangle \, \lambda \, 
    \langle {\bf 1}^{\lambda}
  \right)^{\dot{\alpha}}_{\, \, \, \,  \dot{\beta}} 
  = m_1 \delta^{\dot{\alpha}}_{\dot{\beta}} \, .
\label{eq:completeness4}
\end{align}
The complex conjugates are given by
  \begin{align}
 & (\spa{{\bf 1}^{\lambda_1}}.{{\bf 2}^{\lambda_2}})^*
  =
  \spb{{\bf 2}^{-\lambda_2}}.{{\bf 1}^{-\lambda_1}}
  \, , \nn\\
  &
  (\spb{{\bf 1}^{\lambda_1}}.{{\bf 2}^{\lambda_2}})^*
  =
  \spa{{\bf 2}^{-\lambda_2}}.{{\bf 1}^{-\lambda_1}}
  \, .
  \end{align}

The amplitude ${\cal A}_{n_s+n_f}$ depends on the 
coupling functions at the vacuum, {\it{e.g.}},
\begin{align}
& {V}_{; (123)} \biggr|_0 \, , \quad
 {V}_{; (1234)} \biggr|_0 \, , \quad
 {R}_{1234} \biggr|_0 \, ,\quad \nn\\
&
 {R}_{1234; 5} \biggr|_0 \, , \quad  
 {M}_{\hat{1}\hat{2}; 3} \biggr|_0 , \quad
 {R}_{\hat{1}\hat{2}34} \biggr|_0 , \quad
 \cdots \, .
\end{align}
Hereafter we omit the vertical bar symbols so that
\begin{align}
&
 {V}_{; (123)} \, , \quad
 {V}_{; (1234)} \, , \quad
 {R}_{1234}  \, ,\nn\\
 &
 {R}_{1234; 5} \, , \quad  
 {M}_{\hat{1}\hat{2}; 3} , \quad
 {R}_{\hat{1}\hat{2}34} , \quad
 \cdots \, ,
\end{align}
are understood to be evaluated at the vacuum.

%%%%%%%%%%%%%%%%%%%%%
\subsection{Three scalars}
%---------
\begin{figure}
	\centering
	\includegraphics[width=2cm,clip]{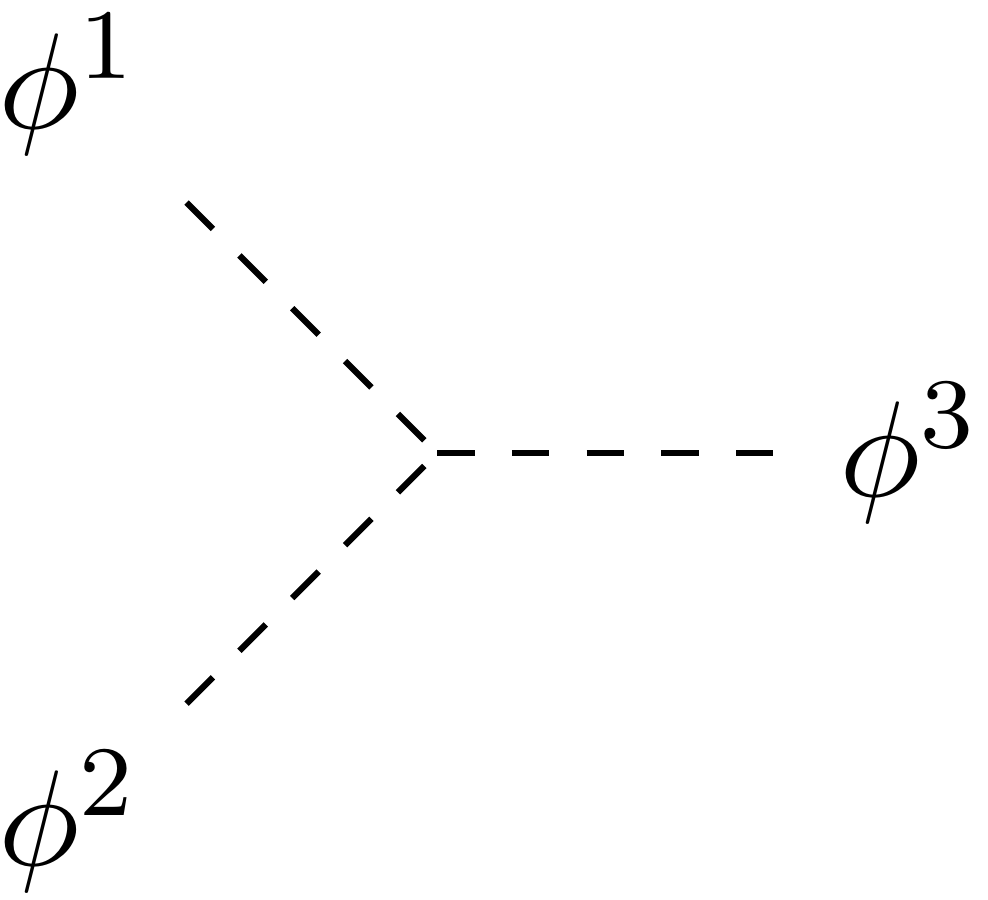}
	\caption{Feynman diagram for $\mathcal{A}_3({{1}},{{2}},{{3}})$.}
	\label{fig:sss}
\end{figure}
%---------

We start with a three-point scalar amplitude $\mathcal{A}_3({{1}},{{2}},{{3}})$, which is given by the contact diagram shown in Fig.~\ref{fig:sss}. 
We have calculated the on-shell three-point scalar amplitude in 
Ref.~\cite{Nagai:2019tgi}. 
The amplitude is simply given by
\be
i\mathcal{A}_3({{1}},{{2}},{{3}})
=
-i{V}_{;(123)}
\,.
\label{eq:scalar-three-point-amplitude}
\ee
As we have shown in Ref.~\cite{Nagai:2019tgi}, 
if we do not use the normal coordinate, 
we need to perform much involved computations 
to get the
final expression of the amplitude (\ref{eq:scalar-three-point-amplitude}).

%%%%%%%%%%%%%%%%%%%%%
\subsection{Two fermions and one scalar}
%---------
\begin{figure}
	\centering
	\includegraphics[width=2cm,clip]{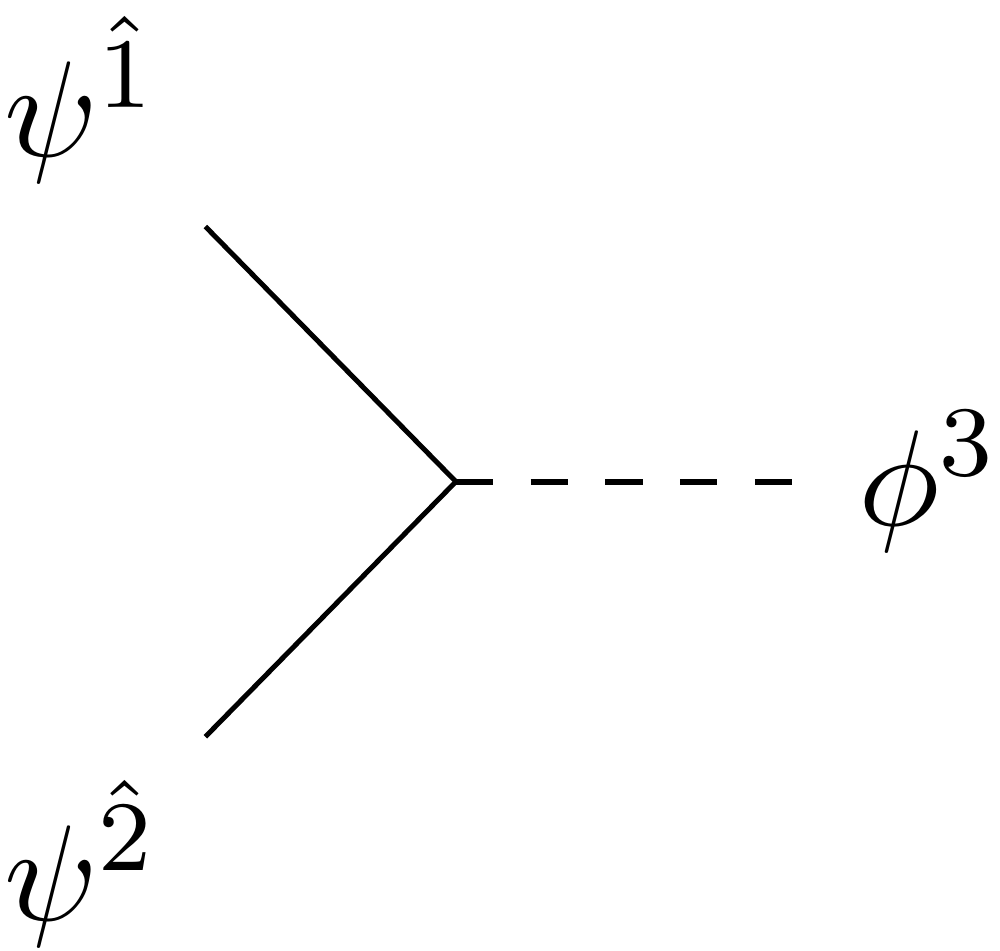}
	\caption{Feynman diagram for $\mathcal{A}_3({\bf{1}}^{\lambda_1},{\bf{2}}^{\lambda_2},{\bf{3}})$. We assign the out-going momenta $p_1$, $p_2$ and $p_3$ to $\psi^{{\h{1}}}$, $\psi^{{\h{2}}}$, and $\phi^3$. }
	\label{fig:ffs}
\end{figure}
%---------
We next consider a three-point amplitude with two fermions and one scalar, 
$\mathcal{A}_3({\bf{1}}^{\lambda_1},{\bf{2}}^{\lambda_2},{{3}})$. 
The amplitude is given by the diagram shown 
in Fig.~\ref{fig:ffs}, 
where the three-point vertex is read from 
the normal coordinate formula (\ref{eq:M-exp}). 
The on-shell amplitude is given as
\be
i\mathcal{A}_3({\bf{1}}^{\lambda_1},{\bf{2}}^{\lambda_2},{{3}})
=
-i{M}_{{\h{1}}{\h{2}};3}
\spb{{\bf{1}}^{\lambda_1}}.{{\bf{2}}^{\lambda_2}}
-
i{M}_{{\h{1}}^*{\h{2}}^*;3}
\spa{{\bf{1}}^{\lambda_1}}.{{\bf{2}}^{\lambda_2}}
\, .
\label{eq:two-fermions-one-scalar-amplitude}
\ee
We have confirmed the formula (\ref{eq:two-fermions-one-scalar-amplitude})
without using the normal coordinate technique.

Once we specify the kinematics and helicities of the external states, 
we can explicitly estimate the spinor inner products in terms of the 
kinematical 
variables. 
See appendix \ref{app:helicitystate} for the explicit expressions. 
For an example, 
we consider the decay process of the scalar $3$ into the fermion pair 
${\bf 1}$ and ${\bf 2}$, 
\be
\phi^3(-p_3)\to \psi^{{\h{1}}}(p_1)\,\psi^{{\h{2}}}(p_2)
\,.
\ee 
Initial state momentum is assigned to be $-p_3$ in our 
notation of the amplitude.

We evaluate the decay amplitudes in the rest frame of $\phi^3$.
Note that 
the final state angular momentum should vanish in this process,
since the interaction vertex does not contain derivatives.
The conservation of the total angular momentum implies that
the final state spin momentum should also be zero.
We therefore expect
\be
\mathcal{A}_3({\bf{1}}^{\pm},{\bf{2}}^{\mp},{{3}})
=
0
\, ,
\ee
which actually is confirmed in our explicit computation, 
since 
$\spb{{\bf{1}}^\pm}.{{\bf{2}}^\mp}=\spa{{\bf{1}}^\pm}.{{\bf{2}}^\mp}=0$
in the center-of-mass frame as we show in Appendix~\ref{app:helicitystate}.
On the other hand, 
$\mathcal{A}_3({\bf{1}}^{\pm},{\bf{2}}^{\pm},{{3}})$ can be non-zero.
The masses of fermions and scalar are denoted by $m_{{1}}$,
$m_{{2}}$ and $m_{3}$, 
respectively.
For $m_3 \gg m_1 , m_2$, we find
\begin{align}
  &\spb{{\bf 1}^+}.{{\bf 2}^+} \simeq \spb{1}.{2} = - m_3 \, , \qquad
   \spb{{\bf 1}^-}.{{\bf 2}^-} \simeq 0 \, , 
\\
  &\spa{{\bf 1}^-}.{{\bf 2}^-} \simeq \spa{1}.{2} = + m_3 \, , \qquad
   \spa{{\bf 1}^+}.{{\bf 2}^+} \simeq 0 \, , 
\end{align}
and thus
\begin{align}
  \mathcal{A}_3({\bf{1}}^{+},{\bf{2}}^{+},{{3}})
 & \simeq
  m_3\,{M}_{{\h{1}}{\h{2}};3} \,,\nn\\
  \mathcal{A}_3({\bf{1}}^{-},{\bf{2}}^{-},{{3}})
 & \simeq
  -m_3\,{M}_{{\h{1}}^*{\h{2}}^*;3}  \, ,
\end{align}
where we ignore $\mathcal{O}(m^2_{{1}}/m^2_3, m^2_{{2}}/m^2_3)$ corrections.
%%%%%%%%%%%%%%%%%%%%%
\subsection{Four scalars}
%---------
\begin{figure}
	\centering
	\includegraphics[width=8.5cm,clip]{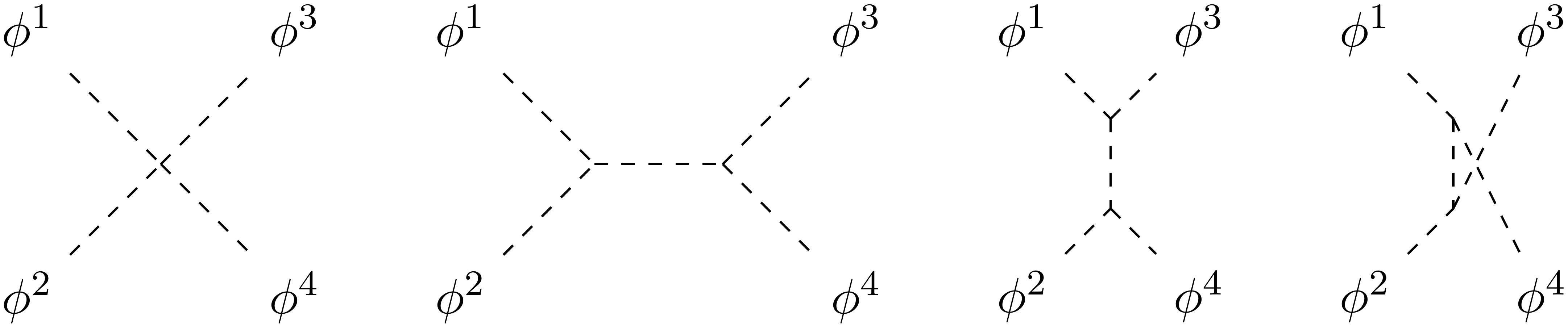}
	\caption{Feynman diagram for $\mathcal{A}_4({{1}},{{2}},{{3}},{{4}})$. We assign the out-going momenta $p_1$, $p_2$, $p_3$ and $p_4$ to $\phi^1$, $\phi^2$, $\phi^3$ and $\phi^4$. }
	\label{fig:ssss}
\end{figure}
%---------
We next consider a four-point scalar amplitude $\mathcal{A}_4({{1}},{{2}},{{3}},{{4}})$. The amplitude is given by the sum of the contact diagram and the scalar-exchange diagrams as shown in Fig. \ref{fig:ssss};
\begin{align}
\mathcal{A}_4({{1}},{{2}},{{3}},{{4}})
&=
\mathcal{A}^{(c)}_4({{1}},{{2}},{{3}},{{4}})
\nn\\
&+
\mathcal{A}^{(\phi)}_4({{1}},{{2}},{{3}},{{4}})
\,.
\end{align}
We have estimated the four-point amplitudes in Ref.~\cite{Nagai:2019tgi}. 
The results are
\bea
i\mathcal{A}^{(c)}_4({{1}},{{2}},{{3}},{{4}})
=
&-&
\frac{2i}{3} {R}_{1(34)2} s_{12}
-
\frac{2i}{3} {R}_{1(24)3} s_{13}
\nn\\
&-&
\frac{2i}{3} {R}_{1(23)4} s_{14}
-{V}_{;(1234)}
\,,
\label{eq:ssss1}
\eea
and  
\begin{align}
i\mathcal{A}^{(\phi)}_4({{1}},{{2}},{{3}},{{4}})
=
&-
\sum_{i,j} {V}_{;(12i)}\,[D(s_{12})]^{ij}\,{V}_{;(34j)}
\nn\\
&-
% -
\sum_{i,j} {V}_{;(13i)}\,[D(s_{13})]^{ij}\,{V}_{;(24j)}
\nn\\
&-
%-
\sum_{i,j} {V}_{;(14i)}\,[D(s_{14})]^{ij}\,{V}_{;(23j)}
\, ,
\label{eq:ssss2}
\end{align}
where $[D(s)]^{ij}$ denotes the scalar propagator, 
\be
[D(s)]^{ij}
:=
\frac{i}{s-m^2_{i}}\,{g}^{ij}
\,.
\label{eq:scalarprop}
\ee
with $m_i$ being the scalar mass.

The scalar four-point amplitude 
diverges in the high energy limit, 
$s=s_{12}\gg m_1^2$, $m_2^2$, $m_3^2$, $m_4^2$.
For an example, we consider 
\be
\phi^1(-p_1)\,\phi^2(-p_2)\,\to\phi^3(p_3)\,\phi^4(p_4) \, .
\ee 
In the high-energy limit, 
the corresponding scattering amplitude behaves as
\be
\mathcal{A}_4({{1}},{{2}},{{3}},{{4}})
\simeq
{R}_{1423}\,s
+\frac{1}{2}{R}_{1234}\,s\,(1+\cos\theta)
\,,
\label{eq:ssss}
\ee
with 
$\theta$ being
the scattering angle in the center of mass.
This result implies that, with ${R}_{1423} \ne 0$ 
or ${R}_{1234} \ne 0$, 
the perturbative unitarity  is violated in the high energy scattering amplitude 
among the scalar bosons. 
This observation indicates that the longitudinally 
polarized gauge boson scattering amplitudes violates the 
perturbative unitarity if the scalar manifold is 
curved~\cite{Alonso:2015fsp,Alonso:2016oah,Nagai:2019tgi}. 
The Lee-Quigg-Thacker sum rules~\cite{Lee:1977yc,Lee:1977eg}
for the perturbative unitarity 
in the $W_L W_L \to W_L W_L$ amplitude can thus be regarded 
as conditions on the scalar curvature tensor $R_{\pi\pi\pi\pi}=0$.

%%%%%%%%%%%%%%%%%%%%%
\subsection{Two fermions and two scalars}
%---------
\begin{figure}
	\centering
	\includegraphics[width=8.5cm,clip]{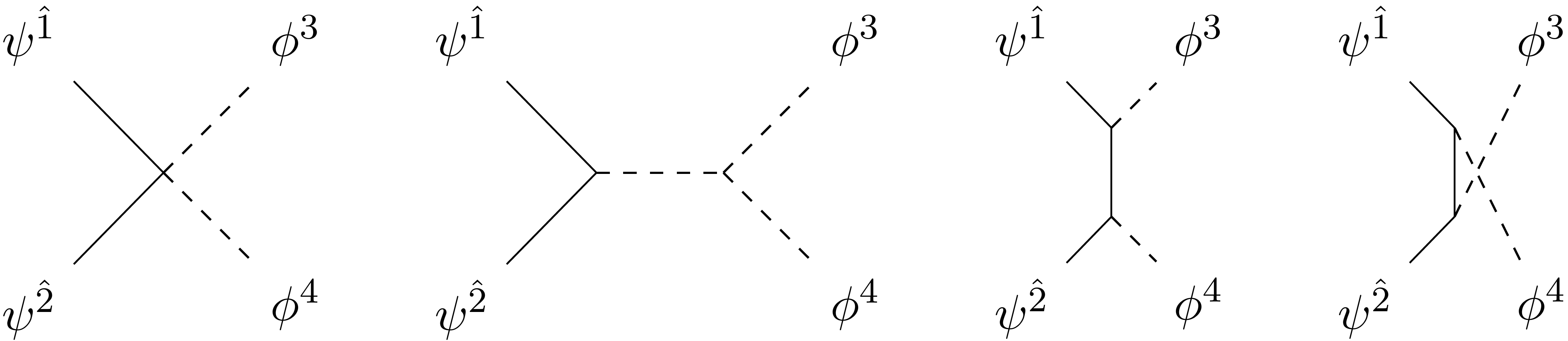}
	\caption{Feynman diagram for $\mathcal{A}_4({\bf{1}}^{\lambda_1},{\bf{2}}^{\lambda_2},{{3}},{{4}})$. We assign the out-going momenta $p_1$, $p_2$, $p_3$ and $p_4$ to $\psi^{{\h{1}}}$, $\psi^{{\h{2}}}$, $\phi^3$ and $\phi^4$. }
	\label{fig:ffss}
\end{figure}
%---------
We next consider a four-point amplitude with two fermions and two scalars, 
$\mathcal{A}_4({\bf{1}}^{\lambda_1},{\bf{2}}^{\lambda_2},{{3}},{{4}})$. 
As shown in Fig.~\ref{fig:ffss}, 
the amplitude consists of 
contact, scalar-exchange, and fermion-exchange diagrams,  
\begin{align}
\mathcal{A}_4({\bf{1}}^{\lambda_1},{\bf{2}}^{\lambda_2},{{3}},{{4}})
&=
\mathcal{A}^{(c)}_4({\bf{1}}^{\lambda_1},{\bf{2}}^{\lambda_2},{{3}},{{4}})
\nn\\
&+
\mathcal{A}^{(\phi)}_4({\bf{1}}^{\lambda_1},{\bf{2}}^{\lambda_2},{{3}},{{4}})
\nn\\
&+\mathcal{A}^{(\psi)}_4({\bf{1}}^{\lambda_1},{\bf{2}}^{\lambda_2},{{3}},{{4}})
\,.
\end{align}

We first focus on the contact diagram $\mathcal{A}^{(c)}_4$, 
which appears from the vertices, $R_{\hat{i}\hat{j}^* kl}$,
$M_{\hat{i}\hat{j}; \, (kl)}$ and  $M_{\hat{i}^*\hat{j}^*; \, (kl)}$
in the normal coordinate. We find
\begin{align}
\lefteqn{
i\mathcal{A}^{(c)}_4({\bf{1}}^{\lambda_1},{\bf{2}}^{\lambda_2},{{3}},{{4}})}\nn\\
&= -
\frac{i}{2}{R}_{{\h{2}}{\h{1}}^*34}
(p_3-p_4)_\mu\,
x^\dag_{\dot{\alpha}}(\vec{p}_1,\lambda_1)\,
(\bar{\sigma}^\mu)^{\dot{\alpha}\beta}\,
y_\beta(\vec{p}_2,\lambda_2)
\nn\\
& \quad -
\frac{i}{2}{R}_{{\h{2}}^*{\h{1}}34}
(p_3-p_4)_\mu\,
y^{\alpha}(\vec{p}_1,\lambda_1)\,
({\sigma}^\mu)_{\alpha\dot{\beta}}\,
x^{\dag\dot{\beta}}(\vec{p}_2,\lambda_2)
\nn\\
& \quad -
i{M}_{{\h{1}}^*{\h{2}}^*;(34)}
x^\dag_{\dot{\alpha}}(\vec{p}_1,\lambda_1)\,
x^{\dag{\dot{\alpha}}}(\vec{p}_2,\lambda_2)
\nn\\
& \quad -
i{M}_{{\h{1}}{\h{2}};(34)}
y^\alpha(\vec{p}_1,\lambda_1)\,
y_\alpha(\vec{p}_2,\lambda_2)
\, .
\end{align}
Rewriting the amplitude in terms of the angle/square 
spinors (\ref{eq:defmassivebraket1})-(\ref{eq:defmassivebraket4}), 
the spinor wavefunction structure becomes clearer
  \begin{align}
  \lefteqn{
i\mathcal{A}^{(c)}_4({\bf{1}}^{\lambda_1},{\bf{2}}^{\lambda_2},{{3}},{{4}})}\nn\\
&= -
\frac{i}{2}{R}_{{\h{2}}{\h{1}}^*34}
\left(
\sum_{\lambda_3} 
\spa{{\bf{1}}^{\lambda_1}}.{{\bf{3}}^{-\lambda_3}} \lambda_3
\spb{{\bf{3}}^{\lambda_3}}.{{\bf{2}}^{\lambda_2}}
\r.\nn\\
&\qquad \qquad  \qquad \l.
-\sum_{\lambda_4}
\spa{{\bf{1}}^{\lambda_1}}.{{\bf{4}}^{-\lambda_4}} \lambda_4
\spb{{\bf{4}}^{\lambda_4}}.{{\bf{2}}^{\lambda_2}}
\right)
\nn\\
&\quad -
\frac{i}{2}{R}_{{\h{2}}^*{\h{1}}34}
\left(
\sum_{\lambda_3}
\spb{{\bf{1}}^{\lambda_1}}.{{\bf{3}}^{\lambda_3}} \lambda_3
\spa{{\bf{3}}^{-\lambda_3}}.{{\bf{2}}^{\lambda_2}}
\r.\nn\\
&\qquad \qquad  \qquad \l.
-\sum_{\lambda_4}
\spb{{\bf{1}}^{\lambda_1}}.{{\bf{4}}^{\lambda_4}} \lambda_4
\spa{{\bf{4}}^{-\lambda_4}}.{{\bf{2}}^{\lambda_2}}
\right)
\nn\\
&\quad -
i{M}_{{\h{1}}^*{\h{2}}^*;(34)}
\spa{{\bf{1}}^{\lambda_1}}.{{\bf{2}}^{\lambda_2}}
-
i{M}_{{\h{1}}{\h{2}};(34)}
\spb{{\bf{1}}^{\lambda_1}}.{{\bf{2}}^{\lambda_2}}
\, , 
\label{eq:four-point-2f2b-contact}
  \end{align}
where we decomposed 
$(p_{3} -p_4)_\mu \bar{\sigma}^\mu$ and
$(p_{3} -p_4)_\mu {\sigma}^\mu$
into products of massive spinor wavefunctions
using Eqs.\,(\ref{eq:completeness1})-(\ref{eq:completeness4}).

The computation of the scalar-exchange amplitude is easy.
It is given by
  \begin{align}
\lefteqn{i\mathcal{A}^{(\phi)}_4({\bf{1}}^{\lambda_1},{\bf{2}}^{\lambda_2},{{3}},{{4}})}\nn\\
=
&-
\sum_{i,j}
{V}_{;(34i)}\,[D(s_{12})]^{ij}\, \biggl( 
{M}_{12;j} 
\spb{{\bf{1}}^{\lambda_1}}.{{\bf{2}}^{\lambda_2}}
\nn\\ 
&
\qquad \qquad \qquad \qquad \qquad
+
{M}_{1^*2^*;j}
\spa{{\bf{1}}^{\lambda_1}}.{{\bf{2}}^{\lambda_2}}
\biggr) \, , 
  \end{align}
where $[D(s)]^{ij}$ is defined in Eq.\,(\ref{eq:scalarprop}).

On the other hand, 
the computation of the fermion exchange amplitude $\mathcal{A}^{(\psi)}_4$
\begin{align}
i\mathcal{A}^{(\psi)}_4({\bf{1}}^{\lambda_1},{\bf{2}}^{\lambda_2},{{3}},{{4}})
&= i \mathcal{A}^{[ \,  ]}_4({\bf{1}}^{\lambda_1},{\bf{2}}^{\lambda_2},{{3}},{{4}}) 
\nonumber\\
& \quad
  +i \mathcal{A}^{\langle \,  \rangle}_4({\bf{1}}^{\lambda_1},{\bf{2}}^{\lambda_2},{{3}},{{4}}) 
\nonumber\\
& \quad
  +i \mathcal{A}^{[ \,  \rangle}_4({\bf{1}}^{\lambda_1},{\bf{2}}^{\lambda_2},{{3}},{{4}}) 
  \nonumber\\
& \quad
  +i \mathcal{A}^{\langle \,  ]}_4({\bf{1}}^{\lambda_1},{\bf{2}}^{\lambda_2},{{3}},{{4}}) 
\label{eq:scalar-exchange-amp1}
\end{align}
is a bit involved.
Here we organized the amplitude in accord with the spinor structure, {\it{i.e.}}, 
\begin{align}
\lefteqn{\mathcal{A}^{[\, ]}_4({\bf{1}}^{\lambda_1},{\bf{2}}^{\lambda_2},{{3}},{{4}}) }\nn\\
&= 
    \sum_{\hat{i}, \hat{j}} M_{\hat{1}\hat{i}; \, 3} \, 
     [{\bf 1}^{\lambda_1} \, D^{\hat{i}\hat{j}}_{][}(p_{13}) \, {\bf 2}^{\lambda_2} ] 
     \, M_{\hat{j} \hat{2}; \, 4} 
\nonumber\\
&\quad
   +\sum_{\hat{i}, \hat{j}} M_{\hat{1}\hat{i}; \, 4} \, 
     [{\bf 1}^{\lambda_1} \, D^{\hat{i}\hat{j}}_{][}(p_{14}) \, {\bf 2}^{\lambda_2} ] 
     \, M_{\hat{j} \hat{2}; \, 3} \, , 
\label{eq:scalar-exchange-amp2}
\\
\lefteqn{
\mathcal{A}^{\langle \, \rangle}_4({\bf{1}}^{\lambda_1},{\bf{2}}^{\lambda_2},{{3}},{{4}}) }\nn\\
&= \sum_{\hat{i}^*, \hat{j}^*} M_{\hat{1}^*\hat{i}^*; \, 3} \, 
    \langle {\bf 1}^{\lambda_1} \, D^{\hat{i}^*\hat{j}^*}_{\rangle\langle}(p_{13}) \, 
    {\bf 2}^{\lambda_2} \rangle  M_{\hat{j}^*\hat{2}^*; \, 4}
\nonumber\\
&\quad
  +\sum_{\hat{i}^*, \hat{j}^*} M_{\hat{1}^*\hat{i}^*; \, 4} \, 
    \langle {\bf 1}^{\lambda_1} \, D^{\hat{i}^*\hat{j}^*}_{\rangle\langle}(p_{14}) \, 
    {\bf 2}^{\lambda_2} \rangle  M_{\hat{j}^*\hat{2}^*; \, 3} \, ,
\label{eq:scalar-exchange-amp3}
\end{align}
\begin{align}
\lefteqn{
\mathcal{A}^{[ \, \rangle}_4({\bf{1}}^{\lambda_1},{\bf{2}}^{\lambda_2},{{3}},{{4}}) }\nn\\
&= \sum_{\hat{i}, \hat{j}^*} M_{\hat{1}\hat{i}; \, 3} \, 
   [{\bf 1}^{\lambda_1} \, D^{\hat{i}\hat{j}^*}_{]\langle}(p_{13}) \, 
   {\bf 2}^{\lambda_2} \rangle \, M_{\hat{j}^*\hat{2}^*; \, 4}
\nonumber\\
&\quad
  +\sum_{\hat{i}, \hat{j}^*} M_{\hat{1}\hat{i}; \, 4} \, 
   [{\bf 1}^{\lambda_1} \, D^{\hat{i}\hat{j}^*}_{]\langle}(p_{14}) \, 
   {\bf 2}^{\lambda_2} \rangle \, M_{\hat{j}^*\hat{2}^*; \, 3} \, , 
\label{eq:scalar-exchange-amp4}
\\
\lefteqn{
\mathcal{A}^{\langle \, ]}_4({\bf{1}}^{\lambda_1},{\bf{2}}^{\lambda_2},{{3}},{{4}}) }\nn\\
&=\sum_{\hat{i}^*, \hat{j}} M_{\hat{1}^*\hat{i}^*; \, 3} 
  \langle {\bf 1}^{\lambda_1} \, D^{\hat{i}^*\hat{j}}_{\rangle[}(p_{13}) \, 
  {\bf 2}^{\lambda_2} ] \, M_{\hat{j}\hat{2}; \, 4} 
\nonumber\\
&\quad
  +\sum_{\hat{i}^*, \hat{j}} M_{\hat{1}^*\hat{i}^*; \, 4} 
  \langle {\bf 1}^{\lambda_1} \, D^{\hat{i}^*\hat{j}}_{\rangle[}(p_{14}) \, 
  {\bf 2}^{\lambda_2} ] \, M_{\hat{j}\hat{2}; \, 3}  \, .
\label{eq:scalar-exchange-amp5}
\end{align}
with
\begin{align}
  p_{13} := p_1 + p_3 \, , \qquad
  p_{14} := p_1 + p_4  \, .
\end{align}
In the expressions above, the internal fermion propagators 
are 
\begin{align}
  (\, D_{][}^{\hat{i}\hat{j}}(p) \, )_{\alpha}{}^{\beta}
  &:= \dfrac{i m_{\hat{i}}}{p^2-m_{\hat{i}}^2} \delta^{\hat{i}\hat{j}} \,
      \delta_\alpha^\beta \, , 
\\
  (\, D_{\rangle\langle}^{\hat{i}^*\hat{j}^*}(p) \, )^{\dot{\alpha}}{}_{\dot{\beta}}
  &:= \dfrac{i m_{\hat{i}^*}}{p^2-m_{\hat{i}^*}^2} \delta^{\hat{i}^*\hat{j}^*} \, 
      \delta^{\dot{\alpha}}_{\dot{\beta}} \, , 
\\
  (\, D_{]\langle}^{\hat{i}\hat{j}^*}(p) \, )_{\alpha\dot{\beta}}
  &:= \dfrac{i}{p^2-m_{\hat{i}}^2} \delta^{\hat{i}\hat{j}^*}
      (\, p] \, )_{\alpha} \, (\, \langle p \, )_{\dot{\beta}} \, , 
\\
  (\, D_{\rangle[}^{\hat{i}^*\hat{j}}(p) \, )^{\dot{\alpha}\beta}
  &:= \dfrac{i}{p^2-m_{\hat{i}^*}^2} \delta^{\hat{i}^*\hat{j}}
      (\, p\rangle \, )^{\dot{\alpha}} \, (\, [p \, )^\beta \, . 
\end{align}
The amplitudes (\ref{eq:scalar-exchange-amp2}) and (\ref{eq:scalar-exchange-amp3}) are computed by using
\begin{align}
[{\bf 1}^{\lambda_1} \, D^{\hat{i}\hat{j}}_{][}(p_{13}) \, {\bf 2}^{\lambda_2} ] 
&= ( \, \hat{D}(s_{13}) \, )^{\hat{i}\hat{j}} m_{\hat{i}} 
   \spb{{\bf 1}^{\lambda_1}}.{{\bf 2}^{\lambda_2}}  \, , 
\\
\langle{\bf 1}^{\lambda_1} \, D^{\hat{i}^*\hat{j}^*}_{\rangle\langle}(p_{13}) \, {\bf 2}^{\lambda_2} \rangle
&= ( \, \hat{D}(s_{13}) \, )^{\hat{i}^*\hat{j}^*} m_{\hat{i}^*} 
   \spa{{\bf 1}^{\lambda_1}}.{{\bf 2}^{\lambda_2}}  \, , 
\end{align}
with
\begin{equation}
  ( \, \hat{D}(s) \, )^{\hat{i}\hat{j}}
 := \dfrac{i}{s-m_{\hat{i}}^2} \delta^{\hat{i}\hat{j}} \, , 
 \quad
  ( \, \hat{D}(s) \, )^{\hat{i}^*\hat{j}^*}
 := \dfrac{i}{s-m_{\hat{i}^*}^2} \delta^{\hat{i}^*\hat{j}^*} \, . 
\end{equation}
In the computation of the amplitudes (\ref{eq:scalar-exchange-amp4})
and (\ref{eq:scalar-exchange-amp5}), we use identities on the
spinor wavefunctions, 
\begin{align}
\lefteqn{
[ {\bf 1}^{\lambda_1} \, D^{\hat{i}\hat{j}^*}_{]\langle}(p_{13}) \, 
  {\bf 2}^{\lambda_2} \rangle }\nn\\
&= \dfrac{1}{2}
  ( \, \hat{D}(s_{13}^2) \, )^{\hat{i}\hat{j}^*}
  \biggl(
    m_1 \, \spa{{\bf 1}^{\lambda_1}}.{{\bf 2}^{\lambda_2}}
  + m_2 \, \spb{{\bf 1}^{\lambda_1}}.{{\bf 2}^{\lambda_2}} 
\nonumber\\
& \qquad
  +\sum_{\lambda_3= \pm 1} 
   \spb{{\bf 1}^{\lambda_1}}.{{\bf 3}^{\lambda_3}} \, \lambda_3 
    \, \spa{{\bf 3}^{-\lambda_3}}.{{\bf 2}^{\lambda_2}}
    \nonumber\\
& \qquad
  -\sum_{\lambda_4=\pm 1}
   \spb{{\bf 1}^{\lambda_1}}.{{\bf 4}^{\lambda_4}} \, \lambda_4
    \, \spa{{\bf 4}^{-\lambda_4}}.{{\bf 2}^{\lambda_2}}
  \biggr) \, , 
\\
\lefteqn{
\langle {\bf 1}^{\lambda_1} \, D^{\hat{i}^*\hat{j}}_{\rangle[}(p_{13}) \, 
  {\bf 2}^{\lambda_2} ] }\nn\\
&= \dfrac{1}{2}
  ( \, \hat{D}(s_{13}^2) \, )^{\hat{i}^*\hat{j}}
  \biggl(
    m_1 \, \spb{{\bf 1}^{\lambda_1}}.{{\bf 2}^{\lambda_2}}
  + m_2 \, \spa{{\bf 1}^{\lambda_1}}.{{\bf 2}^{\lambda_2}} 
\nonumber\\
& \qquad
  +\sum_{\lambda_3=\pm 1} 
   \spa{{\bf 1}^{\lambda_1}}.{{\bf 3}^{-\lambda_3}} \, \lambda_3 \, 
   \spb{{\bf 3}^{\lambda_3}}.{{\bf 2}^{\lambda_2}}
   \nonumber\\
& \qquad
  -\sum_{\lambda_4=\pm 1}
   \spa{{\bf 1}^{\lambda_1}}.{{\bf 4}^{-\lambda_4}} \, \lambda_4 \, 
   \spb{{\bf 4}^{\lambda_4}}.{{\bf 2}^{\lambda_2}}
  \biggr) \, , 
\end{align}
with
\begin{equation}
  ( \, \hat{D}(s) \, )^{\hat{i}\hat{j}^*}
 := \dfrac{i}{s-m_{\hat{i}}^2} \delta^{\hat{i}\hat{j}^*} \, , 
 \quad
  ( \, \hat{D}(s) \, )^{\hat{i}^*\hat{j}}
 := \dfrac{i}{s-m_{\hat{i}^*}^2} \delta^{\hat{i}^*\hat{j}} \, . 
\end{equation}
These identities are derived from
\begin{align}
  p_{13} &= \dfrac{1}{2} ( p_1 - p_2 + p_3 - p_4 ) \, , \nn\\
  p_{14} &= \dfrac{1}{2} ( p_1 - p_2 - p_3 + p_4 ) \, , 
\end{align}
and
\begin{align}
  &[ {\bf 1}^{\lambda_1} \, (p_{1\mu} \sigma^\mu) 
  = m_1 \langle {\bf 1}^{\lambda_1}  \, , 
  &( p_{2\mu} \bar{\sigma}^\mu) \, {\bf 2}^{\lambda_2} ]
  =  - {\bf 2}^{\lambda_2} \rangle m_2 \, , 
\\
  &\langle {\bf 1}^{\lambda_1} \, (p_{1\mu} \bar{\sigma}^\mu)
  = m_1 [ {\bf 1}^{\lambda_1} \, 
  &( p_{2\mu} \sigma^\mu) \, {\bf 2}^{\lambda_2} \rangle
  =  - {\bf 2}^{\lambda_2} ] m_2 \,   .
\end{align}
We are now ready to compute the fermion-exchange 
amplitude (\ref{eq:scalar-exchange-amp1}).
Combining the above formulas, we obtain
\begin{align}
\lefteqn{
i\mathcal{A}^{(\psi)}_4({\bf{1}}^{\lambda_1},{\bf{2}}^{\lambda_2},{{3}},{{4}})
} \nonumber\\
&=  
i\mathcal{A}^{(\psi,yx)}_4
\left(
\sum_{\lambda_3=\pm 1}
\spb{{\bf{1}}^{\lambda_1}}.{{\bf{3}}^{\lambda_3}} \, \lambda_3 \, 
\spa{{\bf{3}}^{-\lambda_3}}.{{\bf{2}}^{\lambda_2}}
\r.\nonumber\\
&\qquad \qquad \qquad \l.
-\sum_{\lambda_4=\pm 1}
\spb{{\bf{1}}^{\lambda_1}}.{{\bf{4}}^{\lambda_4}} \, \lambda_4 \, 
\spa{{\bf{4}}^{-\lambda_4}}.{{\bf{2}}^{\lambda_2}}
\right)
\nn\\
& \quad  + 
i\mathcal{A}^{(\psi,xy)}_4
\left(
\sum_{\lambda_3=\pm 1}
\spa{{\bf{1}}^{\lambda_1}}.{{\bf{3}}^{-\lambda_3}} \, \lambda_3 \, 
\spb{{\bf{3}}^{\lambda_3}}.{{\bf{2}}^{\lambda_2}}
\r.\nonumber\\
&\qquad \qquad \qquad \l.
-
\sum_{\lambda_4=\pm 1}
\spa{{\bf{1}}^{\lambda_1}}.{{\bf{4}}^{-\lambda_4}} \, \lambda_4 \, 
\spb{{\bf{4}}^{\lambda_4}}.{{\bf{2}}^{\lambda_2}}
\right)
\nn\\
& \quad  +
i\mathcal{A}^{(\psi,yy)}_4
\spb{{\bf{1}}^{\lambda_1}}.{{\bf{2}}^{\lambda_2}}
+
i\mathcal{A}^{(\psi,xx)}_4
\spa{{\bf{1}}^{\lambda_1}}.{{\bf{2}}^{\lambda_2}}
\, , 
\end{align}
with
\begin{align}
i\mathcal{A}^{(\psi,yx)}_4
&=
-\frac{1}{2} \sum_{\h{i}, \h{j}^*} \l(
{M}_{{\h{i}}{\h{1}};3}\,[\hat{D}(s_{13})]^{{\h{i}}{\h{j}}^*}\,{M}_{{\h{j}}^*{\h{2}}^*;4}
\r.\nonumber\\
&\qquad \qquad  \l.
+
{M}_{{\h{i}}{\h{1}};4}\,[\hat{D}(s_{14})]^{{\h{i}}{\h{j}}^*}\,{M}_{{\h{j}}^*{\h{2}}^*;3}
\r)\,,\\
i\mathcal{A}^{(\psi,xy)}_4
&=
-\frac{1}{2} \sum_{\h{i}, \h{j}^*} \l(
{M}_{{\h{j}}^*{\h{1}}^*;3}\,[\hat{D}(s_{13})]^{{\h{j}}^*{\h{i}}}\,{M}_{{\h{i}}{\h{2}};4}
\r.\nonumber\\
&\qquad \qquad  \l.
+
{M}_{{\h{j}}^*{\h{1}}^*;4}\,[\hat{D}(s_{14})]^{{\h{j}}^*{\h{i}}}\,{M}_{{\h{i}}{\h{2}};3}
\r)\,,
\end{align}
and
\begin{align}
i\mathcal{A}^{(\psi,yy)}_4
&=
-\frac{1}{2}
m_1 
\sum_{\hat{i}^*, \hat{j}}
\l(
{M}_{{\h{i}}^*{\h{1}}^*;3}\,[\hat{D}(s_{13})]^{{\h{i}}^*{\h{j}}}\,{M}_{{\h{j}}{\h{2}};4}
\r.\nonumber\\
&\qquad \qquad  \l.
+
{M}_{{\h{i}}^*{\h{1}}^*;4}\,[\hat{D}(s_{14})]^{{\h{i}}^*{\h{j}}}\,{M}_{{\h{j}}{\h{2}};3}
\r)
\nn\\
& \quad -
\frac{1}{2}
m_2
\sum_{\hat{i} , \hat{j}^*}
\l(
{M}_{{\h{i}}{\h{1}};3}\,[\hat{D}(s_{13})]^{{\h{i}}{\h{j}}^*}\,{M}_{{\h{j}}^*{\h{2}}^*;4}
\r.\nonumber\\
&\qquad \qquad  \l.
+
{M}_{{\h{i}}{\h{1}};4}\,[\hat{D}(s_{14})]^{{\h{i}}{\h{j}}^*}\,{M}_{{\h{j}}^*{\h{2}}^*;3}
\r)
\nn\\
& \quad -
\sum_{\hat{i}, \hat{j}}  m_{\hat{i}}
\l(
{M}_{{\h{i}}{\h{1}};3}
[\hat{D}(s_{13})]^{\hat{i}\hat{j}} \, 
{M}_{{\h{j}}{\h{2}};4}
\r.\nonumber\\
&\qquad \qquad  \l.
+
{M}_{{\h{i}}{\h{1}};4}\,
[\hat{D}(s_{14})]^{\hat{i}\hat{j}} \, 
\,{M}_{{\h{j}}{\h{2}};3}
\r)
\,,
\label{eq:ffssf1}
\\
i\mathcal{A}^{(\psi,xx)}_4
&=
-\frac{1}{2}
m_1 
\sum_{\hat{i}, \hat{j}^*}
\l(
{M}_{{\h{i}}{\h{1}};3}\,[\hat{D}(s_{13})]^{{\h{i}}{\h{j}}^*}
\,{M}_{{\h{j}}^*{\h{2}}^*;4}
\r.\nonumber\\
&\qquad \qquad  \l.
+
{M}_{{\h{i}}{\h{1}};4}\,[\hat{D}(s_{14})]^{{\h{i}}{\h{j}}^*}\,
{M}_{{\h{j}}^*{\h{2}}^*;3}
\r)
\nn\\
& \quad -
\frac{1}{2}
m_2
\sum_{\hat{i}^* , \hat{j}}
\l(
{M}_{{\h{i}}^*{\h{1}}^*;3}\,[\hat{D}(s_{13})]^{{\h{i}}^*{\h{j}}}\,
{M}_{{\h{j}}{\h{2}};4}
\r.\nonumber\\
&\qquad \qquad  \l.
+
{M}_{{\h{i}}^*{\h{1}}^*;4}\,[\hat{D}(s_{14})]^{{\h{i}}^*{\h{j}}}\,
{M}_{{\h{j}}{\h{2}};3}
\r)
\nn\\
& \quad -
\sum_{\hat{i}^*, \hat{j}^*}  m_{\hat{i}^*}
\l(
{M}_{{\h{i}}^*{\h{1}}^*;3}
[\hat{D}(s_{13})]^{\hat{i}^*\hat{j}^*} \, 
{M}_{{\h{j}}^*{\h{2}}^*;4}
\r.\nonumber\\
&\qquad \qquad  \l.
+
{M}_{{\h{i}}^*{\h{1}}^*;4}\,
[\hat{D}(s_{14})]^{\hat{i}^*\hat{j}^*} \, 
\,{M}_{{\h{j}}^*{\h{2}}^*;3}
\r)\, .
\label{eq:ffssf2}
\end{align}

We evaluate the scattering amplitude
\begin{align}
  \phi^3(-p_3) \phi^4(-p_4) \to \psi^{\hat{1}}(p_1) \psi^{\hat{2}}(p_2)
\end{align}
in the center-of-mass frame, which implies
\begin{align}
  &\spb{{\bf 1}^+}.{{\bf 2}^-} = 0 \, , \qquad
   \spb{{\bf 1}^-}.{{\bf 2}^+} = 0 \, , 
\\
  &\spa{{\bf 1}^+}.{{\bf 2}^-} = 0 \, , \qquad
   \spa{{\bf 1}^-}.{{\bf 2}^+} = 0 \, . 
\end{align}
The contact amplitudes (\ref{eq:four-point-2f2b-contact}) dominate in 
the computation for the high-energy limit
$s=s_{12} \gg m_1^2, m_2^2, m_3^2, m_4^2$.
We find
\begin{align}
  &\spb{{\bf 1}^+}.{{\bf 2}^+} \simeq \spb{1}.{2} = - \sqrt{s} \, , \qquad
   \spb{{\bf 1}^-}.{{\bf 2}^-} \simeq 0 \, , 
\\
  &\spa{{\bf 1}^-}.{{\bf 2}^-} \simeq \spa{1}.{2} = + \sqrt{s} \, , \qquad
   \spa{{\bf 1}^+}.{{\bf 2}^+} \simeq 0 \, , 
\end{align}
and
\begin{align}
   \sum_{\lambda_3} 
  \spa{{\bf 1}^-}.{{\bf 3}^{-\lambda_3}} \, \lambda_3 \,
  \spb{{\bf 3}^{\lambda_3}}.{{\bf 2}^+}
  &\simeq
  \spa{1}.{3} \, \spb{3}.{2}
  \nn\\
  &
  = -\spa{1}.{3} \, \spb{2}.{3}
   \nn\\
  &
  = \sqrt{s_{13} \, s_{23}} \, , 
\\
  \sum_{\lambda_4} 
  \spa{{\bf 1}^-}.{{\bf 4}^{-\lambda_4}} \, \lambda_4 \,
  \spb{{\bf 4}^{\lambda_3}}.{{\bf 2}^+}
  &\simeq
  \spa{1}.{4} \, \spb{4}.{2}
    \nn\\
  &
  =-\sqrt{s_{14} \, s_{42}} \, , 
\end{align}
\begin{align}
&
 \sum_{\lambda_3} 
  \spa{{\bf 1}^+}.{{\bf 3}^{-\lambda_3}} \, \lambda_3 \,
  \spb{{\bf 3}^{\lambda_3}}.{{\bf 2}^+}
-\sum_{\lambda_4} 
  \spa{{\bf 1}^+}.{{\bf 4}^{-\lambda_4}} \, \lambda_4 \,
  \spb{{\bf 4}^{\lambda_4}}.{{\bf 2}^+}
  \nn\\
  &
\simeq 
  -\,m_1 \dfrac{s_{13} - s_{14} }{\sqrt{s_{12}}} \, . 
\end{align}
We obtain
\begin{align}
\mathcal{A}_4({\bf{1}}^{+},{\bf{2}}^{+},{{3}},{{4}})
&\simeq
\sqrt{s}\,
\biggl(
{M}_{{\h{1}}{\h{2}};(34)}
\nn\\
& \hspace{-0.5cm}
-\frac{1}{2}(
m_{2}{R}_{{\h{1}}{\h{2}}^*34}
+
m_{1}{R}_{{\h{1}}^*{\h{2}}34})
\cos\theta
\biggr)
\,,
\\
\mathcal{A}_4({\bf{1}}^{+},{\bf{2}}^{-},{{3}},{{4}})
&\simeq
\frac{1}{2}{R}_{{\h{1}}{\h{2}}^*34}\,
s\,\sin\theta
\,,
\end{align}
with $\theta$ being the scattering angle in the center-of-mass frame, 
\begin{align}
  \sin\theta \simeq 2 \dfrac{\sqrt{s_{13} s_{14}}}{s_{12}} \, , \qquad
  \cos\theta \simeq  \dfrac{s_{13}-s_{14}}{s_{12}} \, .
\end{align}
It is also straightforward to compute 
$\mathcal{A}_4({\bf{1}}^{-},{\bf{2}}^{-},{{3}},{{4}})$
and
$\mathcal{A}_4({\bf{1}}^{-},{\bf{2}}^{+},{{3}},{{4}})$
amplitudes.
Unless 
$M_{\hat{1}\hat{2}; (34)}=M_{\hat{1}^*\hat{2}^*; (34)}=0$ and
$R_{\hat{1}\hat{2}^*34}=R_{\hat{1}^*\hat{2}34}=0$, 
these amplitudes eventually violate the perturbative unitarity at the
high-energy scale. 
Considering the equivalence theorem, these results indicate that the scattering amplitudes of the fermion pair scattering to longitudinally polarized gauge-boson violate the perturbative unitarity at a certain high-energy scale unless ${M}_{{\h{1}}{\h{2}};(34)}=0$ and ${R}_{{\h{1}}{\h{2}}^*34}=0$. 
The Appelquist-Chanowitz sum rules~\cite{Chanowitz:1978uj, Chanowitz:1978mv, Appelquist:1987cf,Maltoni:2001dc,Dicus:2004rg,Chivukula:2007gse} for the perturbative unitarity 
in the $W_L W_L \to t \bar{t}$ amplitude can thus be regarded as 
conditions on $R_{t\bar{t} \pi\pi}$ and $M_{\bar{t}t; (\pi\pi)}$.

%%%%%%%%%%%%%%%%%%%%%
\subsection{Four fermions}
%---------
\begin{figure}
	\centering
	\includegraphics[width=8.5cm,clip]{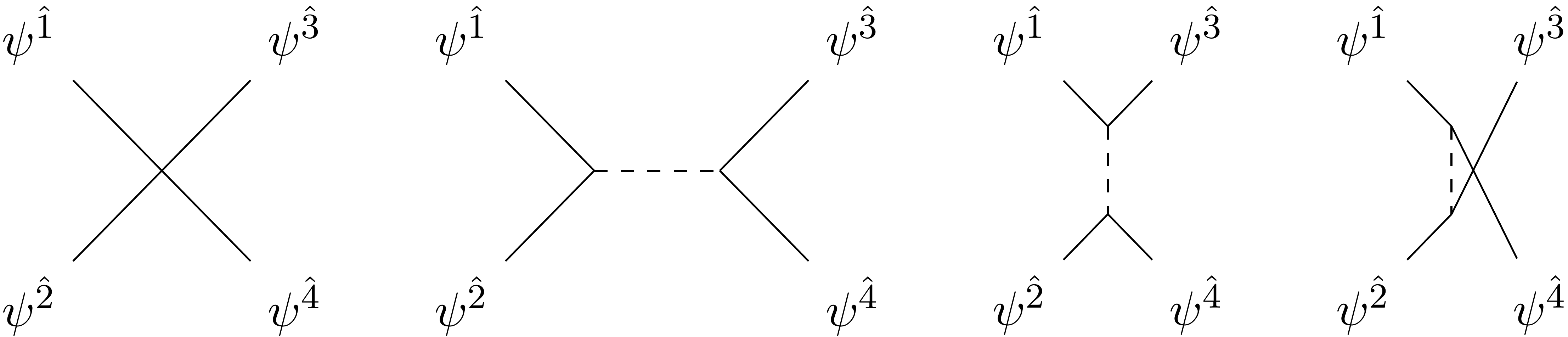}
	\caption{Feynman diagram for $\mathcal{A}_4({\bf{1}}^{\lambda_1},{\bf{2}}^{\lambda_2},{\bf{3}}^{\lambda_3},{\bf{4}}^{\lambda_4})$. We assign the out-going momenta $p_1$, $p_2$, $p_3$ and $p_4$ to $\psi^{{\h{1}}}$, $\psi^{{\h{2}}}$, $\psi^{{\h{3}}}$ and $\psi^{{\h{4}}}$. }
	\label{fig:ffff}
\end{figure}
%---------
The four-points fermion amplitude 
is computed from the Feynman diagram Fig.~\ref{fig:ffff}. 
We decompose the amplitude into three categories;
\begin{align}
\mathcal{A}_4({\bf{1}}^{\lambda_1},{\bf{2}}^{\lambda_2},{\bf{3}}^{\lambda_3},{\bf{4}}^{\lambda_4})
&=
\mathcal{A}^{(c)}_4({\bf{1}}^{\lambda_1},{\bf{2}}^{\lambda_2},{\bf{3}}^{\lambda_3},{\bf{4}}^{\lambda_4})
\nn\\
&+
\mathcal{A}^{(c')}_4({\bf{1}}^{\lambda_1},{\bf{2}}^{\lambda_2},{\bf{3}}^{\lambda_3},{\bf{4}}^{\lambda_4})
\nn\\
&+
\mathcal{A}^{(\phi)}_4({\bf{1}}^{\lambda_1},{\bf{2}}^{\lambda_2},{\bf{3}}^{\lambda_3},{\bf{4}}^{\lambda_4})
\,,
\end{align}
where $\mathcal{A}^{(c)}$ and $\mathcal{A}^{(c')}$ respectively denote the holomolophic and non-holomolophic contact diagrams, and $\mathcal{A}^{(\phi)}$ denotes the scalar-exchange diagram.

Let us first focus on the contact diagram 
induced from the holomolophic contact interactions 
($\psi\psi\psi\psi$ and $\psi^\dag\psi^\dag\psi^\dag\psi^\dag$). 
The amplitude is given as
\begin{align}
\lefteqn{
i\mathcal{A}^{(c)}_4({\bf{1}}^{\lambda_1},{\bf{2}}^{\lambda_2},{\bf{3}}^{\lambda_3},{\bf{4}}^{\lambda_4})}\nn\\
&=
i{S}_{{\h{1}}{\h{2}}{\h{3}}{\h{4}}}
\spb{{\bf{1}}^{\lambda_1}}.{{\bf{2}}^{\lambda_2}}
\spb{{\bf{3}}^{\lambda_3}}.{{\bf{4}}^{\lambda_4}}
\nn\\
&\quad +
i{S}_{{\h{1}}{\h{3}}{\h{4}}{\h{2}}}
\spb{{\bf{1}}^{\lambda_1}}.{{\bf{3}}^{\lambda_3}}
\spb{{\bf{4}}^{\lambda_4}}.{{\bf{2}}^{\lambda_2}}
\nn\\
&\quad +
i{S}_{{\h{1}}{\h{4}}{\h{2}}{\h{3}}}
\spb{{\bf{1}}^{\lambda_1}}.{{\bf{4}}^{\lambda_4}}
\spb{{\bf{2}}^{\lambda_2}}.{{\bf{3}}^{\lambda_3}}
\nn\\
&\quad +
i{S}_{{\h{1}}^*{\h{2}}^*{\h{3}}^*{\h{4}}^*}
\spa{{\bf{1}}^{\lambda_1}}.{{\bf{2}}^{\lambda_2}}
\spa{{\bf{3}}^{\lambda_3}}.{{\bf{4}}^{\lambda_4}}
\nn\\
&\quad +
i{S}_{{\h{1}}^*{\h{3}}^*{\h{4}}^*{\h{2}}^*}
\spa{{\bf{1}}^{\lambda_1}}.{{\bf{3}}^{\lambda_3}}
\spa{{\bf{4}}^{\lambda_4}}.{{\bf{2}}^{\lambda_2}}
\nn\\
&\quad +
i{S}_{{\h{1}}^*{\h{4}}^*{\h{2}}^*{\h{3}}^*}
\spa{{\bf{1}}^{\lambda_1}}.{{\bf{4}}^{\lambda_4}}
\spa{{\bf{2}}^{\lambda_2}}.{{\bf{3}}^{\lambda_3}}
\,,
\label{eq:Affff1}
\end{align}
Using the Schouten identities (\ref{eq:bracket-schouten1}) and (\ref{eq:bracket-schouten2}), 
the amplitude are expressed in terms of non-redundant 
parameters $R_{\hat{i}\hat{j}\hat{k}\hat{l}}$, 
\begin{align}
\lefteqn{i\mathcal{A}^{(c)}_4({\bf{1}}^{\lambda_1},{\bf{2}}^{\lambda_2},{\bf{3}}^{\lambda_3},{\bf{4}}^{\lambda_4})}\nn\\
&=-
\frac{2i}{3}
{R}_{{\h{1}}({\h{3}}{\h{4}}){\h{2}}}
\spb{{\bf{1}}^{\lambda_1}}.{{\bf{2}}^{\lambda_2}}
\spb{{\bf{3}}^{\lambda_3}}.{{\bf{4}}^{\lambda_4}}
\nn\\
&\quad -
\frac{2i}{3}
{R}_{{\h{1}}({\h{2}}{\h{4}}){\h{3}}}
\spb{{\bf{1}}^{\lambda_1}}.{{\bf{3}}^{\lambda_3}}
\spb{{\bf{4}}^{\lambda_4}}.{{\bf{2}}^{\lambda_2}}
\nn\\
&\quad -
\frac{2i}{3}
{R}_{{\h{1}}({\h{2}}{\h{3}}){\h{4}}}
\spb{{\bf{1}}^{\lambda_1}}.{{\bf{4}}^{\lambda_4}}
\spb{{\bf{2}}^{\lambda_2}}.{{\bf{3}}^{\lambda_3}}
\nn\\
&\quad -
\frac{2i}{3} 
{R}_{{\h{1}}^*({\h{3}}^*{\h{4}}^*){\h{2}}^*}
\spa{{\bf{1}}^{\lambda_1}}.{{\bf{2}}^{\lambda_2}}
\spa{{\bf{3}}^{\lambda_3}}.{{\bf{4}}^{\lambda_4}}
\nn\\
&\quad -
\frac{2i}{3}{R}_{{\h{1}}^*({\h{2}}^*{\h{4}}^*){\h{3}}^*}
\spa{{\bf{1}}^{\lambda_1}}.{{\bf{3}}^{\lambda_3}}
\spa{{\bf{4}}^{\lambda_4}}.{{\bf{2}}^{\lambda_2}}
\nn\\
&\quad -
\frac{2i}{3} 
{R}_{ {\h{1}}^* ({\h{2}}^* {\h{3}}^*) {\h{4}}^*} 
\spa{{\bf{1}}^{\lambda_1}}.{{\bf{4}}^{\lambda_4}}
\spa{{\bf{2}}^{\lambda_2}}.{{\bf{3}}^{\lambda_3}}
\, .
\label{eq:Affff2}
\end{align}
with $R_{\hat{i}\hat{j}\hat{k}\hat{l}}$ being ``curvature'' tensors
as defined in (\ref{eq:new-coeff}).
Similary, 
the amplitude from the non-holomorophic contact 
interaction ($\psi\psi\psi^\dag\psi^\dag$) is given by 
\begin{align}
\lefteqn{i\mathcal{A}^{(c')}_4({\bf{1}}^{\lambda_1},{\bf{2}}^{\lambda_2},{\bf{3}}^{\lambda_3},{\bf{4}}^{\lambda_4})}\nn\\
&=
i{R}_{{\h{1}}{\h{3}}^*{\h{2}}{\h{4}}^*}
\spb{{\bf{1}}^{\lambda_1}}.{{\bf{2}}^{\lambda_2}}
\spa{{\bf{3}}^{\lambda_3}}.{{\bf{4}}^{\lambda_4}}
\nn\\
&\quad +
i{R}_{{\h{1}}{\h{4}}^*{\h{3}}{\h{2}}^*}
\spb{{\bf{1}}^{\lambda_1}}.{{\bf{3}}^{\lambda_3}}
\spa{{\bf{4}}^{\lambda_4}}.{{\bf{2}}^{\lambda_2}}
\nn\\
&\quad +
i{R}_{{\h{1}}{\h{2}}^*{\h{4}}{\h{3}}^*}
\spb{{\bf{1}}^{\lambda_1}}.{{\bf{4}}^{\lambda_4}}
\spa{{\bf{2}}^{\lambda_2}}.{{\bf{3}}^{\lambda_3}}
\nn\\
&\quad +
i{R}_{{\h{3}}{\h{1}}^*{\h{4}}{\h{2}}^*}
\spa{{\bf{1}}^{\lambda_1}}.{{\bf{2}}^{\lambda_2}}
\spb{{\bf{3}}^{\lambda_3}}.{{\bf{4}}^{\lambda_4}}
\nn\\
&\quad +
i{R}_{{\h{4}}{\h{1}}^*{\h{2}}{\h{3}}^*}
\spa{{\bf{1}}^{\lambda_1}}.{{\bf{3}}^{\lambda_3}}
\spb{{\bf{4}}^{\lambda_4}}.{{\bf{2}}^{\lambda_2}}
\nn\\
&\quad +
i{R}_{{\h{2}}{\h{1}}^*{\h{3}}{\h{4}}^*}
\spa{{\bf{1}}^{\lambda_1}}.{{\bf{4}}^{\lambda_4}}
\spb{{\bf{2}}^{\lambda_2}}.{{\bf{3}}^{\lambda_3}}
\,.
\end{align}
Here the ``curvature'' tensor $R_{\hat{i}\hat{j}^*\hat{k}\hat{l}^*}$ 
is defined in (\ref{eq:Rnonh}).
The scalar-exchange amplitude is computed as
\begin{align}
\lefteqn{
i\mathcal{A}^{(\phi)}_4({\bf{1}}^{\lambda_1},{\bf{2}}^{\lambda_2},{\bf{3}}^{\lambda_3},{\bf{4}}^{\lambda_4})
} \nonumber\\
 &
 = - \sum_{i, j}
     \biggl( 
      {M}_{{\h{1}}{\h{2}};i}
      \spb{{\bf{1}}^{\lambda_1}}.{{\bf{2}}^{\lambda_2}}
     +{M}_{{\h{1}}^*{\h{2}}^*;i}
      \spa{{\bf{1}}^{\lambda_1}}.{{\bf{2}}^{\lambda_2}}
     \biggr) 
     \nn\\
     & \qquad 
\,[D(s_{12})]^{ij}\,
     \biggl( 
      {M}_{{\h{3}}{\h{4}};i}
      \spb{{\bf{3}}^{\lambda_3}}.{{\bf{4}}^{\lambda_4}}
     +{M}_{{\h{3}}^*{\h{4}}^*;i}
      \spa{{\bf{3}}^{\lambda_3}}.{{\bf{4}}^{\lambda_4}}
     \biggr) 
\nn\\
  & \quad
   - \sum_{i, j}
     \biggl( 
      {M}_{{\h{1}}{\h{3}};i}
      \spb{{\bf{1}}^{\lambda_1}}.{{\bf{3}}^{\lambda_3}}
     +{M}_{{\h{1}}^*{\h{3}}^*;i}
      \spa{{\bf{1}}^{\lambda_1}}.{{\bf{3}}^{\lambda_3}}
     \biggr) 
       \nn\\
     & \qquad 
\,[D(s_{13})]^{ij}\,
     \biggl( 
      {M}_{{\h{2}}{\h{4}};i}
      \spb{{\bf{2}}^{\lambda_2}}.{{\bf{4}}^{\lambda_4}}
     +{M}_{{\h{2}}^*{\h{4}}^*;i}
      \spa{{\bf{2}}^{\lambda_2}}.{{\bf{4}}^{\lambda_4}}
     \biggr) 
\nn\\
  & \quad 
  - \sum_{i, j}
     \biggl( 
      {M}_{{\h{1}}{\h{4}};i}
      \spb{{\bf{1}}^{\lambda_1}}.{{\bf{4}}^{\lambda_4}}
     +{M}_{{\h{1}}^*{\h{4}}^*;i}
      \spa{{\bf{1}}^{\lambda_1}}.{{\bf{4}}^{\lambda_4}}
     \biggr) 
     \nn\\
     & \qquad 
\,[D(s_{14})]^{ij}\,
     \biggl( 
      {M}_{{\h{2}}{\h{3}};i}
      \spb{{\bf{2}}^{\lambda_2}}.{{\bf{3}}^{\lambda_3}}
     +{M}_{{\h{2}}^*{\h{3}}^*;i}
      \spa{{\bf{2}}^{\lambda_2}}.{{\bf{3}}^{\lambda_3}}
     \biggr) 
\,,
\nn\\
\label{eq:Affff3}
\end{align}
where $[D(s)]^{ij}$ denotes the scalar propagator (\ref{eq:scalarprop}).

We are now ready to discuss the high-energy behavior of the 
four-fermion amplitude, 
\bea
\psi^{{\h{1}}}(-p_1)\,\psi^{{\h{2}}}(-p_2)
\to
\psi^{{\h{3}}}(p_3)\,\psi^{{\h{4}}}(p_4)
\, ,
\eea
in the center-of-mass frame.
Taking the high-energy limit, $s=s_{12}\gg m_1^2$, $m_2^2$, $m_3^2$, $m_4^2$, 
we find the eight helicity amplitudes
\begin{align}
\mathcal{A}_{4}
({\bf{1}}^+,{\bf{2}}^+,{\bf{3}}^+,{\bf{4}}^+) \, , 
\quad
\mathcal{A}_{4}
({\bf{1}}^-,{\bf{2}}^-,{\bf{3}}^-,{\bf{4}}^-) \, , 
\end{align}
and
\begin{align}
&
\mathcal{A}_{4}
({\bf{1}}^+,{\bf{2}}^+,{\bf{3}}^-,{\bf{4}}^-) \, , 
\quad
\mathcal{A}_{4}
({\bf{1}}^+,{\bf{2}}^-,{\bf{3}}^+,{\bf{4}}^-) \, , 
\nn\\
&
\mathcal{A}_{4}
({\bf{1}}^+,{\bf{2}}^-,{\bf{3}}^-,{\bf{4}}^+) \, , 
\quad
\mathcal{A}_{4}({\bf{1}}^-,{\bf{2}}^-,{\bf{3}}^+,{\bf{4}}^+) \, , 
\nn\\
&
\mathcal{A}_{4}
({\bf{1}}^-,{\bf{2}}^+,{\bf{3}}^-,{\bf{4}}^+) \, , 
\quad
\mathcal{A}_{4}
({\bf{1}}^-,{\bf{2}}^+,{\bf{3}}^+,{\bf{4}}^-) \, , 
\end{align}
grow as energy squared. 
We obtain
\begin{align}
\mathcal{A}_{4}
({\bf{1}}^+,{\bf{2}}^+,{\bf{3}}^+,{\bf{4}}^+)
&\simeq
{R}_{{\h{1}}{\h{4}}{\h{2}}{\h{3}}}\,s
+\frac{1}{2}{R}_{{\h{1}}{\h{2}}{\h{3}}{\h{4}}}\,s\,(1+\cos\theta)
\,,
\label{eq:ffff++++}
\,\\
\mathcal{A}_{4}
({\bf{1}}^+,{\bf{2}}^+,{\bf{3}}^-,{\bf{4}}^-)
&\simeq
{R}_{{\h{1}}{\h{3}}^*{\h{2}}{\h{4}}^*}\,s
\label{eq:ffff++--}
\, .
\end{align}
We also find that the eight of sixteen helicity amplitudes 
\begin{align}
&
\mathcal{A}_{4}
({\bf{1}}^-,{\bf{2}}^+,{\bf{3}}^+,{\bf{4}}^+) \, , 
\quad
\mathcal{A}_{4}
({\bf{1}}^+,{\bf{2}}^-,{\bf{3}}^+,{\bf{4}}^+) \, , 
\nn\\
&
\mathcal{A}_{4}
({\bf{1}}^+,{\bf{2}}^+,{\bf{3}}^-,{\bf{4}}^+) \, , 
\quad
\mathcal{A}_{4}
({\bf{1}}^+,{\bf{2}}^+,{\bf{3}}^+,{\bf{4}}^-) \, , 
\nonumber\\
&
\mathcal{A}_{4}
({\bf{1}}^+,{\bf{2}}^-,{\bf{3}}^-,{\bf{4}}^-) \, , 
\quad
\mathcal{A}_{4}
({\bf{1}}^-,{\bf{2}}^+,{\bf{3}}^-,{\bf{4}}^-) \, , 
\nonumber\\
&
\mathcal{A}_{4}
({\bf{1}}^-,{\bf{2}}^-,{\bf{3}}^+,{\bf{4}}^-) \, , 
\quad
\mathcal{A}_{4}
({\bf{1}}^-,{\bf{2}}^-,{\bf{3}}^-,{\bf{4}}^+) \, , 
\end{align}
behave as $\sqrt{s}$ in high-energy limit.
For an example, we obtain
\begin{align}
\lefteqn{\mathcal{A}_{4}
({\bf{1}}^-,{\bf{2}}^+,{\bf{3}}^+,{\bf{4}}^+)}\nn\\
&\simeq
\frac{\sqrt{s}}{2}
\,\sin\theta
\biggl(
m_{{1}}{R}_{{\h{1}}{\h{2}}{\h{3}}{\h{4}}}
-
m_{{3}}{R}_{{\h{2}}{\h{1}}^*{\h{4}}{\h{3}}^*}
+
m_{{4}}{R}_{{\h{2}}{\h{1}}^*{\h{3}}{\h{4}}^*}
\biggr)
\, .
\end{align}
Note that the helicity structure determines
the difference of the high-energy behaviors.

%%%%%%%%%%%%%%%%%%%%%%%
\section{Summary}
\label{sec:summary}
We have formulated an extension of Higgs effective field theory 
(Generalized HEFT; GHEFT) which includes arbitrary number of spin-0 and 
spin-1/2 particles with arbitrary electric and chromoelectric charges. 
These particles include the SM quarks and leptons, and also BSM
Higgs bosons and fermions.
GHEFT can therefore describe the amplitude involving these non-SM particles.
This is in contrast to the usual EFT frameworks such as SMEFT and HEFT,
which cannot compute the cross sections and decay-widths of these new 
particles,
because these new particles are integrated out in these EFTs.
The leading order GHEFT Lagrangian has been expressed in accord with
the GHEFT chiral order counting rule, which clarifies
the relationship between the loop expansion and the operator expansion.

The $S$-matrix of a quantum field theory is unchanged by field 
redefinitions.
This fact, known as the Kamefuchi-O'Raifeartaigh-Salam (KOS) theorem,
tells us that seemingly different effective Lagrangians 
connected through the field coordinate transformations can describe the
identical scattering amplitudes.
The parametrization of the effective Lagrangian is therefore not unique.
In this paper, we have proposed to use the geometric quantities 
such as the 
curvature of field space in the GHEFT Lagrangian to resolve the 
redundancy.
We have also shown that, 
by introducing a useful coordinate (Normal coordinate) in the field space 
manifold, the computations of the scattering amplitudes are
significantly simplified.

We have also estimated tree-level on-shell amplitudes 
in section \ref{sec:amplitude}. 
These on-shell amplitudes are expressed in terms of the 
square and angle bracket notation of the spinor wavefunction.
The high-energy behaviors of the on-shell scattering amplitudes
are computed.
We found that the four-point scattering amplitudes grow as 
the scattering energy, and the coefficient of the energy-growing terms 
relate with the covariant tensors such as 
the curvature tensors on the field space. 
Perturbative unitarity in the scattering amplitudes requires 
the flatness in the scalar/fermion field space around the vacuum, {\it{i.e.}}, 
\begin{align}
&R_{1234}\biggr|_0 = R_{\hat{1}\hat{2}^* 34}\biggr|_0 = 
R_{\hat{1}\hat{2}\hat{3}\hat{4}}\biggr|_0 \nn\\
&= 
R_{\hat{1}^*\hat{2}^*\hat{3}^*\hat{4}^*}\biggr|_0 = 
R_{\hat{1}\hat{2}^* \hat{3}\hat{4}^*} \biggr|_0
=0 \, \nonumber.
\end{align}

The GHEFT framework should be studied further.
In order to apply the geometrical formulation in 
phenomenological studies, we need to compute the on-shell 
amplitudes involving the SM spin-1 particles $W$
and $Z$ in a geometrical language.
It should also be emphasized extra spin-1 particles often appear in 
models beyond the SM. 
For example, extra gauge bosons exist in the extensions of the
SM gauge group.
Spin-1 resonances like techni-$\rho$ may appear
in the strong dynamics models of the electroweak symmetry breaking.
These spin-1 particles have been studied in the electroweak resonance 
chiral perturbation theories.
It will be illuminating to investigate the geometrical formulations
for these spin-1 resonances in the GHEFT framework.

Radiative corrections should also be incorporated.
In order to compute the $\gamma\gamma$ decays of the Higgs particles, 
for an example, we need to investigate radiative corrections
in the GHEFT framework.
As we have shown in this paper,
the chiral order counting rule provides a basis
to compute these radiative corrections.
A geometrical formulation for the next-to-leading operators 
will also be useful in such a computation.

%%%%%%%%%%%%%%%%%%%%%%%%%%%%%%%%%%%%
\section*{Acknowledgments}
We thank Keisuke Izumi for valuable comments on the manuscript.
This work was supported by JSPS KAKENHI Grant Numbers 
JP19K14701 (RN), JP19K03846 (MT), and JP18H05543 (KT). The work of R.N. was also supported by the University of Padua through the ``New Theoretical Tools to Look at the Invisible Universe'' project and by Istituto Nazionale di Fisica Nucleare (INFN) through the ``Theoretical Astroparticle Physics'' (TAsP) project. 

%%%%%%%%%%%%%%%%%%%%%%%
\appendix
\section{Higgs effective field theory}
\label{app:HEFT}
Higgs effective field theory (HEFT) \refheft~is one of low-energy effective field theories for electroweak symmetry breaking. In the gaugeless limit, the leading order HEFT Lagrangian is defined as
\begin{align}
\mathcal{L}_{\rm{HEFT}}
&=
\mathcal{L}_{\rm{HEFT,boson}}
+
\mathcal{L}_{\rm{HEFT,fermion}}\,,
\label{eq:HEFT1}
\end{align}
where $\mathcal{L}_{\rm{HEFT,boson}}$ is found in Eq.\,(\ref{eq:LHEFTboson}), while $\mathcal{L}_{\rm{HEFT,fermion}}$ is given as
\begin{align}
\lefteqn{\mathcal{L}_{\rm{HEFT,fermion}}}\nn\\
&=
q^\dag_L i\bar{\sigma}^\mu \partial_\mu q_L
+
q^\dag_R i{\sigma}^\mu \partial_\mu q_R
\nn\\
&\quad
+
l^\dag_L i\bar{\sigma}^\mu \partial_\mu l_L
+
l^\dag_R i{\sigma}^\mu \partial_\mu l_R
\nn\\
&\quad+
\biggl[
q^\dag_L U 
\biggl(\mathcal{F}_{Y_q}(h){\bf{1}}_2
+
\mathcal{F}_{\hat{Y}_q}(h)\tau^3
\biggr)q_R
+h.c.
\biggr]\nn\\
&\quad+
\biggl[
l^\dag_L U 
\biggl(\mathcal{F}_{Y_l}(h){\bf{1}}_2
+
\mathcal{F}_{\hat{Y}_l}(h)\tau^3
\biggr)l_R
+h.c.
\biggr]\,,
\end{align}
with ${\bf{1}}_2$ being a $2\times 2$ unit matrix. $h$ denotes the 125\,GeV Higgs boson. $\mathcal{F}$'s are arbitrary functions of $h$. 
$q_L$ and $l_L$ denote the $SU(2)_W$ doublet SM quark and lepton fields, respectively. $q_R$ and $l_R$ are vectors defined as
\be
q_R
=
\left(
\begin{array}{ccc}
u_R\\
d_R\\
\end{array}
\right)
\,,~~~
l_R
=
\left(
\begin{array}{ccc}
0\\
e_R\\
\end{array}
\right)
\,,
\ee
where $u_R$, $d_R$, $e_R$ are the $SU(2)_W$ singlet up-quark, down-quark, and electron, respectively. Here we consider only one generation just for simplicity. It is straightforward to introduce the other generations.

Under the $G=SU(2)_W\times U(1)_Y$ transformation, the fields in Eq.\,(\ref{eq:HEFT1}) transforms as
\bea
U&\to& \fg_W U \fg^\dag_Y\,,\\
h&\to& h\,,\\
q_L&\to& e^{\frac{i}{6}\theta_Y}\fg_W q_L\,,~~~
q_R\to e^{\frac{i}{6}\theta_Y}\fg_Y q_R\,,\\
l_L&\to& e^{-\frac{i}{2}\theta_Y}\fg_W l_L\,,~~~
l_R\to e^{-\frac{i}{2}\theta_Y}\fg_Y l_R\,,
\eea
where $\fg_W\in SU(2)_W$ and $\fg_Y\in U(1)_Y$.
We can easily check that the HEFT Lagrangian (\ref{eq:HEFT1}) respects $G=SU(2)_W\times U(1)_Y$ invariance.

Let us introduce $\hat{u}_{L,R}$,  $\hat{e}_{L,R}$, and $\hat{\nu}_L$ as
\bea
\left(
\begin{array}{ccc}
\hat{u}_{L}\\
\hat{d}_{L}\\
\end{array}
\right)
&:=&
\exp\biggl({\frac{i}{6}\pi^3(x)}\biggr)\xi^\dag_W q_L\,,
\\
\left(
\begin{array}{ccc}
\hat{u}_{R}\\
\hat{d}_{R}\\
\end{array}
\right)
&:=&
\exp\biggl({\frac{i}{6}\pi^3(x)}\biggr)\xi_Y q_R\,,\\
\left(
\begin{array}{ccc}
\hat{\nu}_{L}\\
\hat{e}_{L}\\
\end{array}
\right)
&:=&
\exp\biggl(-{\frac{i}{2}\pi^3(x)}\biggr)\xi^\dag_W l_L\,,
\\
\left(
\begin{array}{ccc}
0\\
\hat{e}_{R}\\
\end{array}
\right)
&:=&
\exp\biggl(-{\frac{i}{2}\pi^3(x)}\biggr)\xi_Y l_R\,,
\eea
where $\xi_W$ and $\xi_Y$ are introduced in Eqs.~(\ref{eq:xiW}) and (\ref{eq:xiY}). We note that, under the $G$-transformation, the hatted fields transform as like $\psi^{\h{i}}$ we introduced;
\bea
\hat{u}_{L,R}
&\to&
\exp\biggl(i q_u \theta_Y(\pi,\fg_W,\fg_Y) \biggr)
\hat{u}_{L,R}\,,\\
\hat{d}_{L,R}
&\to&
\exp\biggl(i q_d \theta_Y(\pi,\fg_W,\fg_Y) \biggr)
\hat{d}_{L,R}\,,\\
\hat{\nu}_{L}
&\to&
\hat{\nu}_{L}\,,\\
\hat{e}_{L,R}
&\to&
\exp\biggl(i q_e \theta_Y(\pi,\fg_W,\fg_Y) \biggr)
\hat{e}_{L,R}\,,
\eea
where $(q_u,q_d,q_e)=(2/3,-1/3,-1)$. 

It is now easy to see that the GHEFT Lagrangian (\ref{eq:LGHEFT-sym}) reproduces the HEFT Lagrangian. The matter particle content of the HEFT corresponds to
\bea
\phi^i
&=&
(\pi^1,\pi^2,\pi^3,h)
\,,\\
\psi^{\h{i}}
&=&
(\hat{u}_L,\hat{d}_L,\hat{u}^\dag_R,\hat{d}^\dag_R,\hat{\nu}_L,\hat{e}_L,0,\hat{e}^\dag_R)
\,,
\label{eq:HEFTpsi}
\eea
with $h$ being $U(1)_{\rm{em}}$ neutral scalar, and the $U(1)_{\rm{em}}$ charges for $\psi$ are assigned as $q_{\h{i}}=(q_u,q_d,-q_u,-q_d,0,q_e,0,-q_e)$. 
We introduce a zero component (``0'') in Eq.\.(\ref{eq:HEFTpsi}) for latter convenience. 
We find that, in HEFT, $G_{aI}$, $G_{IJ}$,  $G_{{\h{i}}{\h{j}}^*}$, and $M_{{\h{i}}{\h{j}}}$  are taken as
\bea
G_{ah}
&=&
0\,,\\
G_{hh}
&=&
1\,,\\
G_{{\h{i}}{\h{j}}^*}
&=&
\delta_{{\h{i}}{\h{j}}^*}\,,
\\
M_{{\h{i}}{\h{j}}}
&=&
\left(
\begin{array}{cccc}
{\bf{0}} & \mathcal{F}_{q}(h) & {\bf{0}} & {\bf{0}}\\
\mathcal{F}_{q}(h) &  {\bf{0}} & {\bf{0}} & {\bf{0}}\\
{\bf{0}} & {\bf{0}} & {\bf{0}} & \mathcal{F}_{l}(h) \\
{\bf{0}} & {\bf{0}} & \mathcal{F}_{l}(h) & {\bf{0}} \\
\end{array}
\right)
\,,
\eea
where ${\bf{0}}$ denotes $2\times 2$ zero matrix and $\mathcal{F}_{q,l}=\mathcal{F}_{Y_{q,l}}{\bf{1}}_2+\mathcal{F}_{\hat{Y}_{q,l}}\tau^3$.
The other parameters are tuned to be
\bea
&&
G_{11}
=
G_{22}
=
G(h)\,,
~~~
G_{33}
=
G_Z(h)\,,\\
&&
G_{ab}
=
0\,~~\mbox{for}~~a\neq b\,,
\\
&&
V_{{\h{i}}{\h{j}}^*a}
=
-\left(
\begin{array}{cccc}
\tau^a & {\bf{0}} & {\bf{0}} & {\bf{0}}\\
{\bf{0}} & {\bf{0}} & {\bf{0}} & {\bf{0}}\\
{\bf{0}} & {\bf{0}} & \tau^a & {\bf{0}}\\
{\bf{0}} & {\bf{0}} & {\bf{0}} & {\bf{0}}\\
\end{array}
\right)
\,,~~~(a=1,2)\\
&&
V_{{\h{i}}{\h{j}}^*3}
=
-\left(
\begin{array}{cccc}
\tau^3 & {\bf{0}} & {\bf{0}} & {\bf{0}}\\
{\bf{0}} & {\bf{0}} & {\bf{0}} & {\bf{0}}\\
{\bf{0}} & {\bf{0}} & \tau^3 & {\bf{0}}\\
{\bf{0}} & {\bf{0}} & {\bf{0}} & {\bf{0}}\\
\end{array}
\right)\nn\\
&&
\qquad \quad
-c\,\mbox{diag}(q_u,q_d,-q_u,-q_d,0,q_e,0,-q_e)
\,,\\
&&
V_{{\h{i}}{\h{j}}^*h}
=
0
\,,
\eea
where $c$ is an arbitral parameter which appeared in Eq.\,(\ref{eq:covdel}). Furthermore, the four-fermion operators are assumed to be next-leading order in the HEFT. Namely, it is assumed that
\bea
&&S_{{\h{1}}{\h{2}}{\h{3}}{\h{4}}}=0\,,\\
&&S_{{\h{1}}^*{\h{2}}^*{\h{3}}^*{\h{4}}^*}=0\,,\\
&&S_{{\h{1}}{\h{2}}{\h{3}}^*{\h{4}}^*}=0\,,
\eea
at the leading order.
%%%%%%%%%%%%%%%%%%%%%%%
\section{Helicity eigenstate}
\label{app:helicitystate}
We consider a spin 1/2 field carrying a four momentum 
$p^{\mu} = (E,\vec{p} \, )$ 
where the direction of $\vec{p}$ is given by $\hat{p}=(\sin\theta_p\cos\phi_p,\sin\theta_p\sin\phi_p,\cos\theta_p)$. 
The two-component helicity spinor wave functions are given as \cite{Dreiner:2008tw}
\begin{align}
&x_\alpha(\vec{p},\lambda)
=
\omega_{-\lambda}(\vec{p}\,)\,\chi_\lambda (\hat{p})
\,,\\
&
x^\alpha(\vec{p},\lambda)
=
-\lambda\,\omega_{-\lambda}(\vec{p}\,)\,\chi^\dag_{-\lambda} (\hat{p})
\,,\\
&y_\alpha(\vec{p},\lambda)
=
\lambda\,\omega_{\lambda}(\vec{p}\,)\,\chi_{-\lambda} (\hat{p})
\,,\\
&
y^\alpha(\vec{p},\lambda)
=
\omega_{\lambda}(\vec{p}\,)\,\chi^\dag_{\lambda} (\hat{p})
\,,\\
&x^{\dag\dot{\alpha}}(\vec{p},\lambda)
=
-\lambda\,\omega_{-\lambda}(\vec{p}\,)\,\chi_{-\lambda} (\hat{p})
\,,\\
&
x^{\dag}_{\dot{\alpha}}(\vec{p},\lambda)
=
\omega_{-\lambda}(\vec{p}\,)\,\chi^\dag_{\lambda} (\hat{p})
\,,\\
&y^{\dag\dot{\alpha}}(\vec{p},\lambda)
=
\omega_{\lambda}(\vec{p}\,)\,\chi_{\lambda} (\hat{p})
\,,\\
&y^{\dag}_{\dot{\alpha}}(\vec{p},\lambda)
=
\lambda\,\omega_{\lambda}(\vec{p}\,)\,\chi^\dag_{-\lambda} (\hat{p})
\,,
\end{align}  
where 
\begin{align}
\chi_+(\hat{p})
&=
\left(
\begin{array}{ccc}
\cos\frac{\theta_p}{2}\\
e^{i\phi_p}\sin\frac{\theta_p}{2}\\
\end{array}
\right)\,,\\
\chi_-(\hat{p})
&=
\left(
\begin{array}{ccc}
-e^{-i\phi_p}\sin\frac{\theta_p}{2}\\
\cos\frac{\theta_p}{2}\\
\end{array}
\right)
\,,
\end{align}
and
\bea
\omega_{\pm}(\vec{p})
=
\sqrt{E\pm |\vec{p}\, |}\,,
\quad
E=\sqrt{|\vec{p}\, |^2+m^2}
\,.
\eea 

The spinor inner products (\ref{eq:defspb}) and (\ref{eq:defspa}) are 
obtained as
\begin{align}
\spb{{\bf{1}}^{+}}.{{\bf{2}}^{+}}
&=
\omega_{+}(\vec{p}_1)\,
\omega_{+}(\vec{p}_2)
\nn\\
&\quad
\biggl(
e^{-i\phi_{1}}\sin\frac{\theta_1}{2}\cos\frac{\theta_2}{2}
-
e^{-i\phi_{2}}\sin\frac{\theta_2}{2}\cos\frac{\theta_1}{2}
\biggr)
\,,\\
\spb{{\bf{1}}^{-}}.{{\bf{2}}^{-}}
&=
\omega_{-}(\vec{p}_1)\,
\omega_{-}(\vec{p}_2)
\nn\\
&\quad
\biggl(
e^{i\phi_{1}}\sin\frac{\theta_1}{2}\cos\frac{\theta_2}{2}
-
e^{i\phi_{2}}\sin\frac{\theta_2}{2}\cos\frac{\theta_1}{2}
\biggr)
\,,\\
\spb{{\bf{1}}^{\pm}}.{{\bf{2}}^{\mp}}
&=
\mp\,
\omega_{\pm}(\vec{p}_1)\,
\omega_{\mp}(\vec{p}_2)
\nn\\
&\quad
\biggl(
\cos\frac{\theta_1}{2}\cos\frac{\theta_2}{2}
+
e^{\mp i(\phi_{1}-\phi_{2})}\sin\frac{\theta_2}{2}\sin\frac{\theta_1}{2}
\biggr)
\,,\\
\spa{{\bf{1}}^{+}}.{{\bf{2}}^{+}}
&=
-
\omega_{-}(\vec{p}_1)\,
\omega_{-}(\vec{p}_2)
\nn\\
&\quad
\biggl(
e^{-i\phi_{1}}\sin\frac{\theta_1}{2}\cos\frac{\theta_2}{2}
-e^{-i\phi_{2}}\sin\frac{\theta_2}{2}\cos\frac{\theta_1}{2}
\biggr)
\,,\\
\spa{{\bf{1}}^{-}}.{{\bf{2}}^{-}}
&=
-
\omega_{+}(\vec{p}_1)\,
\omega_{+}(\vec{p}_2)
\nn\\
&\quad
\biggl(
e^{i\phi_{1}}\sin\frac{\theta_1}{2}\cos\frac{\theta_2}{2}
-
e^{i\phi_{2}}\sin\frac{\theta_2}{2}\cos\frac{\theta_1}{2}
\biggr)
\,,\\
\spa{{\bf{1}}^{\pm}}.{{\bf{2}}^{\mp}}
&=
\pm\,
\omega_{\mp}(\vec{p}_1)\,
\omega_{\pm}(\vec{p}_2)
\nn\\
&\quad
\biggl(
\cos\frac{\theta_1}{2}\cos\frac{\theta_2}{2}
+
e^{\mp i(\phi_{1}-\phi_{2})}\sin\frac{\theta_2}{2}\sin\frac{\theta_1}{2}
\biggr)\,,
\end{align}
where $\theta_i=\theta_{p_i}$ and $\phi_i=\phi_{p_i}$.

We now confirm that, in the the massless limit $m_1=m_2=0$,  
the spinor inner products become independent of the Lorentz-frame:
\begin{align}
\spb{{\bf{1}}^{+}}.{{\bf{2}}^{+}} \biggr|_{\rm{massless}}
&= -\sqrt{2 p_1 \cdot p_2 }
\, , \\
\spb{{\bf{1}}^{-}}.{{\bf{2}}^{-}}\biggr|_{\rm{massless}}
&=
\spb{{\bf{1}}^{\pm}}.{{\bf{2}}^{\mp}}\biggr|_{\rm{massless}}
=
0\,,\\
\spa{{\bf{1}}^{-}}.{{\bf{2}}^{-}} \biggr|_{\rm{massless}}
&= + \sqrt{2 p_1 \cdot p_2 }
\, ,\\
\spa{{\bf{1}}^{+}}.{{\bf{2}}^{+}}\biggr|_{\rm{massless}}
&=
\spa{{\bf{1}}^{\pm}}.{{\bf{2}}^{\mp}}\biggr|_{\rm{massless}}
=
0\, .
\end{align}
It should be emphasized that massive spinor products depend on 
the Lorentz-frame.
Taking the center of mass (c.o.m) frame 
($\theta_1=0, \theta_2=\pi, \phi_1=\phi_2=0$),  we obtain
\begin{align}
\spb{{\bf{1}}^{+}}.{{\bf{2}}^{+}}\biggr|_{\rm{c.o.m}}
&= -\sqrt{(E_1 + |\vec{p}_1|)(E_2 + |\vec{p}_2|)} \, , \\\spb{{\bf{1}}^{-}}.{{\bf{2}}^{-}}\biggr|_{\rm{c.o.m}}
&= \dfrac{m_1 m_2}{\spb{{\bf{1}}^{+}}.{{\bf{2}}^{+}}} \biggr|_{\rm{c.o.m}} \, ,
\\
\spa{{\bf{1}}^{-}}.{{\bf{2}}^{-}}\biggr|_{\rm{c.o.m}}
&= +\sqrt{(E_1 + |\vec{p}_1|)(E_2 + |\vec{p}_2|)} \, ,\\
\spa{{\bf{1}}^{+}}.{{\bf{2}}^{+}}\biggr|_{\rm{c.o.m}}
&= \dfrac{m_1 m_2}{\spa{{\bf{1}}^{-}}.{{\bf{2}}^{-}}} \biggr|_{\rm{c.o.m}} \, , 
\end{align}
and
\bea
\spb{{\bf{1}}^{\pm}}.{{\bf{2}}^{\mp}}\biggr|_{\rm{c.o.m}}
=
\spa{{\bf{1}}^{\pm}}.{{\bf{2}}^{\mp}}\biggr|_{\rm{c.o.m}}
=
0\,.
\eea

%%%%%%%%%%%%%%%%%%%%
\section{Higher order terms in the normal coordinate}
\label{sec:higerod-nc}
\subsection{Taylor expansion of $g_{ij}(\phi)$}
\label{sec-taylor-gij}

The $G_{ij(123)}$ term in the Taylor expansion
(\ref{eq:taylor-gij}) can be computed from 
the first covariant derivative of the
Riemann curvature tensor
\begin{align}
  R_{i' ijk; \, k_1}
  &= R_{i' ijk,\, k_1}
    -R_{k' ijk} \Gamma^{k'}_{i' k_1}
    \nn\\
    &\quad
    -R_{i' k' jk} \Gamma^{k'}_{i k_1}
    -R_{i' i k'k} \Gamma^{k'}_{j k_1}
    -R_{i' i j k'} \Gamma^{k'}_{k k_1} \, .
\end{align}
Since the Affine connection vanishes in the normal coordinate 
at the vacuum, 
\begin{align}
  \Gamma^{i''}_{ik} \biggr|_0
  &= 0 \, , 
\label{eq:bosonic-affine0}
\end{align}
we find
\begin{align}
R_{i12j;\, 3} \biggr|_0
  &= R_{i12j,\, 3} \biggr|_0
  \nonumber\\
  &= \dfrac{1}{2} \left(
      G_{ij(123)} - G_{1j(i23)} - G_{i2(j13)} + G_{12(ij3)}
     \right) \, .
\label{eq:first-R}
\end{align}
Symmetrizing under the $1 \leftrightarrow 2$ exchange, we obtain
\begin{align}
R_{i(12)j;\, 3} \biggr|_0
  &= \dfrac{1}{4} \left(
      2 G_{ij(123)} - G_{i1(j23)} - G_{i2(j13)}\r.\nn\\
      &\l.\quad - G_{1j(i23)} - G_{2j(i13)}
     +2 G_{12(ij3)} 
     \right) \, .
\label{eq:first-R1}
\end{align}
Note that $R_{i j 1 2 ; \, 3}$ satisfies the second Bianchi identity
\begin{align}
  R_{i j 12; \, 3}
 +R_{i j 23; \, 1}
 +R_{i j 31; \, 2} = 0 \, .
\end{align}
The coefficient $G_{ij (12)3}$ should be expressed in terms 
of covariant tensors in the normal coordinate.  
We here assume a form
\begin{align}
  G_{ij(12) 3}
  &= a R_{i(12)j;\, 3} \biggr|_0 
   + b \left[ R_{i(13)j; \, 2} + R_{i(23)j; \, 1} \right] \biggr|_0  \, .
\label{eq:assumed}
\end{align}
The second Bianchi identity implies
\begin{align}
  R_{i(12)3;j} + R_{3(12)j;i} = 2 R_{i(12)j;3} - \left[
    R_{i(23)j; 1} + R_{i(13)j; 2}
  \right] \, .
\end{align}
Therefore, the fifth order independent covariant tensors symmetric under the 
exchangs
\begin{align}
  i \leftrightarrow j \, , \qquad
  1 \leftrightarrow 2
\end{align}
are exhausted in the assumed form Eq.\,(\ref{eq:assumed}).
The condition (\ref{eq:first-R1}) can be
expressed as
\begin{align}
  &R_{i(12)j; 3} \biggr|_0
  = \dfrac{1}{2} \left[
       G_{ij(123)} + G_{12(ij3)} 
     \right]
     \nn\\
     & \quad
    -\dfrac{1}{4} \left[
      G_{i1(j23)} + G_{i2(j13)} +  G_{j1(i23)} + G_{j2(i13)} 
     \right] \, .
\label{eq:first-R2}
\end{align}
Using
\begin{align}
  G_{ij(123)} &= \dfrac{1}{3} \left[
    G_{ij(12)3} + G_{ij(23)1} + G_{ij(31)2}
  \right] 
\end{align}
and the assumption (\ref{eq:assumed}), 
the condition (\ref{eq:first-R2}) can be rewritten as
\begin{align}
  R_{i(12)j; 3} = (a+2b) R_{i(12)j;3} \, .
\end{align}
We therefore find
\begin{align}
  a+2b=1
\end{align}
and thus
\begin{align}
  G_{ij (12)3} 
  &= (1-2b) R_{i(12) j ; 3} \biggr|_0
   + b \left[
       R_{i(23)j; 1} + R_{i(31)j; 2} 
     \right]\biggr|_0
\label{eq:first-result}
\end{align}
satisfies Eq.\,(\ref{eq:first-R1}) with arbitrary $b$.

We see the dependence on the parameter $b$ disappears in
\begin{equation}
  G_{ij(123)} = \dfrac{1}{3} \left[
      R_{i(12)j; 3} + R_{i(23)j; 1} + R_{i(31)j; 2}  
  \right]  \biggr|_0 \, .
\label{eq:first-result2}
\end{equation}
The third order Taylor expansion coefficient in Eq.\,(\ref{eq:metric-taylor})
is therefore uniquely determined in the normal coordinate.

We next check whether or not Eq.\,(\ref{eq:first-result2}) does
satisfy the condition (\ref{eq:first-R}), which is severer 
than its symmetrized form Eq.\,(\ref{eq:first-R1}). 
Plugging the result Eq.\,(\ref{eq:first-result2}) in the RHS
of Eq.\,(\ref{eq:first-R}), we see
\begin{align}
\lefteqn{
  \dfrac{1}{2} (
    G_{ij(123)}  - G_{1j(i23)} - G_{i2(j13)} + G_{12(ij3)}
  )}
  \nonumber\\
  &= \dfrac{1}{12} \bigl[
      4 R_{i12j; 3} + 2 R_{i21j; 3} - 2R_{ij12; 3}
  \nonumber\\
  & \qquad
     +[R_{i23j} + 2R_{i32j} + R_{ij23}]_{;1}
         \nn\\
      &\qquad
     +[R_{i31j} + 2R_{i13j} + R_{ij31}]_{;2}
  \nonumber\\
  & \qquad
     +[R_{j123} + 2 R_{j213} + R_{j312} ]_{; i}
         \nn\\
      &\qquad
     +[R_{i213} + 2 R_{i123} + R_{i321} ]_{; j}
     \bigr] \biggr|_0 \, .
\label{eq:temp-r5}
\end{align}
The Bianchi identity implies
\begin{align}
  R_{i21j} - R_{ij12} &= R_{i12j} \, , 
  \\
  R_{i23j} + R_{ij23} &= R_{i32j} \, , 
  \\
  R_{i31j} + R_{ij31} &= R_{i13j} \, , 
  \\
  R_{j123} + R_{j312} &= R_{j213} \, ,
  \\
  R_{i213} + R_{i321} &= R_{i123} \, .
\end{align}
Using these results, the terms in the RHS of Eq.\,(\ref{eq:temp-r5})
can be simplified
\begin{align}
\lefteqn{
  \dfrac{1}{2} (
    G_{ij(123)}  - G_{1j(i23)} - G_{i2(j13)} + G_{12(ij3)}
  )}
  \nonumber\\
  &= \dfrac{1}{12} \bigl[
      6 R_{i12j; 3} 
     +3 (R_{j23i; 1} + R_{j213; i} ) 
          \nn\\
     &\qquad
     + 3 (R_{i13j; 2} +  R_{i123; j})
     \bigr] \biggr|_0 \, .
\label{eq:temp2-r5}
\end{align}
Using the implicatiosn of the 
second Bianchi identity 
\begin{align}
  R_{j23i; 1} + R_{j213; i} &= R_{j21i; 3} \, , \\
  R_{i13j; 2} + R_{i123; j} &= R_{i12j; 3}  \, , 
\end{align}
we can simplify Eq.\,(\ref{eq:temp2-r5}) further
\begin{align}
 \dfrac{1}{2} (
    G_{ij(123)}  - G_{1j(i23)} - G_{i2(j13)} + G_{12(ij3)}
  ) = R_{i12j; 3} \biggr|_0 \, .
\end{align}
We see the severer condition (\ref{eq:first-R}) is satisfied with 
the solution (\ref{eq:first-result2}) of the weaker condition (\ref{eq:first-R2}).

We next compute the second covariant derivative
\begin{align}
  \lefteqn{R_{i' ijk; \, k_1 k_2} \biggr|_0}\nn\\
  &= R_{i' ijk, \, k_1 k_2} \biggr|_0
    -R_{k' ijk} \Gamma^{k'}_{i' k_1, \, k_2} \biggr|_0
     -R_{i' k' jk} \, \Gamma^{k'}_{i k_1, \, k_2} \biggr|_0
  \nonumber\\
  &\quad
    -R_{i' i k'k} \, \Gamma^{k'}_{j k_1, \, k_2} \biggr|_0
    -R_{i' i j k'} \, \Gamma^{k'}_{k k_1, \, k_2} \biggr|_0 \, .
\end{align}
Using
\begin{align}
  \Gamma^{i''}_{ik,j} \biggr|_0
= -\dfrac{2}{3} g^{i' i''} R_{i' (ik)j} \biggr|_0 \, ,
\label{eq:bosonic-affine1}
\end{align}
we obtain
\begin{align}
  R_{i' ijk; \, k_1 k_2} \biggr|_0
  &= R_{i' ijk, \, k_1 k_2} \biggr|_0
   \nonumber\\
  & \quad
   +\dfrac{2}{3} R_{k' ijk} g^{k' k''} R_{k'' (i' k_1) k_2} \biggr|_0
     \nonumber\\
  & \quad
   +\dfrac{2}{3} R_{i' k' jk} g^{k' k''} R_{k''(i k_1) k_2} \biggr|_0
    \nonumber\\
  & \quad
   +\dfrac{2}{3} R_{i' i k'k} g^{k' k''} R_{k''(j k_1) k_2} \biggr|_0
      \nonumber\\
  & \quad
   +\dfrac{2}{3} R_{i' i j k'} g^{k' k''} R_{k'' (k k_1) k_2} \biggr|_0 \, .
\label{eq:temp-result1}
\end{align}
On the other hand, from Eq.\,(\ref{eq:curvature-expansion}), we obtain
\begin{align}
\lefteqn{  R_{i' ijk, k_1 k_2}}\nn\\
  &= \dfrac{1}{2} \left(
       G_{i'k(ijk_1 k_2)} - G_{ik(i'jk_1 k_2)}
       \r.\nn\\
       &\l.\qquad \qquad 
       - G_{i'j(ik k_1 k_2)} + G_{ij(i'k k_1 k_2)}
     \right)
  \nonumber\\
  & \quad
    -\dfrac{1}{4} \left(
        G_{i' j'(jk_1)} + G_{jj'(i' k_1)} - G_{i'j(j'k_1)}
    \right)
      \nonumber\\
  & \qquad \qquad
   \delta^{j' j''} \left(
        G_{j''i (k k_2)} + G_{j'' k (i k_2)} - G_{ik(j'' k_2)}
    \right)
  \nonumber\\
  & \quad
    -\dfrac{1}{4} \left(
        G_{i' j'(jk_2)} + G_{jj'(i' k_2)} - G_{i'j(j'k_2)}
    \right) 
          \nonumber\\
  & \qquad \qquad
  \delta^{j' j''} \left(
        G_{j''i (k k_1)} + G_{j'' k (i k_1)} - G_{ik(j'' k_1)}
    \right)
  \nonumber\\
  & \quad
   +\dfrac{1}{4} \left(
        G_{i' j'(k k_1)} + G_{k j'(i' k_1)} - G_{i' k (j'k_1)}
    \right)      
     \nonumber\\
  & \qquad \qquad
  \delta^{j' j''} 
    \left(
        G_{j''i (j k_2)} + G_{j'' j (i k_2)} - G_{i j (j'' k_2)}
    \right) 
  \nonumber\\
  & \quad
   +\dfrac{1}{4} \left(
        G_{i' j'(k k_2)} + G_{k j'(i' k_2)} - G_{i' k (j'k_2)}
    \right) 
          \nonumber\\
  & \qquad \qquad
    \delta^{j' j''} 
    \left(
        G_{j''i (j k_1)} + G_{j'' j (i k_1)} - G_{i j (j'' k_1)}
    \right)  \, .
\end{align}
We then use Eq.\,(\ref{eq:zero-result})
\begin{align}
\lefteqn{R_{i' ijk, k_1 k_2}}\nn\\
  &= \dfrac{1}{2} \left(
       G_{i'k(ijk_1 k_2)} - G_{ik(i'jk_1 k_2)} 
       \r.\nn\\
       &\l.\qquad \qquad 
       - G_{i'j(ik k_1 k_2)} + G_{ij(i'k k_1 k_2)}
     \right)
  \nonumber\\
  & \quad
    -\dfrac{1}{9} \left(
        R_{i' (jk_1) j'} + R_{j(i' k_1) j'} - R_{i' (j' k_1) j}
    \right)
          \nonumber\\
  & \qquad \qquad
     g^{j' j''} \left(
        R_{j'' (k k_2) i} + R_{j'' (i k_2) k} - R_{i (j'' k_2) k} 
    \right) \biggr|_0
  \nonumber\\
  & \quad
    -\dfrac{1}{9} \left(
        R_{i' (jk_2) j'} + R_{j(i' k_2) j'} - R_{i' (j' k_2) j}
    \right) 
          \nonumber\\
  & \qquad \qquad
    g^{j' j''} \left(
        R_{j'' (k k_1) i} + R_{j'' (i k_1) k} - R_{i (j'' k_1) k} 
    \right) \biggr|_0
  \nonumber\\
  & \quad
    +\dfrac{1}{9} \left(
        R_{i' (kk_1) j'} + R_{k(i' k_1) j'} - R_{i' (j' k_1) k}
    \right) 
          \nonumber\\
  & \qquad \qquad
    g^{j' j''} \left(
        R_{j'' (j k_2) i} + R_{j'' (i k_2) j} - R_{i (j'' k_2) j} 
    \right) \biggr|_0
  \nonumber\\
  & \quad
    +\dfrac{1}{9} \left(
        R_{i' (kk_2) j'} + R_{k(i' k_2) j'} - R_{i' (j' k_2) k}
    \right) 
          \nonumber\\
  & \qquad \qquad
    g^{j' j''} \left(
        R_{j'' (j k_1) i} + R_{j'' (i k_1) j} - R_{i (j'' k_1) j} 
    \right) \biggr|_0
\, .
\label{eq:temp-result2}
\end{align}
Combining Eqs.\,(\ref{eq:temp-result1}) and (\ref{eq:temp-result2}), 
we now find
\begin{align}
  \dfrac{1}{2} &\bigl(
       G_{i'k(ijk_1 k_2)} - G_{ik(i'jk_1 k_2)} - G_{i'j(ik k_1 k_2)} + G_{ij(i'k k_1 k_2)}
     \bigr)
  \nonumber\\
  &= R_{i' ijk; k_1 k_2} \biggr|_0
   \nn\\
   & \quad  
    - 
   \dfrac{2}{3} R_{k' ijk} g^{k' k''} R_{k'' (i' k_1) k_2} \biggr|_0
   \nn\\
   & \quad
   - 
   \dfrac{2}{3} R_{i' k' jk} g^{k' k''} R_{k''(i k_1) k_2} \biggr|_0
    \nn\\
   & \quad
   - 
   \dfrac{2}{3} R_{i' i k'k} g^{k' k''} R_{k''(j k_1) k_2} \biggr|_0
     \nn\\
   & \quad
   - 
   \dfrac{2}{3} R_{i' i j k'} g^{k' k''} R_{k'' (k k_1) k_2} \biggr|_0 \, .
  \nonumber\\
  & \quad
    +\dfrac{1}{9} \left(
        R_{i' (jk_1) j'} + R_{j(i' k_1) j'} - R_{i' (j' k_1) j}
    \right) 
          \nonumber\\
  & \qquad \qquad
    g^{j' j''} \left(
        R_{j'' (k k_2) i} + R_{j'' (i k_2) k} - R_{i (j'' k_2) k} 
    \right) \biggr|_0
  \nonumber\\
  & \quad
    +\dfrac{1}{9} \left(
        R_{i' (jk_2) j'} + R_{j(i' k_2) j'} - R_{i' (j' k_2) j}
    \right) 
          \nonumber\\
  & \qquad \qquad
    g^{j' j''} \left(
        R_{j'' (k k_1) i} + R_{j'' (i k_1) k} - R_{i (j'' k_1) k} 
    \right) \biggr|_0
  \nonumber\\
  & \quad
    -\dfrac{1}{9} \left(
        R_{i' (kk_1) j'} + R_{k(i' k_1) j'} - R_{i' (j' k_1) k}
    \right) 
          \nonumber\\
  & \qquad \qquad
    g^{j' j''} \left(
        R_{j'' (j k_2) i} + R_{j'' (i k_2) j} - R_{i (j'' k_2) j} 
    \right) \biggr|_0
  \nonumber\\
  & \quad
    -\dfrac{1}{9} \left(
        R_{i' (kk_2) j'} + R_{k(i' k_2) j'} - R_{i' (j' k_2) k}
    \right) 
          \nonumber\\
  & \qquad \qquad
    g^{j' j''} \left(
        R_{j'' (j k_1) i} + R_{j'' (i k_1) j} - R_{i (j'' k_1) j} 
    \right) \biggr|_0
\, .
\end{align}
Replacing the indices
\begin{equation}
  \label{eq:1}
  i' \to i, \quad
  k \to j, \quad
  i \to 1, \quad
  j \to 2, \quad
  k_1 \to 3, \quad
  k_2 \to 4, 
\end{equation}
the structure of the indices can be made more manifest:
\begin{align}
  \dfrac{1}{2} &\bigl(
       G_{ij(1234)} - G_{1j(i234)} - G_{i2(j134)} + G_{12(ij34)}
     \bigr)
  \nonumber\\
  &= R_{i12j; 34} \biggr|_0
  \nonumber\\
  & \quad
    - 
    \dfrac{2}{3} g^{i' j'} \left[
      R_{i' 12j} R_{j' (i 3) 4} 
    + R_{i' ij2} R_{j' (1 3) 4} 
    \r.\nn\\
    &\l.\qquad
   +  R_{i'j i1} R_{j' (23) 4} 
   +  R_{i'21 i} R_{j'(j3) 4} 
   \right] \biggr|_0 \, .
  \nonumber\\
  & \quad
    +\dfrac{1}{9} \left(
        R_{i (23) i'} + R_{2(i3) i'} - R_{i(i' 3) 2}
    \right) 
          \nonumber\\
  & \qquad \qquad
    g^{i' j'} \left(
        R_{j' (j4) 1} + R_{j' (14) j} - R_{1 (j' 4) j} 
    \right) \biggr|_0
  \nonumber\\
  & \quad
    +\dfrac{1}{9} \left(
        R_{i (24) i'} + R_{2(i4) i'} - R_{i (i' 4) 2}
    \right) 
          \nonumber\\
  & \qquad \qquad
    g^{i' j'} \left(
        R_{j' (j 3) 1} + R_{j' (13) j} - R_{1 (j' 3) j} 
    \right) \biggr|_0
  \nonumber\\
  & \quad
    -\dfrac{1}{9} \left(
        R_{i (j3) i'} + R_{j(i3) i'} - R_{i (i' 3) j}
    \right) 
          \nonumber\\
  & \qquad \qquad
    g^{i' j'} \left(
        R_{j' (24)1} + R_{j' (14) 2} - R_{1 (j' 4) 2} 
    \right) \biggr|_0
  \nonumber\\
  & \quad
    -\dfrac{1}{9} \left(
        R_{i (j4) i'} + R_{j(i 4) i'} - R_{i (i' 4) j}
    \right) 
          \nonumber\\
  & \qquad \qquad
    g^{i' j'} \left(
        R_{j' (2 3) 1} + R_{j' (13) 2} - R_{1 (j' 3) 2} 
    \right) \biggr|_0
\label{eq:temp-r6}
\, .
\end{align}
It is convenient to rewrite the formula (\ref{eq:temp-r6}) as
\begin{align}
  \dfrac{1}{2} &\bigl(
       G_{ij(1234)} - G_{1j(i234)} - G_{i2(j134)} + G_{12(ij34)}
     \bigr)
  \nonumber\\
  &= R_{i12j; 34} \biggr|_0
  \nonumber\\
  & \quad
   - 
   \dfrac{2}{3} g^{i' j'} \left[
      R_{i' 12j} R_{j' (i 3) 4} 
    + R_{i' ij2} R_{j' (1 3) 4} 
    \r.\nn\\
    &\l. \qquad
    +  R_{i'j i1} R_{j' (23) 4} 
   +  R_{i'21 i} R_{j'(j3) 4} 
   \right] \biggr|_0 \, .
  \nonumber\\
  & \quad
    +\dfrac{1}{9} g^{i' j'} \left(
        R_{i' (23) i} + R_{i'(i3) 2} - R_{i'(i2) 3}
    \right) 
              \nonumber\\
  & \qquad \qquad
    \left(
        R_{j' (j4) 1} + R_{j' (14) j} - R_{j' (j1) 4} 
    \right) \biggr|_0
  \nonumber\\
  & \quad
    +\dfrac{1}{9} g^{i' j'} \left(
        R_{i' (24) i} + R_{i'(i4) 2} - R_{i' (i2) 4}
    \right) 
              \nonumber\\
  & \qquad \qquad
    \left(
        R_{j' (j 3) 1} + R_{j' (13) j} - R_{j' (j1) 3} 
    \right) \biggr|_0
  \nonumber\\
  & \quad
    -\dfrac{1}{9} g^{i' j'} \left(
        R_{i' (j3) i} + R_{i'(i3) j} - R_{i' (ij) 3}
    \right) 
              \nonumber\\
  & \qquad \qquad
    \left(
        R_{j' (24)1} + R_{j' (14) 2} - R_{j' (12) 4}
    \right) \biggr|_0
  \nonumber\\
  & \quad
    -\dfrac{1}{9} g^{i' j'} \left(
        R_{i' (j4) i} + R_{i'(i 4) j} - R_{i' (ij) 4}
    \right) 
              \nonumber\\
  & \qquad \qquad
    \left(
        R_{j' (23) 1} + R_{j' (13) 2} - R_{j' (12)3}
    \right) \biggr|_0
\label{eq:temp2-r6}
\, .
\end{align}
We are now ready to solve Eq.\,(\ref{eq:temp2-r6}).
We first symmetrize the indices $(1234)$ by replacing
\begin{align}
  G_{1j(i234)} &\to 
    \dfrac{1}{4} \left[ 
      G_{1j(i234)} + G_{2j(i134)} 
      \r.\nn\\
      &\l.\qquad
      +\, G_{3j(i124)} + G_{4j(i123)}
    \right] \, , 
  \\
  G_{12(ij34)} & \to
    \dfrac{1}{6} \left[
      G_{12(ij34)} + G_{34(ij12)} 
       \r.\nn\\
      &\l.\qquad
      +\,G_{13(ij24)} + G_{24(ij13)} 
       \r.\nn\\
      &\l.\qquad
      +\,
     G_{14(ij23)} + G_{23(ij14)} 
    \right] \, ,  
  \\
  R_{i12j;34} & \to
    \dfrac{1}{6} \left[
      R_{i(12)j;(34)} + R_{i(34)j;(12)} 
      \right.
     \nonumber\\
     &
     \qquad 
     +R_{i(13)j;(24)} + R_{i(24)j;(13)}  
    \nonumber\\
    &
    \qquad 
    \left.
     +R_{i(14)j;(23)} + R_{i(23)j;(14)}  
    \right] \, ,  
  \\
  R_{i'12j} & \to
    R_{i'(12)j} \, , 
  \\
  R_{j'(i3)4} &\to
    -\dfrac{1}{2} R_{j'(34)i} \, , 
\label{eq:symmetrization1}
  \\
  g^{i'j'} R_{i'(12)j} R_{j'(34)i} &\to
  \dfrac{1}{6} g^{i' j'} \left[
    R_{i'(12)j} R_{j'(34)i} 
    \r.\nn\\
    &\l. \qquad \qquad
   +R_{i'(34)j} R_{j'(12)i} 
    \right.
    \nonumber\\
    &
    \qquad \qquad 
   +R_{i'(13)j} R_{j'(24)i} 
    \nn\\
    &\qquad \qquad
   +R_{i'(24)j} R_{j'(13)i} 
    \nonumber\\
    &
    \qquad \qquad 
    \left.
   +R_{i'(14)j} R_{j'(23)i}   
       \r.\nn\\
    &\l.\qquad \qquad
   +R_{i'(23)j} R_{j'(14)i}   
  \right]\, ,
\label{eq:symmetrization2}
  \\ 
  R_{j' (13)4}
  &\to  0 \, , 
  \\
  R_{i'(i3)2} - R_{i'(i2)3}
  &\to 0 \, .
\end{align}
We then obtain a symmetrized form of Eq.\,(\ref{eq:temp2-r6}):
\begin{align}
  & \dfrac{1}{2} G_{ij(1234)}
  \nonumber\\
  &
   - \dfrac{1}{8} [ G_{1j(i234)} + G_{2j(i134)} + G_{3j(i124)} + G_{4j(i123)} ]
  \nonumber\\
  &  - \dfrac{1}{8} [ G_{1i(j234)} + G_{2i(j134)} + G_{3i(j124)} + G_{4i(j123)} ]
  \nonumber\\
  &  + \dfrac{1}{12}  [ 
        G_{12(ij34)} + G_{34(ij12)} + G_{13(ij24)} +G_{24(ij13)} 
        \nn\\
        &\qquad
       +G_{14(ij23)} + G_{23(ij14)} ]
  \nonumber\\
  &= \dfrac{1}{6} [
       R_{i(12)j; (34)}  + R_{i(34)j; (12)} 
     + R_{i(13)j; (24)}  
         \nn\\
        &\qquad
        + R_{i(24)j; (13)} 
     + R_{i(14)j; (23)}  + R_{i(23)j; (14)} 
     ] \biggr|_0
  \nonumber\\
  & \quad
     +\dfrac{4}{27} 
    g^{i' j'} [
        R_{i' (12) i} R_{j' (34) j} 
       +R_{i' (13) i} R_{j' (24) j}  
           \nn\\
        &\qquad 
       +R_{i' (14) i} R_{j' (23) j}   
       +R_{i' (34) i} R_{j' (12) j} 
           \nn\\
        &\qquad
       +R_{i' (24) i} R_{j' (13) j}   
       +R_{i' (23) i} R_{j' (14) j}  ] \biggr|_0  \, .
\label{eq:temp3-r6}
\end{align}
The factor $4/27$ in the expression above is obtained as
\begin{displaymath}
  \left(
    -\dfrac{2}{3}
  \right)
  \left(
    -\dfrac{1}{2}
  \right)
  \left(
    \dfrac{1}{6}
  \right)
  \times 2
  + 
  \left(\dfrac{1}{9}\right)
  \left(\dfrac{1}{6}\right)
  \times 2 
  = \dfrac{4}{27} \, . 
\end{displaymath}
Here the factors $-2/3$ and $1/9$ are from Eq.\,(\ref{eq:temp2-r6}), 
and
the factor $-1/2$ from the symmetrization (\ref{eq:symmetrization1}), 
the factor $1/6$ from (\ref{eq:symmetrization2}).
We here assume a form
\begin{align}
\lefteqn{
  G_{ij(1234)}}\nn\\
  &= \dfrac{a}{6} \left[
       R_{i(12)j; (34)}  
     + R_{i(34)j; (12)} 
     + R_{i(13)j; (24)}  
       \r.\nn\\
       &\l. \qquad
     + R_{i(24)j; (13)} 
     + R_{i(14)j; (23)}  
     + R_{i(23)j; (14)} 
     \right] \biggr|_0
  \nonumber\\
  & \quad 
  + \dfrac{b}{6} g^{i' j'} \left[
       R_{i'(12)i} R_{j'(34)j} 
       + R_{i'(13)i} R_{j'(24)j} 
         \r.\nn\\
       &\l. \qquad
       + R_{i'(14)i} R_{j'(23)j}
      +R_{i'(34)i} R_{j'(12)j} 
             \r.\nn\\
       &\l. \qquad
      + R_{i'(24)i} R_{j'(13)j} 
      + R_{i'(23)i} R_{j'(14)j}
     \right] \biggr|_0 .
\label{eq:assumption4}
\end{align}
Using the Bianchi identity and the symmetries of the curvature tensor, 
it can be shown that
\begin{align}
  &G_{1j(i234)} + G_{2j(i134)} 
  \nn\\
  &\qquad
  + G_{3j(i124)} + G_{4j(i123)} 
  = -G_{ij(1234)} \, , \\
  &G_{1i(j234)} + G_{2i(j134)} 
   \nn\\
  &\qquad
  + G_{3i(j124)} + G_{4i(j123)} 
  = -G_{ij(1234)} \, , \\
  &G_{12(ij34)} + G_{34(ij12)} + G_{13(ij24)} \nn\\
  &\quad
  + G_{24(ij13)} + 
  G_{14(ij23)} + G_{23(ij14)} 
  = G_{ij(1234)} 
\end{align}
under the assumption (\ref{eq:assumption4}).
Eq.\,(\ref{eq:temp3-r6}) then reads
\begin{align}
  \lefteqn{\left( \dfrac{1}{2} + \dfrac{1}{8} + \dfrac{1}{8} + \dfrac{1}{12} \right)
  G_{ij(1234)}}
  \nonumber\\
  &
  = \dfrac{1}{6} [
       R_{i(12)j; (34)}  + R_{i(34)j; (12)} 
     + R_{i(13)j; (24)}  
     \nn\\
     & \qquad
     + R_{i(24)j; (13)} 
     + R_{i(14)j; (23)}  + R_{i(23)j; (14)} 
     ] \biggr|_0
  \nonumber\\
  & \qquad \quad
    +\dfrac{4}{27} g^{i' j'} [
        R_{i' (12) i} R_{j' (34) j} 
       +R_{i' (13) i} R_{j' (24) j}   
        \nonumber\\
  & \qquad \qquad\qquad
       +R_{i' (14) i} R_{j' (23) j}   
       +R_{i' (34) i} R_{j' (12) j} 
   \nonumber\\
  & \qquad \qquad\qquad     
       +R_{i' (24) i} R_{j' (13) j}   
       +R_{i' (23) i} R_{j' (14) j}  ] \biggr|_0  
\end{align}
and 
\begin{align}
  a = \dfrac{6}{5} \, , \qquad
  b = 
        \dfrac{16}{15} 
      \, .
\end{align}
We finally obtain
\begin{align}
  G_{ij(1234)}
  &= \dfrac{1}{5} \left[
       R_{i(12)j; (34)}  + R_{i(34)j; (12)} 
       \r. \nn\\
       &\l. \quad \quad
     + R_{i(13)j; (24)}  + R_{i(24)j; (13)} 
       \r. \nn\\
       &\l. \quad \quad
     + R_{i(14)j; (23)}  + R_{i(23)j; (14)} 
     \right] \biggr|_0
  \nonumber\\
  & \quad 
   + \dfrac{8}{45}
          g^{i' j'} \left[
       R_{i'(12)i} R_{j'(34)j} + R_{i'(13)i} R_{j'(24)j} 
         \r. \nonumber\\
  & \qquad \qquad
      + R_{i'(14)i} R_{j'(23)j}
      +R_{i'(34)i} R_{j'(12)j} 
     \nonumber\\
  & \l.\qquad \qquad     
      + R_{i'(24)i} R_{j'(13)j} + R_{i'(23)i} R_{j'(14)j}
     \right] \biggr|_0 .
\end{align}
The result is consistent with the Riemann Normal Coordinate
coefficient
\begin{align}
  G_{ij(k_1 k_2) (k_3 k_4)} 
  &= \dfrac{6}{5} R_{i(k_1 k_2) j ; (k_3 k_4)} \biggr|_0 
   \nn\\
  &\qquad
    +\dfrac{16}{15} R_{i(k_1 k_2) l} R^l{}_{(k_3 k_4) j} \biggr|_0 \, .
\end{align}

\subsection{Taylor expansion of $v_{{\h{i}}{\h{j}}^*i}(\phi)$}
\label{sec-taylor-vnni}

The Taylor expansion coefficient $A_{{\h{i}}{\h{j}}^* i12}$ in Eq.\,(\ref{eq:taylor-expansion-vnni}) can be computed from 
the first covariant derivative of the half-fermionic
Riemann tensor:
\begin{align}
  R_{{\h{i}}{\h{j}}^* ij;\, k_1}
  &= R_{{\h{i}}{\h{j}}^* ij, \, k_1}
    -R_{{\h{i}}'{\h{j}}^* ij} \Gamma^{{\h{i}}'}_{k_1 {\h{i}}} 
    -R_{{\h{i}}{\h{j}}'^* ij} \Gamma^{{\h{j}}'^*}_{k_1 {\h{j}}^*} 
  \nonumber\\
  & \qquad 
    -R_{{\h{i}}{\h{j}}^* i'j} \Gamma^{i'}_{k_1 i} 
    -R_{{\h{i}}{\h{j}}^* ij'} \Gamma^{j'}_{k_1 j} \, .
\end{align}
Since the Affine connections vanish at the vacuum in the normal coordinate, 
we see
\begin{align}
  R_{{\h{i}}{\h{j}}^* ij; \, k_1} \biggr|_{0}
  &= R_{{\h{i}}{\h{j}}^* ij,\, k_1} \biggr|_{0}\, .
\label{eq:temp2}
\end{align}
We then use Eq.\,(\ref{eq:fermion-curvature}) to obtain
\begin{align}
  \lefteqn{R_{{\h{i}}{\h{j}}^* ij,\, k_1} }\nn\\
  &= i A_{{\h{i}}{\h{j}}^* j i,\, k_1} + i A_{{\h{i}}{\h{j}}^* jk_1 , \, i}
     -i A_{{\h{i}}{\h{j}}^* ij,\, k_1} - i A_{{\h{i}}{\h{j}}^* ik_1 , \, j}
  \nonumber\\
  &\quad + i (A_{{\h{i}}{\h{j}}^* jk_2,\, ik_1}-A_{{\h{i}}{\h{j}}^* ik_2,\, jk_1}) \phi^{k_2}
  \nonumber\\
  &\quad 
     + g^{{\h{i}}' {\h{j}}'^*} (A_{{\h{i}}' {\h{j}}^* ik_1} A_{{\h{i}}{\h{j}}'^* jk_2}
                 +A_{{\h{i}}' {\h{j}}^* ik_2} A_{{\h{i}}{\h{j}}'^* jk_1})\phi^{k_2}
  \nonumber\\
  &\quad 
     - g^{{\h{i}}' {\h{j}}'^*} (A_{{\h{i}}' {\h{j}}^* jk_1} A_{{\h{i}}{\h{j}}'^* ik_2}
                 +A_{{\h{i}}' {\h{j}}^* jk_2} A_{{\h{i}}{\h{j}}'^* ik_1}) \phi^{k_2}
  \nonumber\\
  &\quad 
     + \bigl[ g^{{\h{i}}' {\h{j}}'^*} (A_{{\h{i}}' {\h{j}}^* ik_3} A_{{\h{i}}{\h{j}}'^* jk_2}
                 +A_{{\h{i}}' {\h{j}}^* ik_2} A_{{\h{i}}{\h{j}}'^* jk_3}) \bigr]_{,k_1}
       \phi^{k_2}\phi^{k_3}
  \nonumber\\
  &\quad 
     - \bigl[ g^{{\h{i}}' {\h{j}}'^*} (A_{{\h{i}}' {\h{j}}^* jk_3} A_{{\h{i}}{\h{j}}'^* ik_2}
                 +A_{{\h{i}}' {\h{j}}^* jk_2} A_{{\h{i}}{\h{j}}'^* ik_3}) \bigr]_{,k_1}
        \phi^{k_2}\phi^{k_3} \, .
\end{align}
Now we are ready to find a relation between $R_{{\h{i}}{\h{j}}^* ij, k_1}$ and
$A_{{\h{i}}{\h{j}}^* ij, k_1}$ at the vacuum,
\begin{align}
  R_{{\h{i}}{\h{j}}^* ij,\, k_1} \biggr|_0
  &= i [ A_{{\h{i}}{\h{j}}^* ji, \, k_1} + A_{{\h{i}}{\h{j}}^* jk_1, \, i}
   \nn\\
  &\qquad
        -A_{{\h{i}}{\h{j}}^* ij, \, k_1} - A_{{\h{i}}{\h{j}}^* ik_1, \, j}
       ] \biggr|_0 \, .
\end{align}
Combining the result with Eq.\,(\ref{eq:temp2}) we obtain
\begin{align}
  R_{{\h{i}}{\h{j}}^* ij;\, k_1} \biggr|_0
   &= i [ A_{{\h{i}}{\h{j}}^* jk_1, \, i} + A_{{\h{i}}{\h{j}}^* ji, \, k_1}  
    \nn\\
  &\qquad
         -A_{{\h{i}}{\h{j}}^* ij, \, k_1} + A_{{\h{i}}{\h{j}}^* k_1 i, \, j}
       ] \biggr|_0  \, .
\label{eq:temp3}
\end{align}
Since the function $A_{{\h{i}}{\h{j}}^* ij}(\phi)$ is anti-symmetric under the 
exchange of $i\leftrightarrow j$, Eq.\,(\ref{eq:temp3}) can be 
expressed as
\begin{align}
  R_{{\h{i}}{\h{j}}^* ij;\, k_1} \biggr|_0
   &= i [ A_{{\h{i}}{\h{j}}^* jk_1, \, i} + A_{{\h{i}}{\h{j}}^* k_1 i, \, j} -2A_{{\h{i}}{\h{j}}^* ij, \, k_1}  
       ] \biggr|_0  \, .
\label{eq:temp4}
\end{align}
We consider a Taylor expansion,
\begin{align}
  v_{{\h{i}}{\h{j}}^* i}(\phi)
  &= A_{{\h{i}}{\h{j}}^* ij} (\phi) \, \phi^j
  \nonumber\\
  &= A_{{\h{i}}{\h{j}}^* i k_1} \biggr|_0 \phi^{k_1} 
    +\dfrac{1}{2} A_{{\h{i}}{\h{j}}^* i (k_1, k_2)} \biggr|_0 \phi^{k_1} \phi^{k_2} 
    \nn\\
    &\quad
    +\dfrac{1}{3!} A_{{\h{i}}{\h{j}}^* i (k_1, k_2 k_3)} \biggr|_0 \phi^{k_1} \phi^{k_2} \phi^{k_3} 
    +\cdots \, .
\end{align}
The expansion coefficients 
$A_{{\h{i}}{\h{j}}^* ik_1}\biggr|_0$, $A_{{\h{i}}{\h{j}}^* i(k_1,k_2)}\biggr|_0$, 
$A_{{\h{i}}{\h{j}}^* i(k_1,k_2 k_3)}\biggr|_0$, $\cdots$,  should be 
written in terms of the covariant tensors in the normal coordinate.
Noting the anti-symmetry under the $i\leftrightarrow j$ exhange,
we assume a form
\begin{align}
  A_{{\h{i}}{\h{j}}^* ij, k_1} \biggr|_0
  &= a R_{{\h{i}}{\h{j}}^* ij; 1} \biggr|_0
    +b \left[
         R_{{\h{i}}{\h{j}}^* jk_1; i} + R_{{\h{i}}{\h{j}}^* k_1 i; j}
       \right] \, .
\label{eq:assume-a2}
\end{align}
Plugging the assumed form Eq.\,(\ref{eq:assume-a2}) 
in the RHS of Eq.\,(\ref{eq:temp4}), we obtain
\begin{align}
  \mbox{RHS}
  &= i(a-b) \left[ R_{{\h{i}}{\h{j}}^* j1; i} + R_{{\h{i}}{\h{j}}^* 1i; j} - 2 R_{{\h{i}}{\h{j}}^* ij, 1}
        \right] \biggr|_0 \, .
\end{align}
We note that the $R_{{\h{i}}{\h{j}}^* ij; k}$ satisfies the Bianchi identity
\begin{align}
  R_{{\h{i}}{\h{j}}^* j1; i} + R_{{\h{i}}{\h{j}}^* 1i; j} 
  &= - R_{{\h{i}}{\h{j}}^* ij; 1} \, ,
\end{align}
as we will discuss in \S~\ref{app:Bianchi}. 
Using the Bianchi identity, 
the expression can be simplified further 
\begin{align}
  \mbox{RHS} &= -3i(a-b) R_{{\h{i}}{\h{j}}^* ij; 1} \biggr|_0\, .
\end{align}
Comparing the result with Eq.\,(\ref{eq:temp4}), we now obtain
\begin{align}
  a-b &= \dfrac{i}{3}
\end{align}
and thus
\begin{align}
  A_{{\h{i}}{\h{j}}^* ij, 1} \biggr|_0
  &= \left( \dfrac{i}{3} + b \right) R_{{\h{i}}{\h{j}}^* i j; 1} \biggr|_0
    +b \left[
         R_{{\h{i}}{\h{j}}^* j1; i} + R_{{\h{i}}{\h{j}}^* 1i; j}
       \right] \biggr|_0  \, .
\end{align}
Note that the $b$ dependence is canceled in the the Taylor expansion 
coefficient $A_{{\h{i}}{\h{j}}^* i(1,2)} \biggr|_0$, 
\begin{align}
  A_{{\h{i}}{\h{j}}^* i (1, 2)} \biggr|_0
  &= \left( \dfrac{i}{3} + b \right) R_{{\h{i}}{\h{j}}^* i (1; 2)} \biggr|_0
  \nn\\
  &\qquad
    +b \left[
         R_{{\h{i}}{\h{j}}^* (12); i} - R_{{\h{i}}{\h{j}}^* i (1; 2)}
       \right]
  \nonumber\\
  &= \dfrac{i}{3} R_{{\h{i}}{\h{j}}^* i (1; 2)} \biggr|_0 \, .
\label{eq:A-result}
\end{align}
In the last line of Eq.\,(\ref{eq:A-result}), we used the Riemann tensor
symmetry
\begin{align}
  R_{{\h{i}}{\h{j}}^* 12} + R_{{\h{i}}{\h{j}}^* 21} = 0 \, .
\end{align}

We next consider the second covariant derivative of the
half-fermionic Riemann curvature tensor,
\begin{align}
  R_{{\h{i}}{\h{j}}^* ij;\, k_1 k_2}
  &= (R_{{\h{i}}{\h{j}}^* ij;\, k_1})_{,\, k_2}
    -R_{{\h{i}}'{\h{j}}^* ij;\, k_1} \Gamma^{{\h{i}}'}_{k_2 {\h{i}}} 
     \nonumber\\
  & \quad 
    -R_{{\h{i}}{\h{j}}'^* ij;\, k_1} \Gamma^{{\h{j}}'^*}_{k_2 {\h{j}}^*} 
     -R_{{\h{i}}{\h{j}}^* i'j;\, k_1} \Gamma^{i'}_{k_2 i} 
      \nonumber\\
  & \quad 
    -R_{{\h{i}}{\h{j}}^* ij';\, k_1} \Gamma^{j'}_{k_2 j} 
    -R_{{\h{i}}{\h{j}}^* i'j;\, k'_1} \Gamma^{k'_1}_{k_2 k'_1}  \, ,
\end{align}
which can be computed at the vacuum
\begin{align}
  R_{{\h{i}}{\h{j}}^* ij; \, k_1 k_2} \biggr|_{0}
  &= R_{{\h{i}}{\h{j}}^* ij, \, k_1 k_2} \biggr|_{0}
  \nonumber\\
  & \qquad
      -\dfrac{1}{2} g^{{\h{i}}' {\h{j}}'^*} R_{{\h{i}}' {\h{j}}^* ij} R_{{\h{i}}{\h{j}}'^* k_1 k_2} \biggr|_0
       \nonumber\\
  & \qquad
      +\dfrac{1}{2} g^{{\h{i}}' {\h{j}}'^*} R_{{\h{i}} {\h{j}}'^* ij} R_{{\h{i}}' {\h{j}}^* k_1 k_2} \biggr|_0
  \nonumber\\
&
\qquad
     +\dfrac{1}{3} g^{i' j'} R_{{\h{i}}{\h{j}}^* i' j} \left(
         R_{j' k_1ik_2} + R_{j' i k_1 k_2}
      \right)
       \nonumber\\
  & \qquad
     +\dfrac{1}{3} g^{i' j'} R_{{\h{i}}{\h{j}}^* i i'} \left(
         R_{j' k_1jk_2} + R_{j' j k_1 k_2}
      \right) \, .
\end{align}
Here we used 
Eqs. (\ref{eq:bosonic-affine0}), (\ref{eq:bosonic-affine1}), (\ref{eq:fermionic-affine02}), and (\ref{eq:fermionic-affine12}).
We then obtain
\begin{align}
  R_{{\h{i}}{\h{j}}^* ij;\, (k_1 k_2)} \biggr|_0
  &= R_{{\h{i}}{\h{j}}^* ij,\, (k_1 k_2)} \biggr|_0
  \nonumber\\
  & \qquad
    -\dfrac{1}{3} g^{i' j'} R_{{\h{i}}{\h{j}}^* i' j} R_{j' (k_1 k_2) i} \biggr|_0
     \nonumber\\
  & \qquad
    -\dfrac{1}{3} g^{i' j'} R_{{\h{i}}{\h{j}}^* i i'} R_{j' (k_1 k_2) j} \biggr|_0\, .
\label{eq:temp6}
\end{align}
We next use Eq.\,(\ref{eq:fermion-curvature}) to find
\begin{align}
\lefteqn{
  R_{{\h{i}}{\h{j}}^* ij,\, k_1 k_2} \biggr|_0}\nn\\
  &= i [ A_{{\h{i}}{\h{j}}^* ji,\, k_1 k_2} + 2A_{{\h{i}}{\h{j}}^* j (k_1, \, k_2) i}
  \nn\\
  &\qquad
        -A_{{\h{i}}{\h{j}}^* ij,\, k_1 k_2} - 2A_{{\h{i}}{\h{j}}^* i (k_1, \, k_2) j} ] \biggr|_0
  \nonumber\\
  & \qquad
    - \dfrac{1}{4} g^{{\h{i}}' {\h{j}}'^*} [ R_{{\h{i}}' {\h{j}}^* ik_1} R_{{\h{i}}{\h{j}}'^* jk_2}
                 +R_{{\h{i}}' {\h{j}}^* ik_2} R_{{\h{i}}{\h{j}}'^* jk_1} ] \biggr|_0
  \nonumber\\
  & \qquad
    + \dfrac{1}{4} g^{{\h{i}}' {\h{j}}'^*} [ R_{{\h{i}}' {\h{j}}^* jk_1} R_{{\h{i}}{\h{j}}'^* ik_2}
                 +R_{{\h{i}}' {\h{j}}^* jk_2} R_{{\h{i}}{\h{j}}'^* ik_1} ] \biggr|_0  \, .
\label{eq:temp7}
\end{align}
Combining Eqs.\,(\ref{eq:temp6}) and (\ref{eq:temp7}), we now
find a formula which relates the function $A_{\h{i}\h{j}^* ij}$ with the
Riemann curvature tensor,
\begin{align}
  iA  \biggr|_0 
  &= - R_{{\h{i}}{\h{j}}^* ij;\, (k_1 k_2)} \biggr|_0 
   \nonumber\\
  & \quad
     - \dfrac{1}{3} g^{i' j'} R_{{\h{i}}{\h{j}}^* i' j} R_{j'(k_1 k_2)i} \biggr|_0
      \nonumber\\
  & \quad
     + \dfrac{1}{3} g^{i' j'} R_{{\h{i}}{\h{j}}^* i' i} R_{j'(k_1 k_2)j} \biggr|_0
  \nonumber\\
  & \quad -\dfrac{1}{4} g^{{\h{i}}' {\h{j}}'^*} \left(
       R_{{\h{i}}' {\h{j}}^* ik_1} R_{{\h{i}}{\h{j}}'^* jk_2} + R_{{\h{i}}' {\h{j}}^* ik_2} R_{{\h{i}}{\h{j}}'^* jk_1} 
    \right) \biggr|_0
  \nonumber\\
  & \quad +\dfrac{1}{4} g^{{\h{i}}' {\h{j}}'^*} \left(
       R_{{\h{i}}' {\h{j}}^* jk_1} R_{{\h{i}}{\h{j}}'^* ik_2} + R_{{\h{i}}' {\h{j}}^* jk_2} R_{{\h{i}}{\h{j}}'^* ik_1} 
    \right) \biggr|_0 \, .
\label{eq:r2-formula}
\end{align}
Here we introduced a short-hand notation
\begin{align}
  A &:= A_{{\h{i}}{\h{j}}^* ij,\, (k_1 k_2)} - A_{{\h{i}}{\h{j}}^* ji , \, (k_1 k_2)} 
   \nn\\
  &\qquad
      + 2  A_{{\h{i}}{\h{j}}^* i (k_1,\, k_2)j} - 2 A_{{\h{i}}{\h{j}}^* j (k_1,\, k_2)i} \, .
\label{eq:defA}
\end{align}

The function $A_{{\h{i}}{\h{j}}^* ij}(\phi)$ should be expressed in a covariant form 
in the normal coordinate.  We therefore assume
\begin{align}
\lefteqn{
  A_{{\h{i}}{\h{j}}^* ij, \, k_1 k_2} \biggr|_0
  }\nn\\
  &= a R_{{\h{i}}{\h{j}}^* ij; \, (k_1 k_2)}
       \biggr|_0
  \nonumber\\
  & \quad
    + b g^{i' j'} \left[
        R_{{\h{i}}{\h{j}}^* i' i} R_{j' (k_1 k_2) j} 
      - R_{{\h{i}}{\h{j}}^* i' j} R_{j' (k_1 k_2) i} 
      \right] \biggr|_0
  \nonumber\\
  & \quad
    + c g^{i' j'} \left[
        R_{{\h{i}}{\h{j}}^* i' k_1} R_{j' k_2 ij}
       +R_{{\h{i}}{\h{j}}^* i' k_2} R_{j' k_1 ij}
      \right] \biggr|_0
  \nonumber\\
  & \quad
   + d g^{{\h{i}}' {\h{j}}'^*} \left( 
       R_{{\h{i}}' {\h{j}}^* ik_1} R_{{\h{i}}{\h{j}}'^* jk_2} + R_{{\h{i}}' {\h{j}}^* ik_2} R_{{\h{i}}{\h{j}}'^* jk_1} 
     \right) \biggr|_0
  \nonumber\\
  & \quad
   - d g^{{\h{i}}' {\h{j}}'^*} \left( 
       R_{{\h{i}}' {\h{j}}^* jk_1} R_{{\h{i}}{\h{j}}'^* ik_2} + R_{{\h{i}}' {\h{j}}^* jk_2} R_{{\h{i}}{\h{j}}'^* ik_1} 
     \right) \biggr|_0 \, .
\label{eq:temp8}
\end{align}
Here $a$, $b$, $c$, $d$ are constants.  Plugging the form 
of assumption (\ref{eq:temp8}) into the definition of
the short-hand notation (\ref{eq:defA}) and using the symmetry
structure of $R_{{\h{i}}{\h{j}}^* ij}$ and 
\begin{align}
  R_{{\h{i}}{\h{j}}^* ij;\, k_1 k_2} 
 -R_{{\h{i}}{\h{j}}^* ij;\, k_2 k_1} 
  &= R_{{\h{i}}' {\h{j}}^* ij} R^{{\h{i}}'}{}_{{\h{i}} k_1 k_2} 
   \nonumber\\
  & \quad
    +R_{{\h{i}} {\h{j}}'^* ij} R^{{\h{j}}'^*}{}_{{\h{j}}^* k_1 k_2} 
  \nonumber\\
  &\quad
    +R_{{\h{i}}{\h{j}}^* i'j} R^{i'}{}_{ik_1 k_2}
     \nonumber\\
  & \quad
    +R_{{\h{i}}{\h{j}}^* ij'} R^{j'}{}_{jk_1 k_2} \, , 
\end{align}
we find
\begin{align}
  &A \biggr|_0= 4a  R_{{\h{i}}{\h{j}}^* ij;\, (k_1k_2)} \biggr|_0
    \nonumber\\
    &\qquad
      + \left( -a + b + 2c \right)
       g^{i' j'} \left[
        R_{{\h{i}}{\h{j}}^* i' i} R_{j' (k_1 k_2) j} 
        \r.\nn\\
        &\l. \qquad \qquad \qquad \qquad \qquad \qquad
      - R_{{\h{i}}{\h{j}}^* i' j} R_{j' (k_1 k_2) i} 
      \right] \biggr|_0
   \nonumber\\
   &\qquad
      +\left( 
         \dfrac{a}{2} + \dfrac{3}{2} b + 3 c
        \right)
       g^{i' j'} \left[
        R_{{\h{i}}{\h{j}}^* i' k_1} R_{j' k_2 ij}
          \r.\nn\\
        &\l. \qquad \qquad \qquad \qquad \qquad \qquad
       +R_{{\h{i}}{\h{j}}^* i' k_2} R_{j' k_1 ij}
      \right] \biggr|_0
  \nonumber\\
  &\qquad
   +  a 
       g^{{\h{i}}' {\h{j}}'^*} \left( 
       R_{{\h{i}}' {\h{j}}^* ik_1} R_{{\h{i}}{\h{j}}'^* jk_2} + R_{{\h{i}}' {\h{j}}^* ik_2} R_{{\h{i}}{\h{j}}'^* jk_1} 
     \right) \biggr|_0
  \nonumber\\
  & \qquad
   - a 
       g^{{\h{i}}' {\h{j}}'^*} \left( 
       R_{{\h{i}}' {\h{j}}^* jk_1} R_{{\h{i}}{\h{j}}'^* ik_2} + R_{{\h{i}}' {\h{j}}^* jk_2} R_{{\h{i}}{\h{j}}'^* ik_1} 
     \right) \biggr|_0 \, .
\label{eq:temp9}
\end{align}
Comparing Eq.\,(\ref{eq:temp9}) with Eq.\,(\ref{eq:r2-formula}), we obtain
\begin{align}
  4ia &= -1 \, , \\
  i \left( 
   -a + b + 2c 
  \right)  
     &= \dfrac{1}{3} \, , \\
  i \left( 
    \dfrac{a}{2} + \dfrac{3}{2} b + 3c 
  \right)
     &= 0 \, , \\
  i a 
     &= -\dfrac{1}{4} \, ,
\end{align}
which can be solved as
\begin{align}
  a &= \dfrac{i}{4} \, , \\
  b &= -\dfrac{i}{12}  - 2c \, .
\end{align}
Note that the coefficients $c$ and $d$ are left to be undetermined constants.
We thus obtain an expression of $A_{{\h{i}}{\h{j}}^* ij,\, k_1 k_2}$
\begin{align}
\lefteqn{
  A_{{\h{i}}{\h{j}}^* ij, \, k_1 k_2} \biggr|_0}\nn\\
  &= \dfrac{i}{4} R_{{\h{i}}{\h{j}}^* ij; \, (k_1 k_2)}
       \biggr|_0
  \nonumber\\
  & \quad
    -\left( \dfrac{i}{12} + 2c \right) g^{i' j'} \left[
        R_{{\h{i}}{\h{j}}^* i' i} R_{j' (k_1 k_2) j} 
      - R_{{\h{i}}{\h{j}}^* i' j} R_{j' (k_1 k_2) i} 
      \right] \biggr|_0
  \nonumber\\
  & \quad
    + c g^{i' i''} \left[
        R_{{\h{i}}{\h{j}}^* i' k_1} R_{i'' k_2 ij}
       +R_{{\h{i}}{\h{j}}^* i' k_2} R_{i'' k_1 ij}
      \right] \biggr|_0
  \nonumber\\
  & \quad
   + d g^{{\h{i}}' {\h{j}}'^*} \left( 
       R_{{\h{i}}' {\h{j}}^* ik_1} R_{{\h{i}}{\h{j}}'^* jk_2} + R_{{\h{i}}' {\h{j}}^* ik_2} R_{{\h{i}}{\h{j}}'^* jk_1} 
     \right) \biggr|_0
  \nonumber\\
  & \quad
   - d g^{{\h{i}}' {\h{j}}'^*} \left( 
       R_{{\h{i}}' {\h{j}}^* jk_1} R_{{\h{i}}{\h{j}}'^* ik_2} + R_{{\h{i}}' {\h{j}}^* jk_2} R_{{\h{i}}{\h{j}}'^* ik_1} 
     \right) \biggr|_0 \, .
\end{align}

The dependence on parameters $c$ and $d$ is canceled in 
\begin{align}
  \lefteqn{A_{{\h{i}}{\h{j}}^* i(1, 23)} \biggr|_0}\nn\\
  &= \dfrac{i}{4} R_{{\h{i}}{\h{j}}^* i (1; 23)}  \biggr|_0
  \nonumber\\
  & \quad - \left( 
              \dfrac{i}{12} + 2c
            \right) g^{i' j'}
            \biggl[
               R_{{\h{i}}{\h{j}}^* i' i} R_{j' (123)}
             -\dfrac{1}{3} R_{{\h{i}}{\h{j}}^* i' 1} R_{j' (23) i}
  \nonumber\\
  & \qquad\qquad \qquad 
             -\dfrac{1}{3} R_{{\h{i}}{\h{j}}^* i' 2} R_{j' (31) i}
             -\dfrac{1}{3} R_{{\h{i}}{\h{j}}^* i' 3} R_{j' (12) i}
            \biggr] \biggr|_0
 \nonumber\\
  & \quad -\dfrac{2}{3} c g^{i' j'} \left[
          R_{{\h{i}}{\h{j}}^* i' 1} R_{j' (23) i}
          \r.\nn\\
          &\qquad \qquad \l.
        + R_{{\h{i}}{\h{j}}^* i' 2} R_{j' (31) i}
        + R_{{\h{i}}{\h{j}}^* i' 3} R_{j' (12) i}
      \right] \biggr|_0
 \nonumber\\
  &= \dfrac{i}{4} R_{{\h{i}}{\h{j}}^* i (1; 23)} \biggr|_0
    +\dfrac{i}{36} g^{i' j'} \left[
       R_{{\h{i}}{\h{j}}^* i' 1} R_{j' (23) i}
           \r.\nn\\
          &\qquad \qquad \l.
      +R_{{\h{i}}{\h{j}}^* i' 2} R_{j' (31) i}
      +R_{{\h{i}}{\h{j}}^* i' 3} R_{j' (12) i}
     \right] \biggr|_0 \, .
\end{align}
The third order Taylor expansion coefficient in
\begin{align}
  v_{{\h{i}}{\h{j}}^* i}(\phi)
  &= A_{{\h{i}}{\h{j}}^* ij} (\phi) \, \phi^j
  \nonumber\\
  &= A_{{\h{i}}{\h{j}}^* i k_1} \biggr|_0 \phi^{k_1} 
    +\dfrac{1}{2} A_{{\h{i}}{\h{j}}^* i (k_1, k_2)} \biggr|_0 \phi^{k_1} \phi^{k_2} 
    \nonumber\\
  & \quad
    +\dfrac{1}{3!} A_{{\h{i}}{\h{j}}^* i (k_1, k_2 k_3)} \biggr|_0 \phi^{k_1} \phi^{k_2} \phi^{k_3} 
    +\cdots
\end{align}
is therefore uniquely determined in the normal coordinate.

\subsection{A proof on ``half-fermionic Bianchi identity''}
\label{app:Bianchi}

The half-fermionic curvature tensor as defined in 
Eq.\,(\ref{eq:half-fermionic-curvature}) satisfies a Bianchi type identity
\begin{align}
  R^{\h{i}}{}_{{\h{j}} 12; 3} + R^{\h{i}}{}_{{\h{j}} 23; 1} + R^{\h{i}}{}_{{\h{j}} 31; 2} = 0 \, .
\label{eq:bianchi-identity}
\end{align}
We give a proof of Eq.\,(\ref{eq:bianchi-identity}) in this appendix.

We first compute $R^{\h{i}}{}_{\h{j}12;3}$, 
\begin{align}
\lefteqn{
  R^{\h{i}}{}_{{\h{j}}12;3}}\nn\\
  &= (\Gamma^{\h{i}}_{2{\h{j}}, 31} - \Gamma^{\h{i}}_{1{\h{j}}, 23})
  \nonumber\\
  & \quad
     +(\Gamma^{\h{i}}_{1{\h{j}}'} \Gamma^{{\h{j}}'}_{2{\h{j}}, 3} - \Gamma^{\h{i}}_{3{\h{j}}'} \Gamma^{{\h{j}}'}_{1{\h{j}},2})
     +(\Gamma^{\h{i}}_{3{\h{j}}'} \Gamma^{{\h{j}}'}_{2{\h{j}}, 1} - \Gamma^{\h{i}}_{2{\h{j}}'} \Gamma^{{\h{j}}'}_{1{\h{j}},3})
  \nonumber\\
  & \quad
    +(\Gamma^{\h{i}}_{1{\h{j}}',2} \Gamma^{{\h{j}}'}_{3{\h{j}}} - \Gamma^{\h{i}}_{2{\h{j}}',3} \Gamma^{{\h{j}}'}_{1{\h{j}}})
    +(\Gamma^{\h{i}}_{1{\h{j}}',3} \Gamma^{{\h{j}}'}_{2{\h{j}}} - \Gamma^{\h{i}}_{2{\h{j}}',1} \Gamma^{{\h{j}}'}_{3{\h{j}}})
  \nonumber\\
  & \quad
    +(\Gamma^{\h{i}}_{1{\h{j}}, i} \Gamma^i_{23} - \Gamma^{\h{i}}_{2{\h{j}}, i} \Gamma^i_{31})
    +(\Gamma^{\h{i}}_{i{\h{j}}, 2} \Gamma^i_{31} - \Gamma^{\h{i}}_{i{\h{j}}, 1} \Gamma^i_{23})
  \nonumber\\
  & \quad
   +(\Gamma^{\h{i}}_{3{\h{i}}'} \Gamma^{{\h{i}}'}_{1{\h{j}}'} \Gamma^{{\h{j}}'}_{2{\h{j}}} 
   -\Gamma^{\h{i}}_{1{\h{i}}'} \Gamma^{{\h{i}}'}_{2{\h{j}}'} \Gamma^{{\h{j}}'}_{3{\h{j}}} )
     \nonumber\\
  & \quad
   +(\Gamma^{\h{i}}_{2{\h{i}}'} \Gamma^{{\h{i}}'}_{1{\h{j}}'} \Gamma^{{\h{j}}'}_{3{\h{j}}} 
   -\Gamma^{\h{i}}_{3{\h{i}}'} \Gamma^{{\h{i}}'}_{2{\h{j}}'} \Gamma^{{\h{j}}'}_{1{\h{j}}} )
  \nonumber\\
  & \quad
   +\Gamma^{\h{i}}_{i{\h{i}}'} ( 
      \Gamma^i_{23} \Gamma^{{\h{i}}'}_{1{\h{j}}} 
     -\Gamma^i_{31} \Gamma^{{\h{i}}'}_{2{\h{j}}} 
    )
  +\Gamma^{{\h{i}}'}_{i{\h{j}}} (
       \Gamma^{\h{i}}_{2{\h{i}}'} \Gamma^i_{31}
      -\Gamma^{\h{i}}_{1{\h{i}}'} \Gamma^i_{23}
    ) \, .
\label{eq:r123}
\end{align}
The covariant derivatives 
$R^{\h{i}}{}_{{\h{j}}23;1}$ and $R^{\h{i}}{}_{{\h{j}}31;2}$ are obtained by replacing
$123 \to 231$ and $123 \to 312$ in Eq.\,(\ref{eq:r123}).

It is now almost straightforward to see the identity 
(\ref{eq:bianchi-identity}). For an example, we see
in Eq.\,(\ref{eq:r123}), the second derivative terms of
Affine connection 
\begin{equation}
\Gamma^{\h{i}}_{2{\h{j}}, 13} - \Gamma^{\h{i}}_{1{\h{j}}, 23}
\end{equation}
are contained in $R^{\h{i}}{}_{{\h{j}}12;3}$.
Combined with the contributions from $R^{\h{i}}{}_{{\h{j}}23;1}$ and $R^{\h{i}}{}_{{\h{j}}31;2}$,
we see these second derivative terms disapper in 
$R^{\h{i}}{}_{{\h{j}}12;3}+R^{\h{i}}{}_{{\h{j}}31;2}+R^{\h{i}}{}_{{\h{j}}23;1}$, 
\begin{align}
&(\Gamma^{\h{i}}_{2{\h{j}}, 31} - \Gamma^{\h{i}}_{1{\h{j}}, 23})
\nn\\
&\qquad
+(\Gamma^{\h{i}}_{3{\h{j}}, 12} - \Gamma^{\h{i}}_{2{\h{j}}, 31})
\nn\\
&\qquad \qquad
+(\Gamma^{\h{i}}_{1{\h{j}}, 23} - \Gamma^{\h{i}}_{3{\h{j}}, 12})
= 0 \, .  
\end{align}

Note also
\begin{align}
 -R_{{\h{i}}{\h{j}}^* 12; 3} = 
  R_{{\h{j}}^* {\h{i}} 12; 3}
  &= (g_{{\h{k}}{\h{j}}^*} R^{\h{k}}{}_{{\h{i}}12})_{;3}
  \nonumber\\
  &= g_{{\h{k}}{\h{j}}^*; 3} R^{\h{k}}{}_{{\h{i}}12} 
    +g_{{\h{k}}{\h{j}}^*} R^{\h{k}}{}_{{\h{i}}12; 3} 
  \nonumber\\
  &= g_{{\h{k}}{\h{j}}^*} R^{\h{k}}{}_{{\h{i}}12; 3}  \, .
\end{align}
Multiplying $g_{{\h{i}}{\h{j}}^*}$ to Eq.\,(\ref{eq:bianchi-identity}), we therefore
obtain
\begin{align}
  R_{{\h{i}}{\h{j}}^* 12;3}
 +R_{{\h{i}}{\h{j}}^* 23;1}
 +R_{{\h{i}}{\h{j}}^* 31;2} = 0 \, .
\end{align}

%%%%%%%%%%%%%%%
\bibliography{gheft-ref} 

\end{document}